\pdfoutput=1
\documentclass[twocolumn]{article}

\usepackage[latin9]{inputenc}
\usepackage{geometry}
\usepackage{array}
\usepackage{rotating}
\usepackage{float}
\usepackage{units}
\usepackage{mathtools}
\usepackage{multirow}
\usepackage{amsmath}
\usepackage{amssymb}
\usepackage{graphicx}
\usepackage{esint}
\usepackage[font={small}]{caption}
\usepackage{array}
\usepackage{multirow}
\usepackage{algorithm}
\usepackage{algpseudocode}
\usepackage{fancyref}
\usepackage{appendix}
\usepackage{gensymb}

\usepackage[unicode=true,pdfusetitle,
 bookmarks=true,bookmarksnumbered=false,bookmarksopen=false,
 breaklinks=false,pdfborder={0 0 1},backref=false,colorlinks=false]
 {hyperref}
\PassOptionsToPackage{hyphens}{url}\usepackage{hyperref}
 
\usepackage[numbers]{natbib}
\makeatletter

\providecommand{\tabularnewline}{\\}
\floatstyle{ruled}
\newfloat{algorithm}{tbp}{loa}
\providecommand{\algorithmname}{Algorithm}
\floatname{algorithm}{\protect\algorithmname}

\@ifundefined{date}{}{\date{}}

\providecommand{\tabularnewline}{\\}
\floatstyle{ruled}
\newfloat{algorithm}{tbp}{loa}
\providecommand{\algorithmname}{Algorithm}
\floatname{algorithm}{\protect\algorithmname}

\numberwithin{equation}{section}
\numberwithin{figure}{section}

\author{Rui Vieira\\
\small School of Mathematics and Statistics\\
\small Newcastle University, UK\\
\small \texttt{r.vieira2@ncl.ac.uk}\\
\and
Darren J. Wilkinson\\
\small School of Mathematics and Statistics\\
\small Newcastle University, UK\\
\small \texttt{darren.wilkinson@ncl.ac.uk}
}

\@ifundefined{showcaptionsetup}{}{%
 \PassOptionsToPackage{caption=false}{subfig}}
\usepackage{subfig}
\makeatother

\begin{document}

\title{Online state and parameter estimation in Dynamic Generalised Linear
Models}
\maketitle
\begin{abstract}
Inference for streaming time-series is tightly coupled with the problem
of Bayesian on-line state and parameter inference. In this paper we
will introduce Dynamic Generalised Linear Models, the class of models
often chosen to model continuous and discrete time-series data. We
will look at three different approaches which allow on-line estimation
and analyse the results when applied to different real world datasets
related to inference for streaming data. Sufficient statistics based
methods delay known problems, such as particle impoverishment, especially
when applied to long running time-series, while providing reasonable
parameter estimations when compared to exact methods, such as Particle
Marginal Metropolis-Hastings. State and observation forecasts will
also be analysed as a performance metric. By benchmarking against
a ``gold standard'' (off-line) method, we can better understand
the performance of on-line methods in challenging real-world scenarios.
\end{abstract}

\section{Introduction}

With the modern ubiquity of large streaming datasets comes the requirement
of robust real-time inference. A multitude of different data sources,
such as Internet of Things (IoT) devices, server, network and sensor
metrics all exhibiting particular patterns and observation types,
also increase the demand for flexible and computationally cheap solutions.

Some typical analyses performed on such streaming time-series are
forecasting, anomaly detection and seasonal decomposition in order
to perform statistically-based decisions typically under tight time
constraints.

As standard off-line methods, such as Markov Chain Monte Carlo (MCMC),
are not normally suitable when taking into account such constraints,
we analyse in this paper alternatives such as Sequential Monte Carlo
(SMC). Although SMC is well studied in the scientific literature and
quite prevalent in academic research in the last decade, modern analytics
platforms typically still resort to less powerful methods (such as moving averages).
When coupled with Dynamic Generalised Linear Models (DGLMs), which
allow us to specify complex, non-linear time-series patterns, this
enables performing real-time Bayesian estimations in state-space models.

Inference on streaming time-series is tightly coupled with the problem
of Bayesian on-line state and parameter inference. In this paper we
will perform a comprehensive review of some well established methods
for SMC for DGLMs applied to three distinct datasets. We will start
by first introducing the DGLM, the class of state space models chosen
for our data (Section~\ref{subsec:Dynamic-Generalised-Linear}).

We will then look in Section~\ref{sec:Sequential-Monte-Carlo} at the
fundamentals of SMC and in Section~\ref{sec:State-and-Parameter} we
will look at three algorithms which allow us to perform on-line estimation.
Finally in Section~\ref{sec:Results} we will look at applications
and analyse the results when applied to different real world datasets.
We will also focus on topics which are directly relevant to the main
application area which we approach, streaming time-series, such as
the choice of resampler and the accumulation of Monte Carlo errors
in long running series.

\subsection{Dynamic Generalised Linear Models}\label{subsec:Dynamic-Generalised-Linear}

\begin{figure}
\centering
\includegraphics[width=0.75\columnwidth]{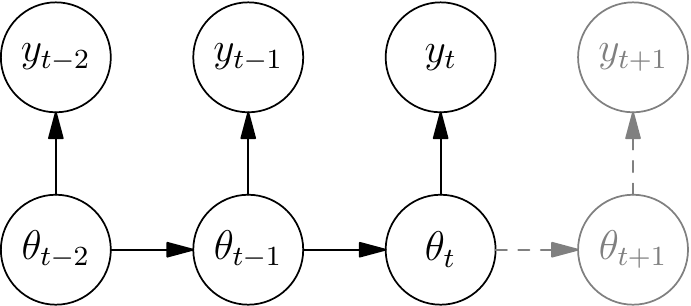} 
\caption{State-Space Model\label{fig:State-Space-Model}}
\end{figure}

To model the data we chose the Dynamic Generalised Linear Model (DGLM)
\cite{west1989Bayfordynmod}, a specific instance of the more general
class of State-Space Models (SSM), illustrated in Figure~\ref{fig:State-Space-Model},
where we have the following relations 
\begin{align}
y_{t}|\boldsymbol{\theta}_{t},\Phi & \sim f\left(y_{t}|\boldsymbol{\theta}_{t},\Phi\right)\label{eq:measurement_model}\\
\boldsymbol{\theta}_{t}|\boldsymbol{\theta}_{t-1},\Phi & \sim g\left(\boldsymbol{\theta}_{t}|\boldsymbol{\theta}_{t-1},\Phi\right).\label{eq:system_model}
\end{align}

Usually~\eqref{eq:measurement_model} is referred to as the \emph{observation
model} and~\eqref{eq:system_model} as the \emph{system model}. We
consider the discrete time case with $t\in\mathcal{\mathbb{N}}$,
the state vector $\boldsymbol{\theta}_{t}\in\mathbb{R}^{m}$ and $\Phi$
as the set of parameters for this model. The sequence of state vectors
$\Theta_{t}$ is a Markov Chain (conditional on $\Phi$) with transition
density $g$, such that 
\[
\Theta_{t}|\left\lbrace \Theta_{t-1}=\boldsymbol{\theta}_{t-1}\right\rbrace \sim g\left(\cdot|\boldsymbol{\theta}_{t-1}\right)
\]
and the sequence of observations $\mathcal{D}_{t}=\left\{ y_{1},\dots,y_{t}\right\} $
is the output of $\Theta_{t}$ such that $Y_{t}|\left\lbrace \Theta_{t}=\boldsymbol{\theta}_{t}\right\rbrace \sim f\left(\cdot|\boldsymbol{\theta}_{t}\right)$.

The second component, the \textit{system model}~\eqref{eq:system_model},
defined by the function $g:\mathbb{R}^{m}\mapsto\mathbb{R}^{m}$ can
be non-linear and specifically in DGLMs will be a linear Gaussian update of the form 
\begin{align}
\boldsymbol{\theta}_{t}|\boldsymbol{\theta}_{t-1},\Phi & \sim p\left(\boldsymbol{\theta}_{t}|\boldsymbol{\theta}_{t-1},\Phi\right)\label{eq:DGLM_state_model}\\
 &\stackrel{\text{\tiny DGLM}}{=}\mathcal{N}\left(\mathsf{G}_{t}\boldsymbol{\theta}_{t-1},\mathsf{W}\right)\label{eq:DGLM_state_model_normal}
\end{align}

where the initial state is assumed to be distributed according to
a normal prior, $\boldsymbol{\theta}_{0}|\boldsymbol{m}_{0},\mathsf{C}_{0}\sim\mathcal{N}\left(\boldsymbol{m}_{0},\mathsf{C}_{0}\right)$.
In DGLMs the \textit{\emph{observation model, characterised by the
density $f:\mathbb{R}^{m}\mapsto\mathbb{R}^{n}$,}} follows an exponential
family distribution in the canonical form of~\eqref{eq:canonical_form}
\begin{align}
y_{t} & \sim p\left(y_{t}|\eta_{t}\right)\label{eq:DGLM_observation_model}\\
 & =\exp\left\{ \frac{z\left(y_{t}\right)\eta_{t}-b\left(\eta_{t}\right)}{a\left(\phi_{t}\right)}+c\left(y_{t},\phi_{t}\right)\right\}\label{eq:canonical_form}.
\end{align}
In the literature $\eta_{t}$ is usually called the \emph{natural
parameter} and $\phi_{t}$ the \emph{dispersion parameter. }We consider
\emph{$a\left(\cdot\right)$} to be twice differentiable in $\eta_{t}$.

We will consider throughout the case where $y_t \in \mathbb{R}$, the continuous univariate case or $y_t \in \mathbb{N}$, the discrete univariate case.

Furthermore, the state vector $\boldsymbol{\theta}_{t}$ is related
to $\eta_{t}$ by a link function $L\left(\cdot\right)$, such that
\[
\eta_{t}=L\left(\mathsf{F}_{t}^{T}\boldsymbol{\theta}_{t}\right)
\]

which will be dependent on the specific distribution used in~\eqref{eq:DGLM_observation_model}.

The factors $\mathsf{F}_{t}$ and $\mathsf{G}_{t}$ are respectively
the \emph{observation} and \emph{system matrices. }They allow us to
specify the structure of our time series. These factors might represent
a \emph{locally constant model}, where the states will represent an
underlying mean, a \emph{locally linear model}, where the states represent
a mean and a trend, or a purely seasonal model, where each component
of the state will represent a seasonality component. A specific way
of representing seasonality is the \emph{reduced form Fourier seasonality}.
Here, we represent cyclical components by a composition of harmonics.
These matrices can vary in time but for the remainder of this text
we will consider them static and known, that is $\mathsf{F}_{t}=\mathsf{F}$
and $\mathsf{G}_{t}=\mathsf{G}$.

It is clear from the above definitions that this class of models possesses
Markovian properties, that is, denoting the sequence of observations
$y_{1:t-1}$ as $\mathcal{D}_{t-1}$: 
\begin{equation}
p\left(\boldsymbol{\theta}_{t}|\boldsymbol{\theta}_{1:t-1},\mathcal{D}_{t-1}\right)=p\left(\boldsymbol{\theta}_{t}|\boldsymbol{\theta}_{t-1}\right).\label{eq:markovian_state}
\end{equation}

DGLMs are a flexible and elegant tool to model streaming data, since
they can represent discrete and continuous data by appropriate selection
of the observation model, as well as providing the means to express
complex time-series behaviour by composing simpler ones. In the remainder
of this paper we will refer to a specific DGLM by classifying it according to the observation model, as detailed below.

\paragraph{Normal DLM}

\label{subsec:Normal-DLM}

A special case of the DGLM is the Normal DLM where the observation
model also consists of a normal distribution, 
\begin{equation}
y_{t}|\boldsymbol{\theta}_{t},\Phi\sim\mathcal{N}\left(\mathsf{F}^{T}\boldsymbol{\theta}_{t},\mathsf{V}\right),\label{eq:normal_dlm_observational}
\end{equation}

where the state model is~\eqref{eq:DGLM_state_model_normal}. In
this case, an analytical solution for the filtering problem exists,
namely the Kalman filter (KF)~\cite{Kalman1960}. However, Kalman
filtering allows solely for state estimation and not, by itself, parameter
estimation.

\paragraph{Poisson DLM}

\label{subsec:Poisson-DLM}

The Poisson DLM is another instance of a DGLM, where the observation
model follows a Poisson distribution
\begin{align*}
y_{t}|\lambda_{t} & \sim\text{Po}\left(\lambda_{t}\right)\text{, where}\\
\log\left(\lambda_{t}\right) & =\mathsf{F}^{T}\boldsymbol{\theta}_{t}.
\end{align*}

\paragraph{Binomial DLM}

In the presence of binary data the Binomial DLM can be used 
\begin{align}
y_{t}|\lambda_{t} & \sim\text{Binom}\left(n,\lambda_{t}\right)\text{, where}\label{eq:binomial_dlm_observational}\\
\text{logit}\left(\lambda_{t}\right) & =\mathsf{F}^{T}\boldsymbol{\theta}_{t}.\nonumber 
\end{align}

\section{Sequential Monte Carlo}\label{sec:Sequential-Monte-Carlo}

In order to perform inference in DGLMs, the main objective is to estimate
the unobserved sequence of states $\boldsymbol{\theta}_{\ensuremath{0:t}}=\left\{ \boldsymbol{\theta}_{0},\dots,\boldsymbol{\theta}_{t}\right\} $
and the parameter set $\Phi=\left\{ \Phi_{1},\dots,\Phi_{n}\right\} $
given the observed data, $\mathcal{D}_{t}=\left\{ y_{1},\dots,y_{t}\right\} $.
That is, we are trying to estimate the joint density 
\begin{equation}
p\left(\boldsymbol{\theta}_{0:t},\Phi|\mathcal{D}_{t}\right)\label{eq:joint_density}
\end{equation}

We will first look at some methods to estimate the state vectors in
an on-line fashion, that is estimating $\boldsymbol{\theta}_{0:t}$
using $\mathcal{D}_{t}$ with $t=1,2,3,\dots$ , while considering
the parameters $\Phi$ known. These methods will provide the fundamental
framework from which extensions can be used to simultaneously estimate
state and parameters in Section~\ref{sec:State-and-Parameter}.

\subsection{State Estimation}

Assuming the set of parameters $\Phi$ to be known, in DGLMs the problem
of estimating the unobserved states $\boldsymbol{\theta}_{0:n}$ can
be expressed as 
\begin{equation}
p\left(\boldsymbol{\theta}_{0:t}|\mathcal{D}_{t}\right)=\frac{p\left(\boldsymbol{\theta}_{0:t},\mathcal{D}_{t}\right)}{p\left(\mathcal{D}_{t}\right)},\label{eq:state_posterior_par_fixed}
\end{equation}
where 
\begin{align}
p\left(\boldsymbol{\theta}_{0:t},\mathcal{D}_{t}\right) & =p\left(\mathcal{D}_{t}|\boldsymbol{\theta}_{0:t}\right)p\left(\boldsymbol{\theta}_{0:t}\right),\label{eq:state_obs_joint}\\
p\left(\mathcal{D}_{t}\right) & =\int p\left(\boldsymbol{\theta}_{0:t},\mathcal{D}_{t}\right)d\boldsymbol{\theta}_{0:t}.\nonumber 
\end{align}

The Markovian nature of the DGLMs can, however, be exploited to provide
a recursive formulation for the state estimation in~\eqref{eq:state_posterior_par_fixed}.
This is crucial in allowing on-line inference in DGLMs, since it provides
us with a tool to perform computations for each time step $t$ separately
from the previous time steps. Considering $\mathcal{D}_{t-1}=y_{1:t-1}$,
the state posterior can then be expressed as a recursive update: 
\[
p\left(\boldsymbol{\theta}_{0:t}|\mathcal{D}_{t}\right)=p\left(\boldsymbol{\theta}_{0:t-1}|\mathcal{D}_{t-1}\right)\frac{p\left(\boldsymbol{\theta}_{t}|\boldsymbol{\theta}_{t-1}\right)p\left(y_{t}|\boldsymbol{\theta}_{t}\right)}{p\left(y_{t}|\mathcal{D}_{t-1}\right)}
\]
where $p\left(y_{t}|\mathcal{D}_{t-1}\right)$ is a normalising constant.
The joint posterior can then be expressed recursively as 

\begin{equation}
p\left(\boldsymbol{\theta}_{0:t},\mathcal{D}_{t}\right)\propto p\left(\boldsymbol{\theta}_{0:t-1},\mathcal{D}_{t-1}\right)
p\left(y_{t}|\boldsymbol{\theta}_{t}\right)
p\left(\boldsymbol{\theta}_{t}|\boldsymbol{\theta}_{t-1}\right)
\end{equation}

However, in order to perform on-line state estimation, we need to perform
the estimation as the observations appear, \emph{i.e.} we need to
estimate the \emph{current state} (conditional on the observations).
This is usually referred in the literature as \emph{Bayesian filtering}
and targets the state's marginal posterior 
\begin{equation}
p\left(\boldsymbol{\theta}_{t}|\mathcal{D}_{t}\right).\label{eq:bayesian_filtering}
\end{equation}
The \emph{filtering} method can be divided into two separate stages,
the \emph{prediction} and the \emph{update} steps. In the prediction
step we calculate the predictive state density given the observations
up to time $t-1$, this is 
\[
p\left(\boldsymbol{\theta}_{t}|\mathcal{D}_{t-1}\right)=\int p\left(\boldsymbol{\theta}_{t}|\boldsymbol{\theta}_{t-1}\right)p\left(\boldsymbol{\theta}_{t-1}|\mathcal{D}_{t-1}\right)d\boldsymbol{\theta}_{t-1}.
\]

State estimation, also commonly referred as \textit{filtering} aims
at determining the density 
\begin{align}
p\left(\boldsymbol{\theta}_{t}|\mathcal{D}_{t}\right) & =\frac{p\left(y_{t}|\boldsymbol{\theta}_{t}\right)p\left(\boldsymbol{\theta}_{t}|\mathcal{D}_{t-1}\right)}{p\left(y_{t}|\mathcal{D}_{t-1}\right)}.\label{eq:filtering_density}
\end{align}

\subsection{Importance Sampling\label{subsec:Importance-Sampling}}

In this context, state estimation can be viewed as the calculation
of arbitrary expectations of the form 
\begin{equation}
\bar{g}=\text{E}\left[g\left(\boldsymbol{\theta}\right)|\mathcal{D}_{t}\right]=\int g\left(\boldsymbol{\theta}\right)p\left(\boldsymbol{\theta}|\mathcal{D}_{t}\right)d\boldsymbol{\theta}\label{eq:posterior_expection}
\end{equation}

Here, $g\left(\cdot\right)$ is an arbitrary function and $p\left(\boldsymbol{\theta}|\mathcal{D}_{t}\right)$
is the state's posterior probability density given the entirety of
the data $\mathcal{D}_{t}=y_{1:t}=\left\lbrace y_{1},\dots,y_{t}\right\rbrace $.
This distribution may be highly complex and with high dimensionality.
The problem with the integral in~\eqref{eq:posterior_expection}
is that typically we cannot solve it analytically. In such cases we
can employ a Monte Carlo approximation by producing samples $s^{(i)}$
from a support distribution, with corresponding weights $w^{(i)}$,
where $\sum_{i=1}^{N}w^{(i)}=1$, such that

\begin{equation}
\sum_{i=1}^{N}g\left(s^{(i)}\right)w^{(i)}\simeq\int g\left(\boldsymbol{\theta}\right)p\left(\boldsymbol{\theta}|\mathcal{D}_{t}\right)d\boldsymbol{\theta},
\end{equation}

an approximation which will converge in probability when $N\rightarrow\infty$.

In the case of Importance Sampling (IS) we assume an \emph{importance
density} $\pi$, having a larger support than $p$, from which we
can easily sample, that is

\begin{equation}
\boldsymbol{\theta}^{(i)}\sim\pi\left(\boldsymbol{\theta}_{0:n}|\mathcal{D}_{t}\right).\label{eq:proposal_distribution}
\end{equation}

In this case, the weights will correspond to $w^{(i)}=A\ p\left(s^{(i)}\right)/\pi\left(s^{(i)}\right)$
with $A^{-1}=\sum_{i=1}^{N}p\left(s^{(i)}\right)/\pi\left(s^{(i)}\right)$
\cite{Cliffordy}, accounting for the difference between the target
and importance densities.

If we consider our target density $p\left(\boldsymbol{\theta}\right)$ 
and our proposal draws $\boldsymbol{\theta}^{\prime}\sim\pi\left(\boldsymbol{\theta}\right)$,
it follows that, starting from~\eqref{eq:posterior_expection}:

\[
\bar{g}=\int g\left(\boldsymbol{\theta}\right)\tilde{W}\left(\boldsymbol{\theta}\right)\pi\left(\boldsymbol{\theta}\right)d\boldsymbol{\theta}
\]

Here, $\tilde{W}\left(\boldsymbol{\theta}\right)$ is the \emph{unnormalised
importance weight }and is given by 
\begin{equation}
\tilde{W}\left(\boldsymbol{\theta}\right)=\frac{p\left(\boldsymbol{\theta}\right)}{\pi\left(\boldsymbol{\theta}\right)}.\label{eq:unnormalised_importance_weight}
\end{equation}

Given~\eqref{eq:unnormalised_importance_weight}, we can then approximate
our expectation in~\eqref{eq:posterior_expection} by

\begin{align*}
\bar{g} & \approx\frac{1}{N}\sum_{i=1}^{N}\frac{p\left(\boldsymbol{\theta}^{(i)}|\mathcal{D}_{t}\right)}{\pi\left(\boldsymbol{\theta}^{(i)}|\mathcal{D}_{t}\right)}g\left(\boldsymbol{\theta}^{(i)}\right)\\
 & =\sum_{i=1}^{N}\tilde{w}^{(i)}g\left(\boldsymbol{\theta}^{(i)}\right)
\end{align*}

Here the weights are defined by 
\begin{equation}
\tilde{w}^{(i)}=\frac{1}{N}\frac{p\left(\boldsymbol{\theta}^{(i)}|\mathcal{D}_{t}\right)}{\pi\left(\boldsymbol{\theta}^{(i)}|\mathcal{D}_{t}\right)}.\label{eq:importance_weights}
\end{equation}

However, in this case we must be able to evaluate $p\left(\boldsymbol{\theta}^{(i)}|\mathcal{D}_{t}\right)$.
Recalling the posterior density in~\eqref{eq:filtering_density}
we can see that the denominator will not be easily calculated. However
if we write the expectation in \eqref{eq:posterior_expection} as
\[
\text{E}\left[g\left(\boldsymbol{\theta}_{t}\right)|\mathcal{D}_{t}\right]\approx\sum_{i=1}^{N}w_{t}^{(i)}g\left(\boldsymbol{\theta}_{t}^{(i)}\right),
\]

this approximation, evaluated at each time point $t=1,\dots,n$, is
defined as the sequential approximation. Sequential importance sampling
works then by approximating the target density's marginal, such that
\[
p\left(\boldsymbol{\theta}_{t}|y_{t}\right)\approx\sum_{i=1}^{N}w_{t}^{(i)}\delta\left(\boldsymbol{\theta}_{t}-\boldsymbol{\theta}_{t}^{(i)}\right)
\]

where $\delta$ is the Dirac $\delta$ function.

Using the Markovian properties of DGLMs as mentioned in Section~\ref{subsec:Dynamic-Generalised-Linear}
we can then write a recursion for the full posterior: 
\begin{equation}
p\left(\boldsymbol{\theta}_{0:t}|\mathcal{D}_{t}\right)\propto p\left(y_{t}|\boldsymbol{\theta}_{t}\right)p\left(\boldsymbol{\theta}_{t}|\boldsymbol{\theta}_{t-1}\right)p\left(\boldsymbol{\theta}_{0:t-1}|\mathcal{D}_{t-1}\right)\label{eq:recursive_decomposition-1}
\end{equation}

If we replace the decomposition \eqref{eq:recursive_decomposition-1}
in the importance weight definition in \eqref{eq:importance_weights}
we have 
\[
w_{t}^{(i)}\propto\frac{p\left(y_{t}|\boldsymbol{\theta}_{t}^{(i)}\right)p\left(\boldsymbol{\theta}_{t}^{(i)}|\boldsymbol{\theta}_{t-1}^{(i)}\right)p\left(\boldsymbol{\theta}_{0:t-1}^{(i)}|\mathcal{D}_{t-1}\right)}{\pi\left(\boldsymbol{\theta}_{0:t}^{(i)}|\mathcal{D}_{t}\right)}
\]

In an analogous way, if we decompose the importance distribution in
a recursive, such that 
\[
\pi\left(\boldsymbol{\theta}_{0:t}|\mathcal{D}_{t}\right)=\pi\left(\boldsymbol{\theta}_{t}|\boldsymbol{\theta}_{0:t-1},\mathcal{D}_{t}\right)\pi\left(\boldsymbol{\theta}_{0:t-1}|\mathcal{D}_{t-1}\right)
\]

and replace in the weights expression, we get 
\[
w_{t}^{(i)}\propto\frac{p\left(y_{t}|\boldsymbol{\theta}_{t}^{(i)}\right)p\left(\boldsymbol{\theta}_{t}^{(i)}|\boldsymbol{\theta}_{t-1}^{(i)}\right)p\left(\boldsymbol{\theta}_{0:t-1}^{(i)}|\mathcal{D}_{t-1}\right)}{\pi\left(\boldsymbol{\theta}_{t}^{(i)}|\boldsymbol{\theta}_{0:t-1}^{(i)},\mathcal{D}_{t}\right)\pi\left(\boldsymbol{\theta}_{0:t-1}^{(i)}|\mathcal{D}_{t-1}\right)}
\]

If we consider the time step at $t-1$, samples can be drawn from
\[
\boldsymbol{\theta}_{0:t-1}^{(i)}\sim\pi\left(\boldsymbol{\theta}_{0:t-1}|\mathcal{D}_{t-1}\right)
\]

and the weights $w_{t-1}^{(i)}$ calculated. Samples $\boldsymbol{\theta}_{\ensuremath{0:t}}^{(i)}$
from the importance distribution $\pi\left(\boldsymbol{\theta}_{0:t}|\mathcal{D}_{t}\right)$
can then be drawn at step $t$ as 
\begin{equation}
\boldsymbol{\theta}_{t}^{(i)}\sim\pi\left(\boldsymbol{\theta}_{t}|\boldsymbol{\theta}_{0:t-1}^{(i)},\mathcal{D}_{t}\right)\label{eq:optimal_importance_density}
\end{equation}
 and the importance weights from the previous step are proportional
to the last term in the previous weights 
\[
w_{t-1}^{(i)}\propto\frac{p\left(\boldsymbol{\theta}_{0:t-1}^{(i)}|\mathcal{D}_{t-1}\right)}{\pi\left(\boldsymbol{\theta}_{0:t-1}^{(i)}|\mathcal{D}_{t-1}\right)}
\]

This allows the weight calculation to satisfy the recursion 
\begin{equation}
w_{t}^{(i)}\propto\frac{p\left(y_{t}|\boldsymbol{\theta}_{t}^{(i)}\right)p\left(\boldsymbol{\theta}_{t}^{(i)}|\boldsymbol{\theta}_{t-1}^{(i)}\right)}{\pi\left(\boldsymbol{\theta}_{t}^{(i)}|\boldsymbol{\theta}_{0:t-1}^{(i)},\mathcal{D}_{t}\right)}w_{t-1}^{(i)}.\label{eq:importance_weight}
\end{equation}

and the discrete approximation of $p\left(\cdot\right)$ is then 
\begin{align}
\hat{p}\left(\boldsymbol{\theta}_{0:t}|\mathcal{D}_{t}\right) & \triangleq\sum_{i=1}^{N_{p}}w_{t}^{(i)}\delta\left(\boldsymbol{\theta}_{0:t}-\boldsymbol{\theta}_{0:t}^{(i)}\right),\label{eq:discrete_approx}\\
\sum_{i=1}^{N_{p}}w_{t}^{(i)} & =1.
\end{align}

Here, $w_{t}^{(i)}$ is the unnormalised \emph{importance weight},
which accounts for the differences between the target distribution
\eqref{eq:filtering_density} and our proposal density~\eqref{eq:proposal_distribution},
given by 
\[
w_{t}^{(i)}\propto\frac{p\left(\boldsymbol{\theta}^{(i)}\right)}{\pi\left(\boldsymbol{\theta}^{(i)}\right)},\qquad\tilde{w}_{t}^{(i)}=\frac{w_{t}^{(i)}}{\sum_{j=1}^{N_{p}}w_{t}^{(j)}}
\]

Ideally the importance density should be chosen as to minimise $\text{Var}\left(w_{t}^{(i)}\right)$.
According to~\cite{Doucet2000} a common choice is the
prior itself, which in the DGLM case is given by~\eqref{eq:DGLM_state_model_normal},
although this is typically far from optimal.

The convergence of this method is guaranteed by the central limit
theorem and the error term is $\mathcal{O}\left(N^{-1/2}\right)$
regardless of the dimensionality of $\boldsymbol{\theta}$~\cite{Liu2008}.

The simplest application of IS for sequential state estimation
is Sequential Importance Sampling (SIS).
If the proposal can be written in the form of~\eqref{eq:recursive_decomposition-1}
importance sampling can be calculated in a sequential manner. Algorithm~\ref{alg:Sequential-Importance-Sampling}
presents a generic method to calculate $p\left(\boldsymbol{\theta}_{0:t}|\mathcal{D}_{t}\right)$
using $N_{p}$ particles.

\begin{algorithm}
\begin{description}
\item [{\caption{Sequential Importance Sampling\label{alg:Sequential-Importance-Sampling}}
}]~
\item [{initialisation}] ($t=0$):
\begin{description}
\item [{for}] $i\leftarrow1$ to $N_{p}$
\begin{description}
\item [{Draw}] $\boldsymbol{\theta}_{0}^{(i)}\sim\mathcal{N}\left(\boldsymbol{m}_{0},\mathsf{C}_{0}\right)$ 
\item [{Set}] $w_{0}^{(i)}=\frac{1}{N_{p}}$
\end{description}
\end{description}
\item [{for}] $t\leftarrow1$ to $k$
\begin{description}
\item [{for}] $i\leftarrow1$ to $N_{p}$
\begin{description}
\item [{Draw}] $\boldsymbol{\theta}_{t}^{(i)}\sim\pi\left(\boldsymbol{\theta}_{t}|\boldsymbol{\theta}_{0:t-1},\mathcal{D}_{t}\right)$
\item [{Calculate}] the \emph{importance weight} using \eqref{eq:importance_weight}.
\item [{Normalise}] weights $w_{t}^{(i)}=\frac{\tilde{w}_{t}^{(i)}}{\sum_{i=1}^{N_{p}}\tilde{w}_{t}^{(i)}}$
\end{description}
\end{description}
\end{description}
\end{algorithm}
The SIS filter will not be included in the subsequent analysis, however
it represents an important framework from which other filters can
be built.

\subsection{Resampling\label{sec:Resampling}}

\begin{figure}
\centering
\includegraphics[width=1\columnwidth]{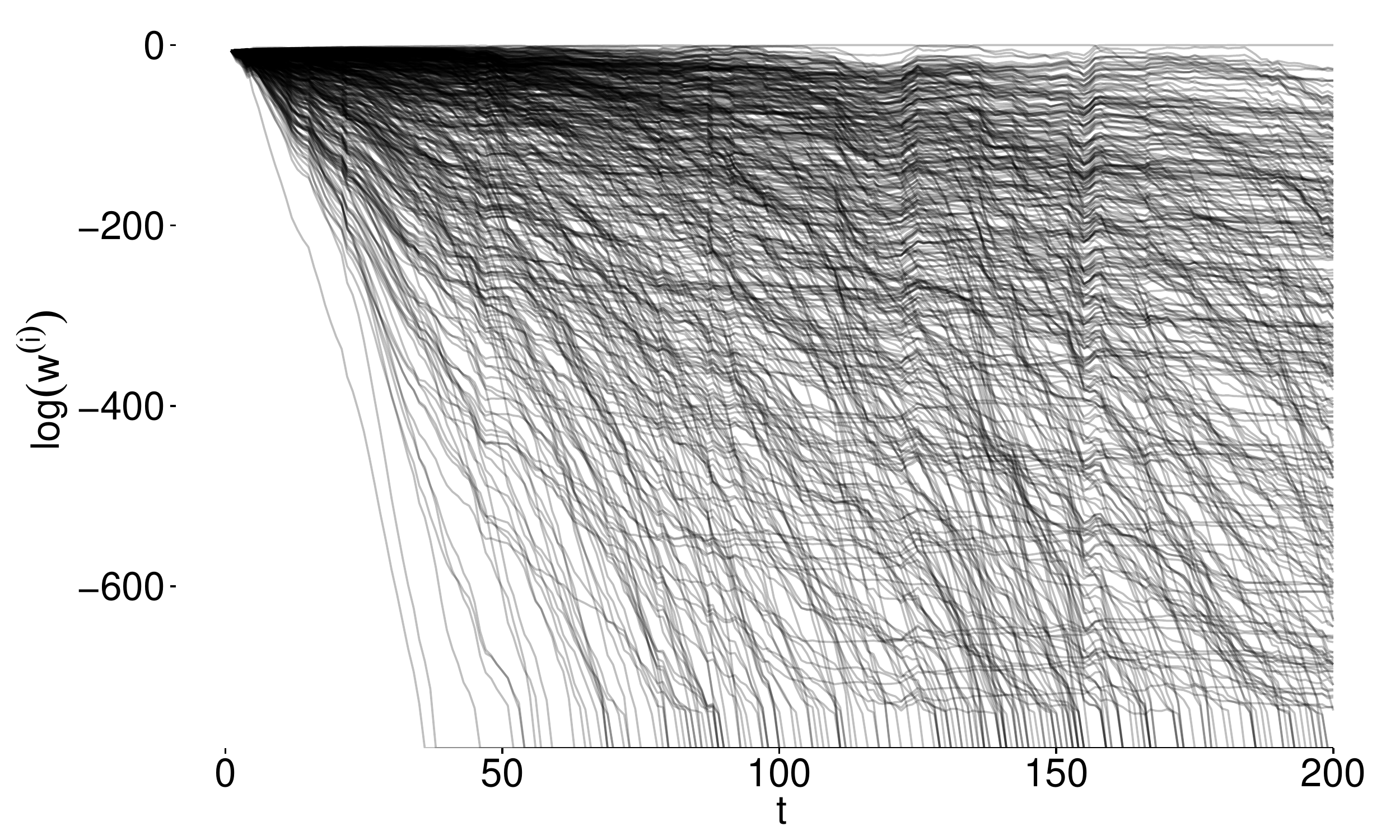}\caption{SIS particle log-weights in a local level Normal DLM for 250 iterations
using simulated data. Initial weights are $\nicefrac{1}{N_{p}}$,
$N_{p}=500$.\label{fig:SIS-particle-log-weights}}
\end{figure}

One of the drawbacks of SIS~\cite{MacEachern1999} is the potential
inaccuracy of the estimation due to the large variance of the importance
weights. Additionally, eventually all except a few particles will have a negligible weight.
This is problematic both from the point of view of accuracy (since
we are constructing the approximation from a few samples) and performance
(computations for particles not contributing are still being executed).
This is the well-known~\cite{Gordon1993a} problem of \emph{weight
degeneracy}, where after a few steps the majority of the weights will
eventually be close to zero (as represented in Figure~\ref{fig:SIS-particle-log-weights}).
The distribution~\eqref{eq:discrete_approx} will then be eventually
approximated by a very small number of particles, becoming inaccurate
and with a large posterior variance. By employing resampling, \emph{i.e.}
choosing particles using their weights as a criterion, the degeneracy
problem can be somewhat mitigated.

The resampling stage, while usually being independent from the state
vector's dimension, is a crucial step regarding the performance of
SMC implementations, impacting both the posterior variance and the
computational speed. It is only natural, then, that research into
resampling methods is an active area with a large variety of available
implementations.

In this paper we choose three of the most common methods (multinomial~\cite{Gordon1993a},
stratified~\cite{Doucet2001a} and systematic~\cite{Kitagawa1996})
to be quantitatively analysed in Section~\ref{subsec:Resampler-benchmarks}.
These selected methods belong to the category of \textit{single-distribution},
\textit{unbiased resamplers}. This \textit{unbiasedness} means that,
for a certain particle $i$, we expect it to be sampled $N_{t}^{(i)}$
times proportional to its weight $w_{t}^{(i)}$. That is

\[
\text{E}\left[N_{t}^{(i)}|w_{t}^{(i)}\right]=Nw_{t}^{(i)}
\]

We also assume that the weights available at each timepoint, prior
to resampling, are normalised, this is $\sum_{i=1}^{N_{p}}w_{t}^{(i)}=1$.

\subsubsection{Resampling methods\label{subsec:Resampling-methods}}

\paragraph{Multinomial resampling\label{subsec:Multinomial-resampling}}

Multinomial resampling is possibly the most common method employed
in the literature. This method samples particles indices from a multinomial
distribution such that
\[
i^{k}\sim\mathcal{MN}\left(N_{p};w^{1},\dots,w^{N_{p}}\right).
\]

Multinomial resampling is not, however, the most efficient resampling
algorithm (as shown in~\cite{Cliffordy}) with a computational complexity
of $\mathcal{O}\left(N_{p}M\right)$.

\paragraph{Stratified resampling\label{subsec:Stratified-resampling}}

Stratified resampling \cite{Kitagawa1996,Doucet2001a} works by generating
$N_{p}$ ordered random numbers 
\[
u_{k}=\frac{\left(k-1\right)+\tilde{u}_{k}}{N_{p}},\qquad\tilde{u}_{k}\sim\mathcal{U}\left[0,1\right)
\]

and drawing the particle indices as
\begin{equation}
i_{k}=\left\{ u_{k}:\sum_{n=1}^{i-1}w_{n}\leq u_{k}\leq\sum_{n=1}^{i}w_{n}\right\} .\label{eq:stratified_indices}
\end{equation}

Stratified resampling runs in $\mathcal{O}\left(N_{p}\right)$ time.

\paragraph{Systematic resampling\label{subsec:Systematic-resampling}}

Systematic resampling \cite{Kitagawa1996} consists of sampling a
single $u_{1}\sim\mathcal{U}\left[0,\frac{1}{N_{p}}\right]$ and setting
$u_{k}=u_{1}+\frac{k-1}{N_{p}}$ for $k=1,\dots,N_{p}.$ We then draw
the particles indices, as in stratified resampling, according to \eqref{eq:stratified_indices}.

Systematic resampling runs in $\mathcal{O}\left(N_{p}\right)$ time.

It is important to note that although resampling schemes alternative
to multinomial do provide a smaller computational burden, they do
not guarantee in practice that we will always have a smaller posterior
variance~\cite{Douc2005}.

\subsubsection{Effective Sample Size}

A standard way of quantifying weight degeneracy is to estimate the
\emph{Effective Sample Size} (ESS). The ESS can be written~\cite{Liu1998,Kong1994,Liu1996}
as
\begin{align*}
ESS & =\frac{N_{p}}{1+N_{p}^{2}\text{Var}\left(w_{t}^{(i)}\right)}.
\end{align*}

From the ESS definition it is clear that this will take values between
the extremes $ESS=N_{p}$, which can be interpreted as all $N_{p}$
particles contributing equally to the density estimation, and $ESS=1$,
interpreted as a single particle contributing to the density estimation.
The usual estimate of the ESS is given by

\begin{equation}
\widehat{ESS_{t}}=\frac{1}{\sum_{i=1}^{N_{p}}\left(w_{t}^{(i)}\right)^{2}}.\label{eq:effective_sample_size}
\end{equation}

This estimate also conforms to the bounds of the ESS, that is $1\leq\widehat{ESS_{t}}\leq N_{p}$.
Using this definition of ESS, we can then express it as a ratio of
the total number of particles (\emph{e.g.}\ half the particles contributing
to the estimation would be equivalent to $\widehat{ESS_{t}}=N_{p}/2$)

\begin{figure}
\centering{}\includegraphics[width=1\columnwidth]{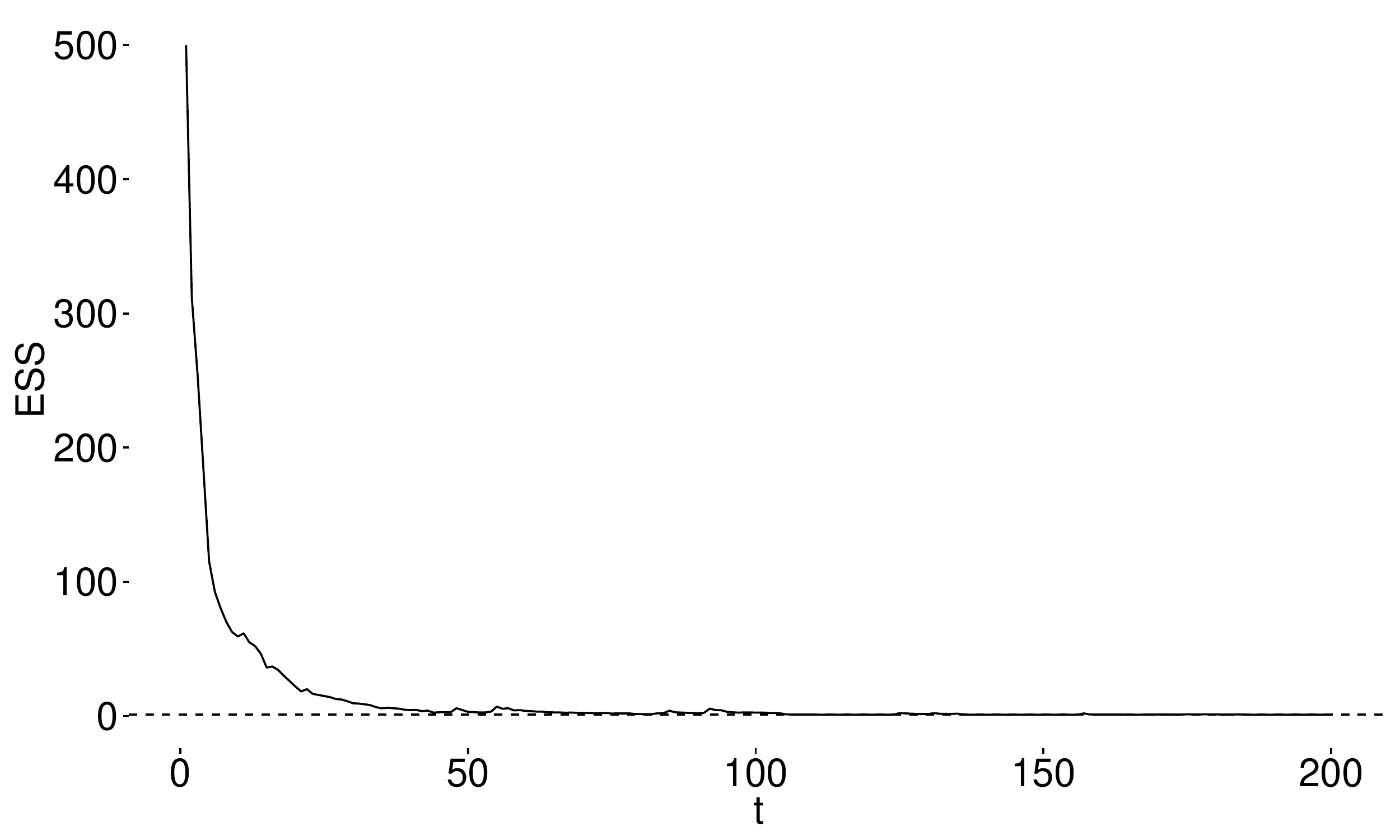}\caption{SIS (with $N_{p}=500$) $\widehat{ESS_{t}}$ in a local
level Normal DLM for 250 iterations.\label{fig:SIS_ESS}}
\end{figure}

Applying the ESS calculation to our SIS previous result, we get the result shown
in Figure~\ref{fig:SIS_ESS}.

From this plot, we can see that the ESS starts from a value of $N_{p}$,
as we have all particles initially with uniform weight of $1/N_{p}$,
decaying to the value of $\widehat{ESS_{t}}\approx1$, interpreted
as a single particle contributing to the estimation.

\subsubsection{Weight degeneracy \emph{vs.} particle impoverishment\label{subsec:Weight-degeneracy-vs.}}

It is important to note that resampling does not completely solve
the degeneracy problem, and in fact introduces a new problem, \emph{particle
impoverishment}. By discarding particles with lower weights we are
also reducing the overall particle diversity and the number of unique
particles trajectories which explore the state space. Resampling at
every time-point $t$ might be inefficient~\cite{Liu1998} and could
be done only when the ESS is below a certain pre-defined threshold
proportional to the number of particles (this was not the approach
in our estimations in Section~\ref{sec:Results} where, for consistency,
all the methods apply resampling at each time $t$).

Resampling also has an impact from the computational implementation
point of view since, where previously in SIS all particles operations
could be computed in parallel, we now have a single point of synchronisation
before proceeding to the next iteration step. There are several \cite{Murray2013,Gong2012}
resampling algorithms which allow for parallel resampling but this
was considered to be outside the scope of this paper.

\subsection{Sequential Importance Resampling}

The \emph{Sequential Importance Resampling} (SIR) method, introduced
by Gordon \emph{et al}. in~\cite{Gordon1993a} and Kitagawa \emph{et
al}.~\cite{Kitagawa1996}, tries to solve the degeneracy problem by
introducing particle resampling.

The SIR algorithm also aims at calculating a discrete approximation
of the state posterior, provided the model parameters are known.

While resampling might help solve the problem of particle degeneracy,
it introduces a different problem, discussed in Section~\ref{subsec:Weight-degeneracy-vs.},
particle impoverishment.

\begin{algorithm}
	\begin{description}
		\item [{\caption{Sequential Importance Resampling}
		}]~
		\item [{initialisation}] ($t=0$): 
		\begin{description}
			\item [{Same}] initialisation as SIS (Algorithm~\ref{alg:Sequential-Importance-Sampling})
		\end{description}
		\item [{for}] $t\leftarrow1$ to $k$
		\begin{description}
			\item [{for}] $i\leftarrow1$ to $N_{p}$
			\begin{description}
				\item [{Draw}] $\boldsymbol{\theta}_{t}^{(i)}\sim\pi\left(\boldsymbol{\theta}_{t}|\boldsymbol{\theta}_{0:t-1},\mathcal{D}_{t}\right)$
				\item [{Calculate}] the \emph{importance weight} using \eqref{eq:importance_weight}.
			\end{description}
			\item [{Normalise}] weights $w_{t}^{(i)}=\frac{\tilde{w}_{t}^{(i)}}{\sum_{i=1}^{N_{p}}\tilde{w}_{t}^{(i)}}$
				\item [{Resample}] according to $p\left(j(i)=l\right)=w_{k}$ (as
				discussed in Section~\ref{sec:Resampling}).
				\item [{for}] $i\leftarrow1$ to $N_{p}$
				\begin{description}
					\item [{Set}] $w_{k}^{(i)}=\frac{1}{N_{p}}$
				\end{description}
		\end{description}
	\end{description}
\end{algorithm}

\subsection{Auxiliary Particle Filter}\label{subsec:Auxiliary-Particle-Filter}

The Auxiliary Particle Filter (APF), first introduced by Pitt \emph{et al}. in~\cite{Pitt1999} is an extension of SIR which aims at partially solving the problem of particle degeneracy by pre-selecting particles
before propagation.

If we consider an auxiliary variable $i$ and define \cite{Pitt1999}
the target joint density to approximate as

\[
p\left(\boldsymbol{\theta}_{t},i|\mathcal{D}_{t}\right)\propto p\left(y_{t}|\boldsymbol{\theta}_{t}\right)p\left(\boldsymbol{\theta}_{t}|\boldsymbol{\theta}_{t-1}^{(i)}\right)w_{t-1}^{(i)},
\]

then, by defining $\mu_{t}^{\left(i\right)}$ as some characterisation
of $\boldsymbol{\theta}_{t}|\boldsymbol{\theta}_{t-1}^{\left(i\right)}$,
which can be the mean, mode, a sample or some other \cite{Pitt1999}
characterisation of $\boldsymbol{\theta}_{t}|\boldsymbol{\theta}_{t-1}^{\left(i\right)}$,
this joint density can be approximated by
\begin{equation}
\pi\left(\boldsymbol{\theta}_{t},i|\mathcal{D}_{t}\right)\propto p\left(y_{t}|\mu_{t}^{(i)}\right)p\left(\boldsymbol{\theta}_{t}|\boldsymbol{\theta}_{t-1}^{(i)}\right)w_{t-1}^{(i)},\label{eq:APF_proposal}
\end{equation}

with the weights proportional to

\[
w_{t}^{(j)}\propto\frac{p\left(y_{t}|\boldsymbol{\theta}_{t}^{(i)}\right)}{p\left(y_{t}|\mu_{t}^{k(i)}\right)}.
\]

The APF allows us to approximate an adapted proposal, that is taking
into account $y_{t}$, when sampling is not possible from the fully
adapted $\pi\left(\boldsymbol{\theta}_{t}|\boldsymbol{\theta}_{0:t-1},\mathcal{D}_{t}\right)$.
The SIR and the APF represent two distinct classes or algorithms,
commonly referred in the literature respectively as the~\emph{sample-resample}
and the~\emph{resample-sample} families, based on the order upon which
the particle selection is computed. Algorithm~\ref{alg:Auxiliary-Particle-Filter},
in the appendix, presents a generic method to implement
the APF.

\section{State and Parameter Estimation}\label{sec:State-and-Parameter}

The methods introduced in Section~\ref{sec:Sequential-Monte-Carlo} allow us to sequentially estimate the state. However, in real world
applications the model's parameters would also be unknown. Extensions to the previous algorithms would then be needed to simultaneously
perform state and parameter estimation, defined by \eqref{eq:joint_density}.
We will introduce additional methods which will enable us to do this
mainly using two different approaches, by either jointly estimating
the states and parameters by incorporating the parameters in the state
space (Liu and West) or by marginalising the parameters using a sufficient
statistics methodology (Storvik and Particle Learning). In the following
sections we will assume that although unknown, the parameters to be
estimated are static, otherwise they could be incorporated into the
state vector.

\subsection{Liu and West's Filter}

The Liu and West particle filter, first described in~\cite{Liu2001},
falls into the category of joint estimation, \emph{i.e}.\ by augmenting
the state-space with the parameters.

One problem of trying to estimate static parameters is that, by definition,
they will not change their value since $t=1$. When estimating the
parameters as a part of the state space, this will cause problems,
namely a degeneracy of the particles.

An initial naive approach could be to add artificial noise to the parameters.
However, this will also lead to an artificial
increase in the variance of the estimates, \emph{i.e.} the posteriors,
and is essentially equivalent to assuming the parameters are slowly
time varying.

The solution proposed by Liu and West~\cite{Liu2001} is to use a
kernel smoothing approximation, with a correction factor to account
for over-dispersion.

According to \cite{Liu2001}, we take the observation density \eqref{eq:measurement_model},
the transition \eqref{eq:system_model} and assume that at time $t+1$
we want to generate a sample from the posterior $p\left(\boldsymbol{\theta}_{t+1}|\mathcal{D}_{t+1}\right)$,
that is, from 
\[
p\left(\boldsymbol{\theta}_{t+1}|\mathcal{D}_{t+1}\right)\propto p\left(y_{t+1}|\boldsymbol{\theta}_{t+1}\right)p\left(\boldsymbol{\theta}_{t+1}|\mathcal{D}_{t}\right).
\]

We can then rewrite the update step by using the discrete approximation to 
$p\left(\boldsymbol{\theta}_{t+1}|\mathcal{D}_{t}\right)$
\[
p\left(\boldsymbol{\theta}_{t+1}|\mathcal{D}_{t+1}\right)\propto p\left(y_{t+1}|\boldsymbol{\theta}_{t+1}\right)\sum_{i=1}^{N_{p}}w_{t}^{(i)}p\left(\boldsymbol{\theta}_{t+1}|\boldsymbol{\theta}_{t}^{(i)}\right)
\]

The Liu and West (LW) filter comprises of a kernel shrinkage step
to help against the variance increase caused by the Gaussian mixture.

Considering that at time $t$ we have a SMC approximation to $p\left(\Phi_{t}|\mathcal{D}_{t}\right)$
given by the draws $\Phi_{t}^{(i)}$ with corresponding weights $w_{t}^{(i)}$,
according to~\cite{West1993}, the smoothed kernel density is given
by

\begin{equation}
p\left(\Phi|\mathcal{D}_{t}\right)\approx\sum_{i=1}^{N_{p}}w_{t}^{(i)}\mathcal{N}\left(\Phi|\boldsymbol{m}_{t}^{(i)},h^{2}V_{t}\right)\label{eq:liuwest_smooth_kernel_density}
\end{equation}

In~\eqref{eq:liuwest_smooth_kernel_density} we have a multivariate
normal distribution in the form $\mathcal{N}\left(\cdot|\boldsymbol{m},\mathsf{C}\right)$
with mean $\boldsymbol{m}$ and covariance $\mathsf{C}$. The sum
results then in a mixture of multivariate normals weighted by their
corresponding weights.

The reasoning, provided by~\cite{Liu2001}, for the shrinkage approach
is that without it, the kernel locations would be $\boldsymbol{m}_{t}^{(i)}=\Phi_{t}^{(i)}$.
This would result in an over-dispersed kernel density relative to
the posterior, since the variance of the mixture will be $\left(1+h^{2}\right)V_{t}$,
always bigger than $V_{t}$. This will lead to accumulation of dispersion,
since an over-dispersed $p\left(\Phi|\mathcal{D}_{t}\right)$ will
lead to even higher over-dispersion in $p\left(\Phi|\mathcal{D}_{t+1}\right)$
approximation.

The kernel's moments are then calculated by 
\begin{equation}
\boldsymbol{m}_{t}^{(i)}=a\Phi^{(i)}+\left(1-a\right)\overline{\Phi}
\label{eq:LW_moments}
\end{equation}

where $a=\sqrt{1-h^{2}}$ and $h>0$ is the smoothing parameter and
the variance by 
\begin{equation}
V_{t}=\sum_{i=1}^{N_{p}}\frac{\left(\Phi^{(i)}-\overline{\Phi}\right)\left(\Phi^{(i)}-\overline{\Phi}\right)^{T}}{N_{p}}
\label{eq:LW-variance}
\end{equation}

with $\overline{\Phi}=\sum_{i=1}^{N}\Phi^{(i)}/N_{p}$. With this
method, the mean $\overline{\Phi}_{t}$ is kept and having correct
variance $V_{t}$, correcting the over-dispersion. A summary for the
steps of the Liu and West filter is presented in Algorithm~\ref{alg:Liu-and-West}
in the appendix.

By propagating the parameter proposals with a MVN impulse we can apply
LW to any class of state-space models.

\subsection{Storvik Filter}

The Storvik filter, first presented in~\cite{Storvik2002a} and related
to~\cite{Fearnhead2002}, unlike Liu \& West does not incorporate
parameters in the state vector but instead works by assuming that
the posterior $p\left(\Phi|\mathcal{D}_{t},\boldsymbol{\theta}_{0:t}\right)$
depends on a set of sufficient statistics (SS) $s_{t}$ with an associated
recursive update. By performing parameter estimation based on this
set of sufficient statistics and separately from the state estimation,
the Storvik filter aims at reducing particle impoverishment while reducing
computational load due to the low-dimensionality of $s_{t}$~\cite{Storvik2002a}. The
deterministic update of $s_{t}$ will depend on the state estimates
and parameter estimates, such that $s_{t}=\mathcal{S}\left(s_{t-1},\theta_{t},\theta_{t-1},y_{t}\right)$.

According to \cite{Storvik2002a}, we use the decomposition
\begin{align}
p\left(\boldsymbol{\theta}_{0:t},\Phi|\mathcal{D}_{t}\right)= & C\cdot p\left(\boldsymbol{\theta}_{0:t-1}|\mathcal{D}_{t-1}\right)p\left(\Phi|s_{t-1}\right)\nonumber \\
 & \times p\left(\boldsymbol{\theta}_{t}|\boldsymbol{\theta}_{t-1},\Phi\right)p\left(y_{t}|\boldsymbol{\theta}_{t},\Phi\right),\label{eq:storvik_decomposition}
\end{align}

where $C=\left[p\left(y_{t}|\mathcal{D}_{t-1}\right)\right]^{-1}$
which is a constant not depending on $\boldsymbol{\theta}_{0:t}$
or $\Phi$.

Simulation from~\ref{eq:storvik_decomposition} can be performed
with the additional step that $\Phi$ also needs to be simulated.
The simplest way to simulate from~\ref{eq:storvik_decomposition}
is to draw from 
\begin{align*}
\boldsymbol{\theta}_{0:t-1} & \sim p\left(\boldsymbol{\theta}_{0:t-1}|\mathcal{D}_{t-1}\right)\\
\Phi & \sim p\left(\Phi|s_{t-1}\right)\\
\boldsymbol{\theta}_{t} & \sim p\left(\boldsymbol{\theta}_{t}|\boldsymbol{\theta}_{0:t-1},\Phi\right)
\end{align*}

and accept with probability $p\left(y_{t}|\boldsymbol{\theta}_{t},\Phi\right)$.
The steps of the Storvik filter are summarised in Algorithm~\ref{alg:Storvik's-algorithm}
in the appendix.

Although Storvik works in a \emph{sample-resample} framework, it can
also be applied within a \emph{resample-sample} framework.

\subsection{Particle Learning}

Particle Learning (PL), first introduced in \cite{Carvalhoa} employs
a similar sufficient statistics mechanism as Storvik, although within
a resample-sample framework. Unlike Storvik, where sufficient statistics
structure is used solely for estimating parameters, in PL the state
can also marginalised if a sufficient statistics structure is available
for the state. This means that prior to sampling from the proposal
distribution, we resample the current state particles and sufficient
statistics taking $y_{t+1}$ into account, using a predictive likelihood.
A general implementation for a Particle Learning is presented in 
Algorithm~\ref{alg:Particle-Learning} in the appendix.

Particle Learning promises to reduce the problem of particle impoverishment,
although in practice it does not solve the problem completely \cite{Chopin2010}.

\subsection{Sufficient Statistics\label{sec:Poisson_DLM_SS}}

An example for sufficient statistics can be given using the Poisson
DLM with a locally constant model evolution. The general model is
given by Section~\ref{subsec:Poisson-DLM}. Considering the model evolution,
where $\mathsf{F}_{t}=\mathsf{F=\begin{bmatrix}1\end{bmatrix}}$ and
$\mathsf{G}_{t}=\mathsf{G=\begin{bmatrix}1\end{bmatrix}}$, we then
have 
\begin{align*}
y_{t}|\Phi,\lambda_{t} & \sim\text{Po}\left(\lambda_{t}\right)\\
\lambda_{t}|\theta_{t} & =\exp\left\{ \theta_{t}\right\} \\
\theta_{t}|\Phi,\theta_{t-1} & \sim\mathcal{N}\left(\theta_{t-1},\mathsf{\sigma^{2}}\right)
\end{align*}

Using an inverse Gamma prior for $\sigma_{0}^{2}\sim\mathcal{IG}\left(\alpha_{0},\beta_{0}\right)$
we have a semi-conjugate update leading to $\sigma^{2}|\theta_{0:n}\sim\mathcal{IG}\left(\alpha_{0}+\frac{n}{2},\beta_{0}+\frac{1}{2}\sum_{i=1}^{n}\left(\theta_{t}-\theta_{t-1}\right)^{2}\right)$
. From this we can extract the necessary quantities as 
\[
s_{t}=\begin{bmatrix}n\\
\left(\theta_{t}-\theta_{t-1}\right)^{2}
\end{bmatrix}
\]
And perform the draws for $\Phi^{(i)}$ as 
\[
\Phi^{(i)}\sim\mathcal{IG}\left(\alpha+\frac{1}{2}s_{0,t}^{(i)},\beta+\frac{1}{2}s_{1,t}^{(i)}\right)
\]
It is important to note that the sufficient statistics method does
not solve entirely the degeneracy and impoverishment problems, since
we are still applying resampling methods to the set $s_{t}$.

\subsection{Forecasting}

We will denote $k$-step ahead forecasting, considering we have observations
until the current time $t$, predicting states or observations up
to time $t+\tau$, where $\tau=1,\dots,k$.

\paragraph{State forecasting}

State forecasting with SMC methods can be performed by carrying the
model forward without performing resampling or reweighting (since we are not in possession
of observations $y_{\ensuremath{t+1}:t+k}$).

If we consider the current marginal posteriors for both the states
and parameters, that is $p\left(\boldsymbol{\theta}_{t}|y_{t},\Phi_{t}^{(i)}\right)$
our aim is then to estimate $p\left(\boldsymbol{\theta}_{t+1:t+k}|y_{t},\Phi_{t}^{(i)}\right)$,
and since we are considering our parameters as static, this is done
according to Algorithm~\ref{alg:State-forecasting}.

\begin{algorithm}
\begin{description}
\item [{\caption{State forecasting\label{alg:State-forecasting}}
}]~
\item [{for}] $\tau\leftarrow1$ to $k$
\begin{description}
\item [{for}] $i\leftarrow1$ to $N_{p}$
\begin{description}
\item [{Sample}] 
\[
\tilde{\boldsymbol{\theta}}_{t+\tau}^{(i)}\sim p\left(\boldsymbol{\theta}_{t+\tau}|\boldsymbol{\theta}_{0:t+\left(\tau-1\right)}^{(i)},\mathcal{D}_{t},\Phi_{t}^{(i)}\right)
\]
\end{description}
\end{description}
\end{description}
\end{algorithm}

It is worth noting that within the proposed DGLM framework, the Algorithm~\ref{alg:State-forecasting} will work directly, since our importance
density $\tilde{\boldsymbol{\theta}}_{t+\tau}^{(i)}\sim p\left(\boldsymbol{\theta}_{t+\tau}|\boldsymbol{\theta}_{0:t+\left(\tau-1\right)}^{(i)},\mathcal{D}_{t},\Phi_{t}^{(i)}\right)$
is in the form of \eqref{eq:DGLM_state_model_normal}. However, fully
adapted proposals will not be directly applicable, since they will
be conditioned on $\mathcal{D}_{t+1:t+\tau}$ , which is not yet available at $t$.

\subsection{Particle Marginal Metropolis-Hastings}

As mentioned previously, these SMC methods will be benchmarked against
a ``gold standard'' off-line method, namely Particle Marginal
Metropolis-Hastings (PMMH) \cite{Andrieu2010}.

If we consider the joint distribution $p\left(\boldsymbol{\theta}_{0:T},\Phi|\mathcal{D}_{t}\right)$,
ideally (if sampling from $p\left(\boldsymbol{\theta}_{0:T}|\mathcal{D}_{t},\Phi\right)$
were possible), we would simply sample from the joint proposal
\[
\pi\left(\left(\boldsymbol{\theta}_{0:T}^{\prime},\Phi^{\prime}\right)|\left(\boldsymbol{\theta}_{0:T},\Phi\right)\right)=\pi\left(\Phi^{\prime}|\Phi\right)p\left(\boldsymbol{\theta}_{0:T}^{\prime}|\mathcal{D}_{t},\Phi^{\prime}\right),
\]

requiring only the specification of a proposal $\pi\left(\Phi^{\prime}|\Phi\right)$
for the construction of the sampler. Since we cannot, generally, sample
directly from $p\left(\boldsymbol{\theta}_{0:T}|y_{1:T},\Phi\right)$ or calculate
$p\left(\mathcal{D}_{t}|\Phi\right)$ directly, PMMH works by using
SMC approximations to these quantities. With the approximation $\hat{p}\left(\mathcal{D}_{t}|\Phi\right)$
and a sampled trajectory $\Theta_{0:T}$ we can then calculate the
Metropolis-Hastings acceptance ratio
\begin{equation}
\min\left\{ \frac{\hat{p}\left(\mathcal{D}_{T}|\Phi^{\prime}\right)\pi\left(\Phi^{\prime}\right)\pi\left(\Phi|\Phi^{\prime}\right)}{\hat{p}\left(\mathcal{D}_{T}|\Phi\right)\pi\left(\Phi\right)\pi\left(\Phi^{\prime}|\Phi\right)}\right\} 
\end{equation}

Pseudo-marginal arguments show that despite the use of approximate estimates, the sampler
nevertheless has the exact posterior as its target.

PMMH is presented in algorithm~\ref{alg:PMMH}, where
$\ell_{0}$ and $\ell_{acc}$ indicate, respectively, the initial
and accepted estimates of $\hat{p}\left(\mathcal{D}_{t}|\Phi\right)$.

\begin{algorithm}
\begin{description}
\item [{\caption{PMMH algorithm\label{alg:PMMH}}
initialisation;}]~
\begin{description}
\item [{With}] initial parameters $\Phi_{0}$ and $\boldsymbol{m}_{0},\mathsf{C}_{0}$
run a SIR filter and store $\left\lbrace \boldsymbol{\theta}_{0:T}^{(k)}\right\rbrace _{0}$,
$\ell_{0}$;
\end{description}
\item [{Set}] $\ell_{acc}\leftarrow\ell_{0}$ and $\Phi_{acc}\leftarrow\Phi_{0}$;
\item [{for}] $n\leftarrow1$ to $N_{iter}$
\begin{description}
\item [{Propose}] new parameters 
\[
\Phi_{n}\sim\mathcal{N}\left(\Phi_{acc},\mathsf{C}_{step}\right)
\]
\item [{Run}] a SIR with parameters $\Phi_{n}$ and store $\left\{ \boldsymbol{\theta}_{0:T}^{(k)}\right\} _{n}$,
$\ell_{n}$
\item [{Draw}] $r\sim\mathcal{U}\left(0,1\right)$
\end{description}
\item [{if}] $\log\left(r\right)<\left(\ell_{n}-\ell_{acc}\right)$
\begin{description}
\item [{$\ell_{acc}\leftarrow\ell_{n}$}]~
\item [{$\Phi_{acc}\leftarrow\Phi_{n}$}]~
\end{description}
\end{description}
\end{algorithm}

\section{Results}\label{sec:Results}

The datasets used for the tests of the algorithms' implementation
aim at covering the three main observation models discussed, Normal,
 Poisson and Binomial.

For the Normal  case we have chosen a continuous measurement dataset,
namely a series of temperature measurements with 5 minute intervals
from the city Austin, Texas (USA) captured by the National Oceanic and Atmospheric
Administration\footnote{\url{http://www1.ncdc.noaa.gov/pub/data/uscrn/products/subhourly01/2015/CRNS0101-05-2015-TX\_Austin\_33\_NW.txt}
{[}Accessed 23/8/2016{]}} (NOAA)~\cite{Diamond2013} shown in Figure~\ref{fig:NOOA-temperature-data}.

For the Poisson  case we have used web server log data, converted
from event time to time series as to represent web hits per second.
The source\footnote{\url{http://ita.ee.lbl.gov/html/contrib/WorldCup.html} {[}Accessed
23/8/2016{]}} of the log data is HTTP requests to the 1998 World Cup server from
April 30, 1998 and July 26, 1998~\cite{Arlitt1996}. A subset corresponding
to May 1998 was used.

The data used for binomial data modelling comes from the US Department
of Transportation's Bureau of Transport Statistics\footnote{\url{http://www.transtats.bts.gov/DL\_SelectFields.asp?Table\_ID=236\&DB\_Short\_Name=On-Time}}
and consist on airport departure times. Since the data consists of scheduled and actual
departure times we dichotomised the dataset into binary data corresponding
to \emph{delayed} and \emph{on-time} flights. A flight was considered
delayed if it departed 30 minutes or more after the scheduled time.
The data was then converted into a time series with intervals of one
minute and missing observations are recorded if no departure happened.
The airport chosen was the JFK airport in New York City and the period
was January 2015.

The Mean Squared Error (MSE) for each single state vector component for each model was compared. 
The state's MSE was calculated using the PMMH estimation values as
\[
\text{MSE}_{i}=\frac{1}{N_{obs}}\sum_{t=1}^{N_{obs}}\left(\overline{\theta}_{i,t}^{PF}-\overline{\theta}_{i,t}^{PMMH}\right)^{2}.
\]

The resampling algorithm used throughout this section
was systematic resampling, as defined in Section~\ref{subsec:Systematic-resampling},
applied at each time $t$.

\subsection{Resampler benchmarks}\label{subsec:Resampler-benchmarks}

\begin{figure}
\centering
\includegraphics[width=1\columnwidth]{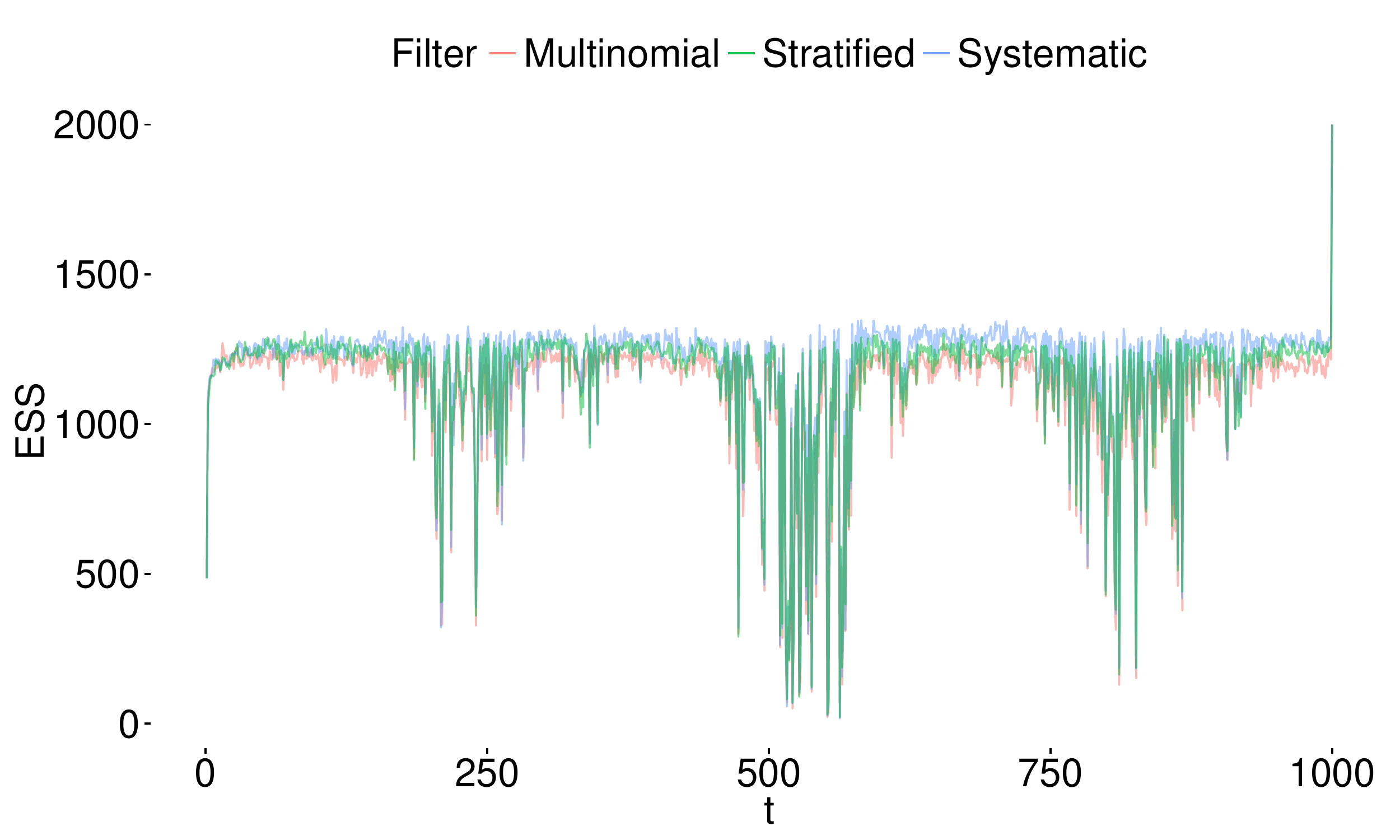} 
\caption{ESS for different resampling methods using Storvik ($N_p=2000$) with a Normal DLM on a subset of the temperature data\label{fig:ESS-for-different}}
\end{figure}

The ESS and posterior variance for different resampling algorithms
(detailed in Section~\ref{subsec:Resampling-methods}) was calculated
using a subset ($N_{obs}=1000$) of the temperature data and a Storvik
filter with $N_{p}=2\times10^{4}$ particles and model as specified
in Section~\ref{subsec:Temperature-data}. 
There was no substantial difference in terms of ESS when using
Systematic, Stratified or Multinomial resamplers (Figure~\ref{fig:ESS-for-different}). The average
ESS values were respectively 1185.349, 1159.503 and 1121.302.

Regarding execution times, systematic and stratified resampling also
had an advantage over multinomial resampling respectively 519.5 and
518.1 seconds, against 639.3 seconds. Following these results we have
chosen to use systematic resampling throughout the subsequent sections.

\subsection{Temperature data\label{subsec:Temperature-data}}

\begin{figure}[H]
	\centering
	\includegraphics[width=1\columnwidth]{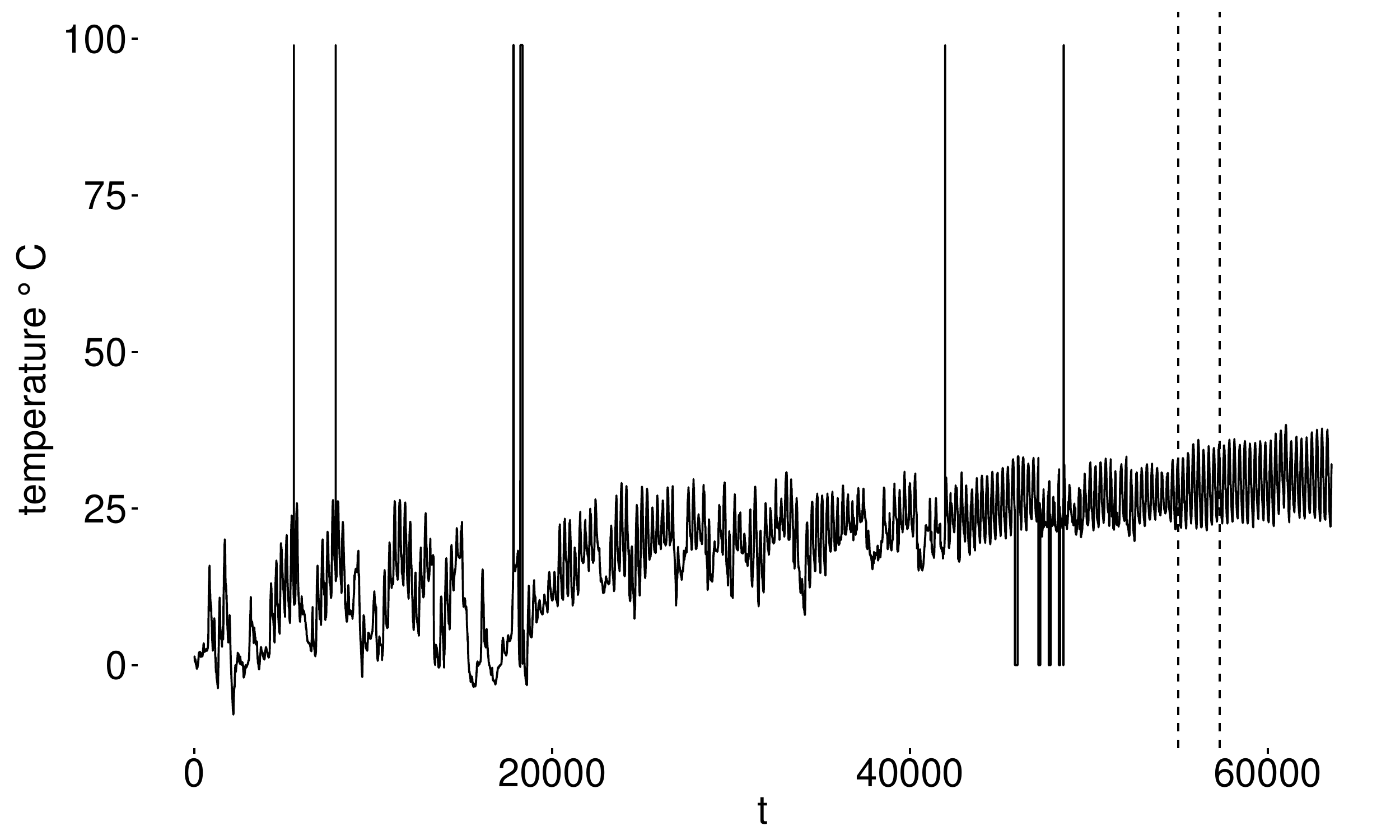} 
	\caption{NOOA temperature data for Austin, Texas (January-July 2015) with 5
		minute sampling interval.\label{fig:NOOA-temperature-data}}
\end{figure}

The temperature dataset includes erroneous measurements of either $100\celsius$
or $0\celsius$, clearly visible in figure \ref{fig:NOOA-temperature-data}.

Below is the estimation for the states and parameters for the temperature
data described in Section~\ref{sec:Results}.

The estimation was performed using three of the filters (LW, Storvik
and PL) with $N_{p}=5000$ and $N_{p}=100$. The dataset consisted
of $N_{obs}=2034\approx7$ days, in a dataset region without the presence of extreme values corresponding to the period between $7^{th}$ and $18^{th}$ July 2015. 

The model used was a Normal DLM, as specified
in Section~\ref{subsec:Normal-DLM}, with a structure consisting
of a locally constant (LC) component with a daily Fourier seasonal
component ($p=288$) with a single harmonic. The corresponding model
structure is
\[
\mathsf{F}=\begin{bmatrix}1\\
1\\
0
\end{bmatrix}^{T}\qquad\mathsf{G}=\begin{bmatrix}1 & 0 & 0\\
0 & \cos\left(2\pi/p\right) & \sin\left(2\pi/p\right)\\
0 & -\sin\left(2\pi/p\right) & \cos\left(2\pi/p\right)
\end{bmatrix},
\]
with a a parameter set $\Phi=\left\lbrace \mathsf{W},V\right\rbrace $,
where 
\[
V=\sigma^{2}\qquad\mathsf{W}=\begin{bmatrix}\tau_{LC}^{2} & 0 & 0\\
0 & \tau_{S1,1}^{2} & 0\\
0 & 0 & \tau_{S1,2}^{2}
\end{bmatrix}.
\]

The observation and state models are

\begin{align*}
p(y_{t}|\eta_{t},\Phi) & =\mathcal{N}\left(\eta_{t},\sigma^{2}\right)\\
\eta_{t} & =\mathsf{F}^{T}\boldsymbol{\theta}_{t}\\
p(\boldsymbol{\theta}_{t}|\boldsymbol{\theta}_{t-1},\Phi) & =\mathcal{N}\left(\mathsf{G}\boldsymbol{\theta}_{t-1},\mathsf{W}\right)
\end{align*}

The state priors were $\boldsymbol{\theta}_{0}\sim\mathcal{N}\left(\left(20,0,0\right)^{T},10\text{\textbf{I}}_{3}\right)$
in order to cover an acceptable range of temperatures for the chosen
period, and the parameter priors where $\sigma_{0}^{2}\sim\mathcal{IG}\left(1,1\right)$
and $\mathsf{W_{0}}\sim\mathcal{IW}\left(3,\mathbf{I}_{3}\right)$.
For LW we have used $\delta=0.98$ as a smoothing parameter, the recommended
general value~\cite{Liu2001}.

\begin{figure}
\centering
\includegraphics[width=1\columnwidth]{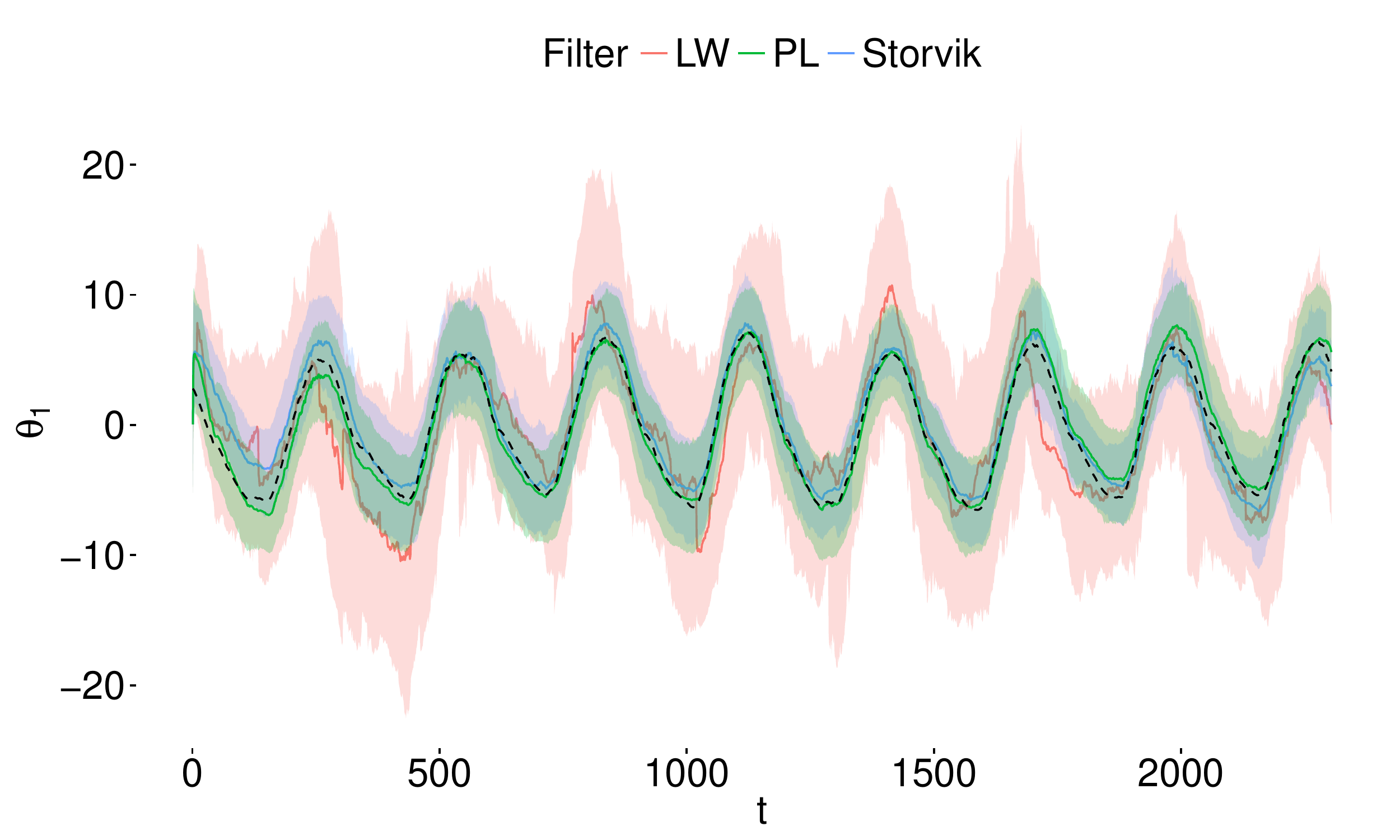}
\caption{$\theta_{1:T}^{2}$ state component estimation for the temperature
data for Storvik, PL and LW filters ($N_p=5000$). Shaded area represents 95\%
coverage. Dashed line represents the PMMH estimation.\label{temperature_state_estimation}}
\end{figure}

\begin{table}
\begin{centering}
\begin{tabular}{c|cccc}
\multicolumn{2}{c}{\emph{\small{}MSE}} & \multicolumn{3}{c}{\emph{\small{}Filter}}\tabularnewline
{\small{}$N_{p}$} & \begin{turn}{90}
\end{turn} & {\small{}LW} & {\small{}Storvik} & {\small{}PL}\tabularnewline
\hline 
\multirow{4}{*}{{\small{}5000}} & {\small{}$\theta^{1}$} & {\small{}6.66} & {\small{}1.511} & {\small{}0.6512}\tabularnewline
 & {\small{}$\theta^{2}$} & {\small{}6.556} & {\small{}1.507} & {\small{}0.6538}\tabularnewline
 & {\small{}$\theta^{3}$} & {\small{}6.442} & {\small{}1.378} & {\small{}1.298}\tabularnewline
\cline{2-5} 
 & {\small{}iteration (}\emph{\small{}ms}{\small{})} & {\small{}8.138} & {\small{}22.75} & {\small{}25.5}\tabularnewline
\hline 
\multirow{4}{*}{{\small{}100}} & {\small{}$\theta^{1}$} & {\small{}199.6} & {\small{}7.014} & {\small{}4.246}\tabularnewline
 & {\small{}$\theta^{2}$} & {\small{}199.0} & {\small{}7.064} & {\small{}4.193}\tabularnewline
 & {\small{}$\theta^{3}$} & {\small{}546.5} & {\small{}7.878} & {\small{}4.192}\tabularnewline
\cline{2-5} 
 & {\small{}iteration (}\emph{\small{}ms}{\small{})} & {\small{}0.4683} & {\small{}0.566} & {\small{}0.6137}\tabularnewline
\cline{2-5} 
\end{tabular}
\par\end{centering}
\caption{State estimation MSE compared to PMMH and computation times for the temperature data
using a Normal DLM\label{tab:MSE-for-state-1}}
\end{table}

Regarding state estimation we see in Table~\ref{tab:MSE-for-state-1}
that both sufficient statistics based methods have a consistently
lower MSE across the state components when compared to LW. The estimation for
the state component $\theta^2_{1:T}$ is on Figure~\ref{temperature_state_estimation}.

In terms of the ESS, PL dominates the other methods, with the mean values for
LW, Storvik and PL being respectively 1202.8, 2839.3 and 4575.9. In
terms of computational times, Storvik and PL have costs in the same
order of magnitude, while LW is the least costly of the three methods.

Additionally, the estimation was performed with a very low number
of particles ($N_{p}=100$) where the difference between the SS based
method accentuates in comparison with LW (Table~\ref{tab:MSE-for-state-1}).
The former still produce an acceptable state estimation whereas the
latter, due to the filter's collapse fails to provide a reasonable
estimation.

\begin{figure}[H]
\centering
\includegraphics[width=1\columnwidth]{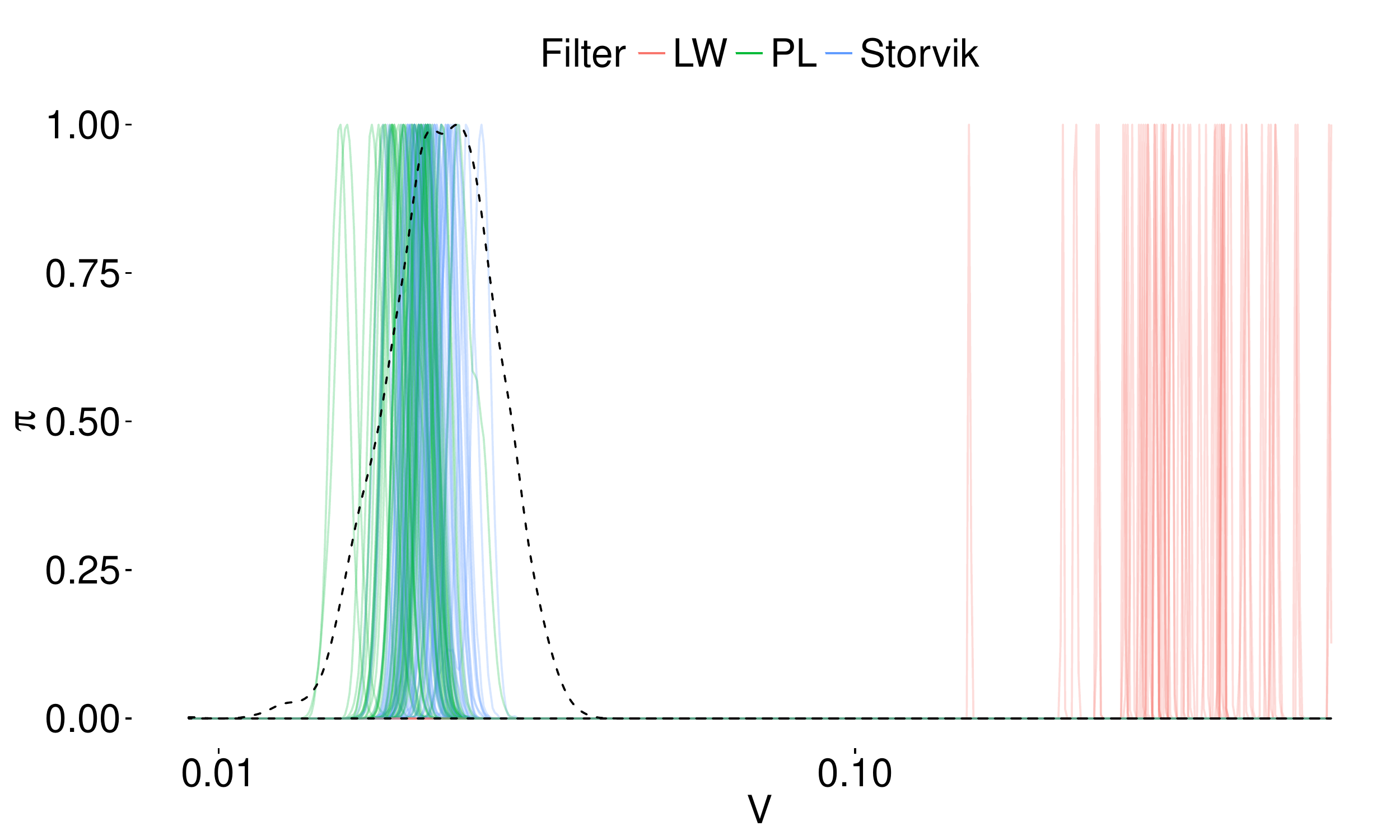}
\caption{$\mathsf{\sigma}^{2}$ posterior at $t=N_{obs}$ (log-scale) for 50 runs 
with the temperature data with $N_{p}=5000$ (LW, Storvik, PL compared to PMMH)\label{fig:temperature-parameter-marginals}}
\end{figure}

\begin{figure}[H]
\includegraphics[width=1\columnwidth]{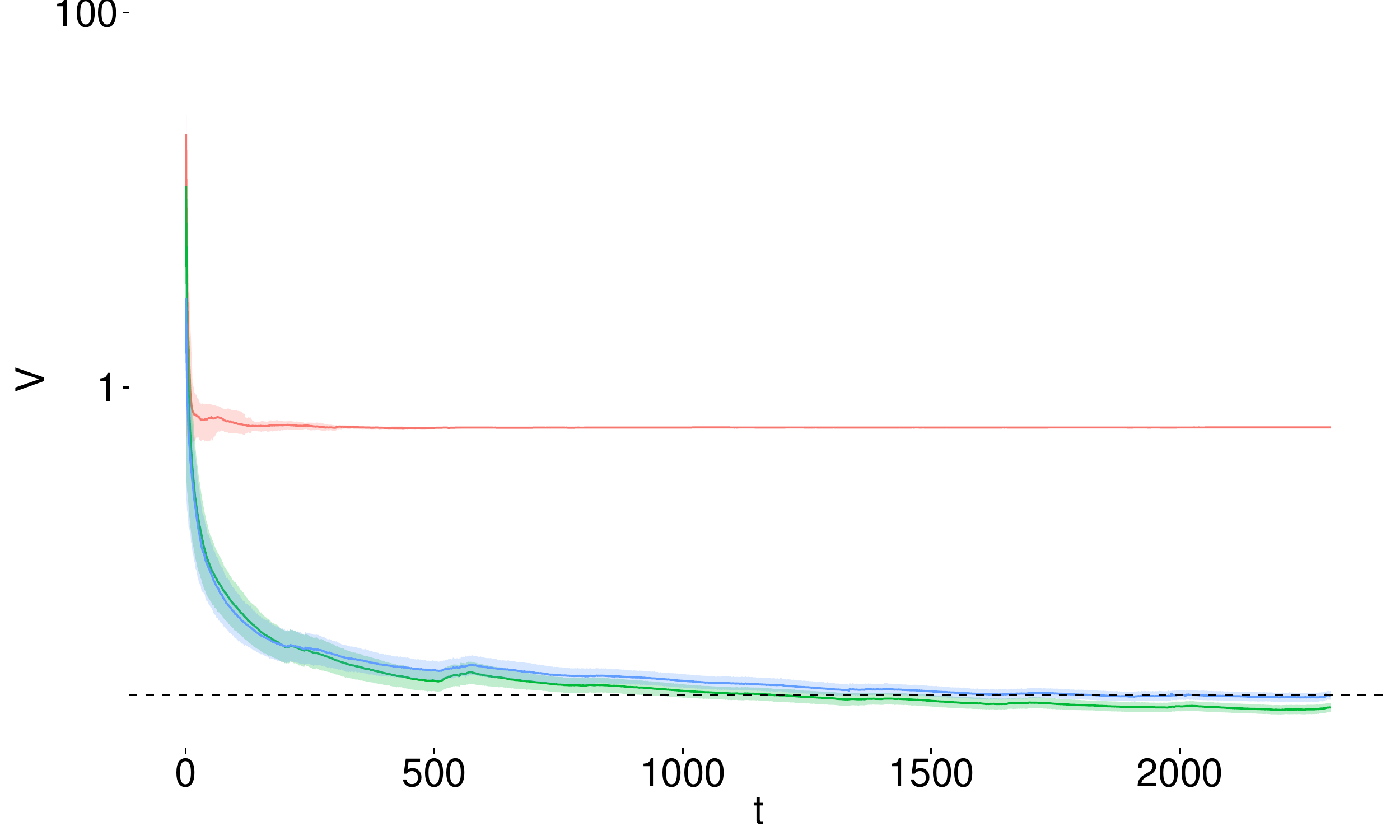} 
\caption{$\mathsf{\sigma}^{2}$ means and 95\% coverage for the temperature data using a Normal DLM ($N_p=5000$). Dashed
line represents PMCMC estimation. \label{fig:temperature-V-history}}
\end{figure}

In figure \ref{fig:temperature-parameter-marginals} we show the $\sigma^{2}$
posterior at $t=N_{obs}$ for 50 runs of each filter where it is visible
that SS based methods fall within the PMMH estimated values but that they grossly
underestimate the true posterior variance. The early collapse of LW is clearly visible in Figure~\ref{fig:temperature-V-history}.

The MSE between the one-step ahead observation forecast and the actual
observation, calculated using $MSE=\nicefrac{1}{N_{obs}}\sum_{t=1}^{N_{obs}}\left(y_{t}-\hat{y}_{t}\right)^{2}$,
was, respectively for LW, Storvik and PL, 0.0582, 0.05795, 0.0578.
We can see in Figure~\ref{fig:Temperature-data-one-step} in the appendix the one-step ahead forecast errors. 

The state (Figure~\ref{fig:state-forecast-for-temperature} in the appendix) and observation (Figure~\ref{fig:temperature-y-forecast})  forecast, when compared respectively to the actual filtered values and observations, fall within the expected range. The
forecast is performed in this case for $k=2500$ steps, roughly equivalent
to 8 days.

\begin{figure}
\centering
\includegraphics[width=1\columnwidth]{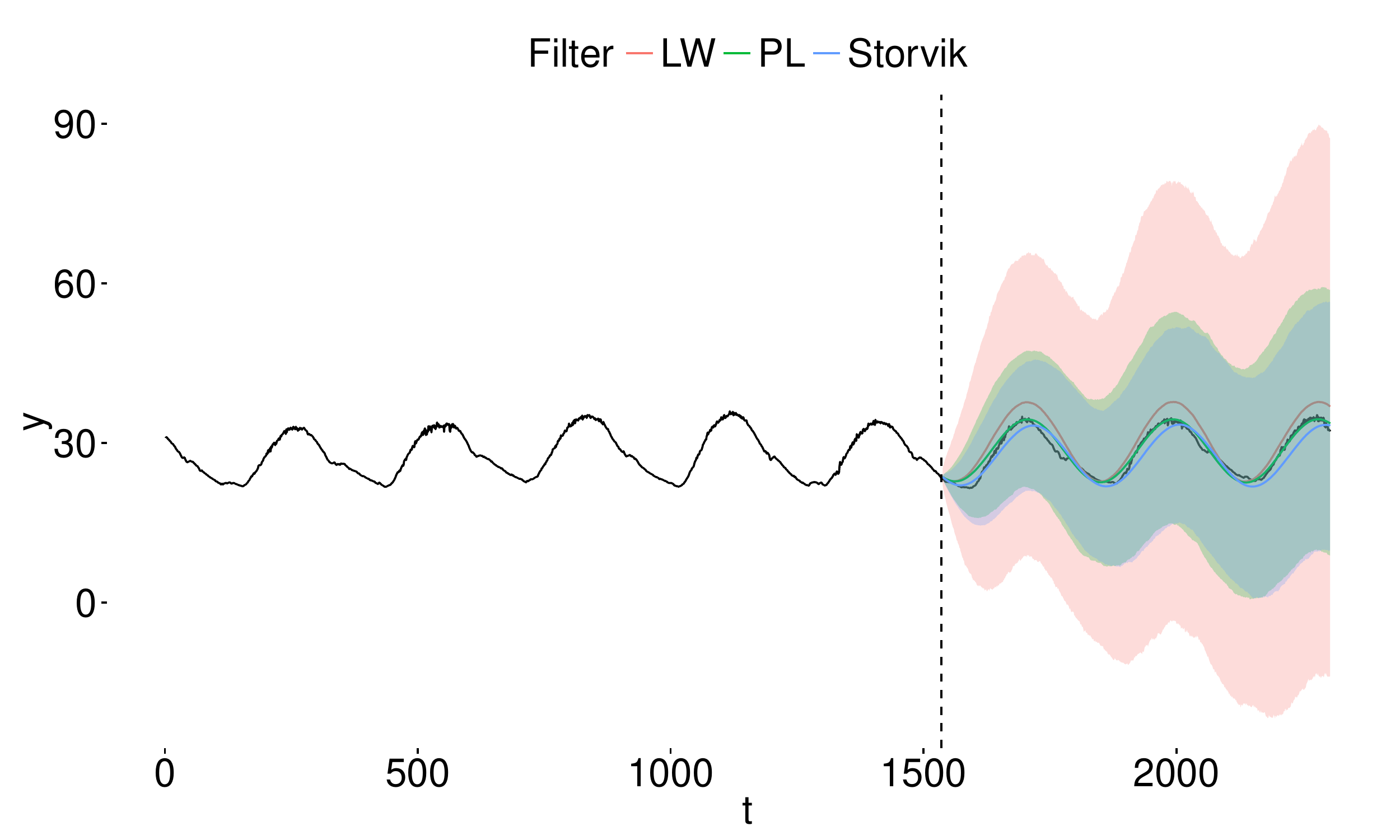} 
\caption{Observation forecast for the temperature dataset (Normal DLM). Shaded
areas represent 95\% coverage for each filter\label{fig:temperature-y-forecast}}
\end{figure}

\begin{figure}
	\centering
	\includegraphics[width=1\columnwidth]{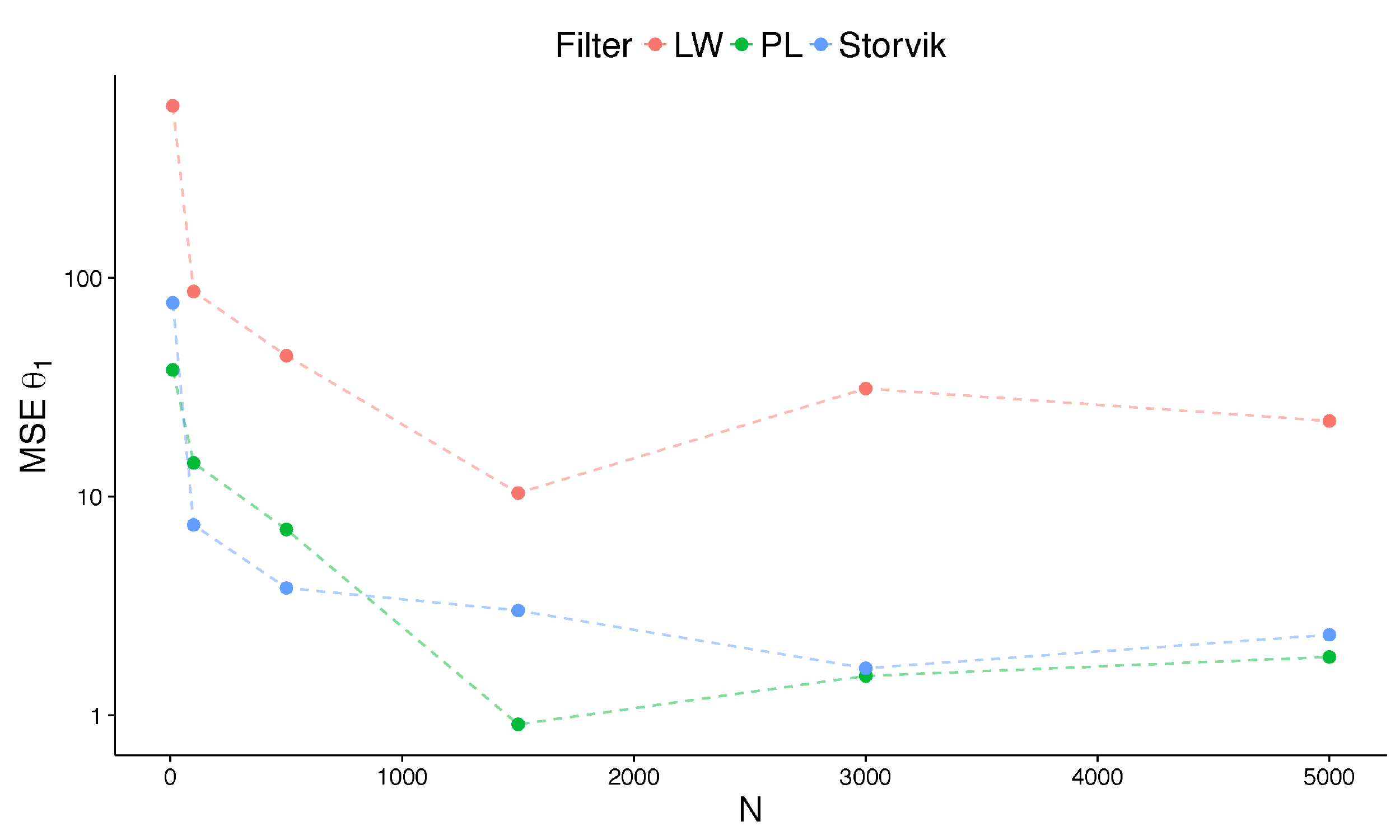}
	\caption{$\theta_{1:T}^{1}$ MSE relative to PMMH for LW, PL and Storvik with
		increasing $N_{p}$.\label{fig:temperature-MSE-relative}}
\end{figure}

Regarding the state estimation's MSE variation with the number of
particles, we can see (Figure~\ref{fig:temperature-MSE-relative})
a sharp decline for low values of $N_{p}$, after which there seems
to be no improvement. This is in conformity to the expected theoretical
Monte Carlo errors which are proportional to $\frac{C_{T}}{\sqrt{N}}$
where $C_{T}$ is a constant dependent on the choice of priors and
parameters \cite{Brockwell2010}. This result was consistent across
the remaining estimations with different datasets and models.

\begin{figure}
	\centering
	\includegraphics[width=1\columnwidth]{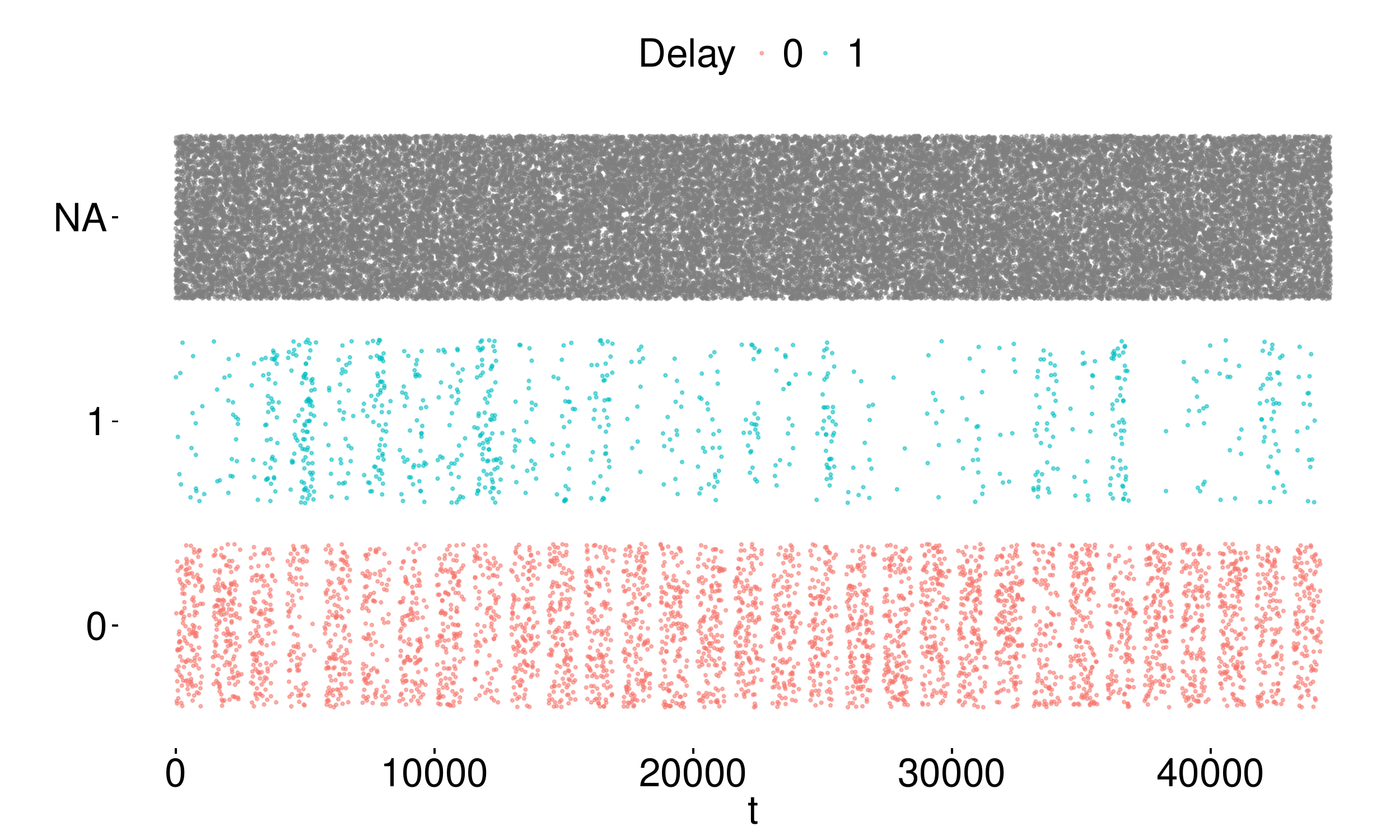} 
	\caption{Airport delay data (January 2015, JFK airport)\label{fig:airport-data}}
\end{figure}

\subsection{Airport flight delay data}

The airport delay data (Figure~\ref{fig:airport-data}) was modelled using a Binomial DLM.\@ The three
filters used were LW, Storvik and PL each with $N_{p}=5000$ and $N_{p}=500$.
The dataset's size was $N_{obs}=4320\approx3$ days. The time-series'
structure consists of a LC component plus a daily seasonality ($p=1440$)
with a single harmonic. The observation and state model correspond
to

\begin{equation}
\mathsf{F}=\begin{bmatrix}1\\
1\\
0
\end{bmatrix}^{T}\qquad\mathsf{G}=\begin{bmatrix}1 & 0 & 0\\
0 & \cos\left(2\pi/p\right) & \sin\left(2\pi/p\right)\\
0 & -\sin\left(2\pi/p\right) & \cos\left(2\pi/p\right)
\end{bmatrix}.
\end{equation}

In this model the parameter set to estimate is $\Phi=\left\lbrace \mathsf{W}\right\rbrace $
where 
\[
\mathsf{W}=\begin{bmatrix}\tau_{LC}^{2} & 0 & 0\\
0 & \tau_{S1,1}^{2} & 0\\
0 & 0 & \tau_{S1,2}^{2}
\end{bmatrix}.
\]

The observation and state models are

\begin{align*}
p(y_{t}|\eta_{t},\Phi) & =\text{B}\left(1,\eta_{t}\right)\\
\eta_{t} & =\text{logit}^{-1}\left(\mathsf{F}^{T}\boldsymbol{\theta}_{t}\right)\\
p(\boldsymbol{\theta}_{t}|\boldsymbol{\theta}_{t-1},\Phi) & =\mathcal{N}\left(\mathsf{G}\boldsymbol{\theta}_{t-1},\mathsf{W}\right)
\end{align*}

The same priors as with the temperature data in~\ref{subsec:Temperature-data}
were used for the parameters with a state prior $\boldsymbol{\theta}_{0}\sim\mathcal{N}\left(\mathbf{0},4\mathbf{I_{3}}\right)$.

\begin{figure}
\centering
\includegraphics[width=1\columnwidth]{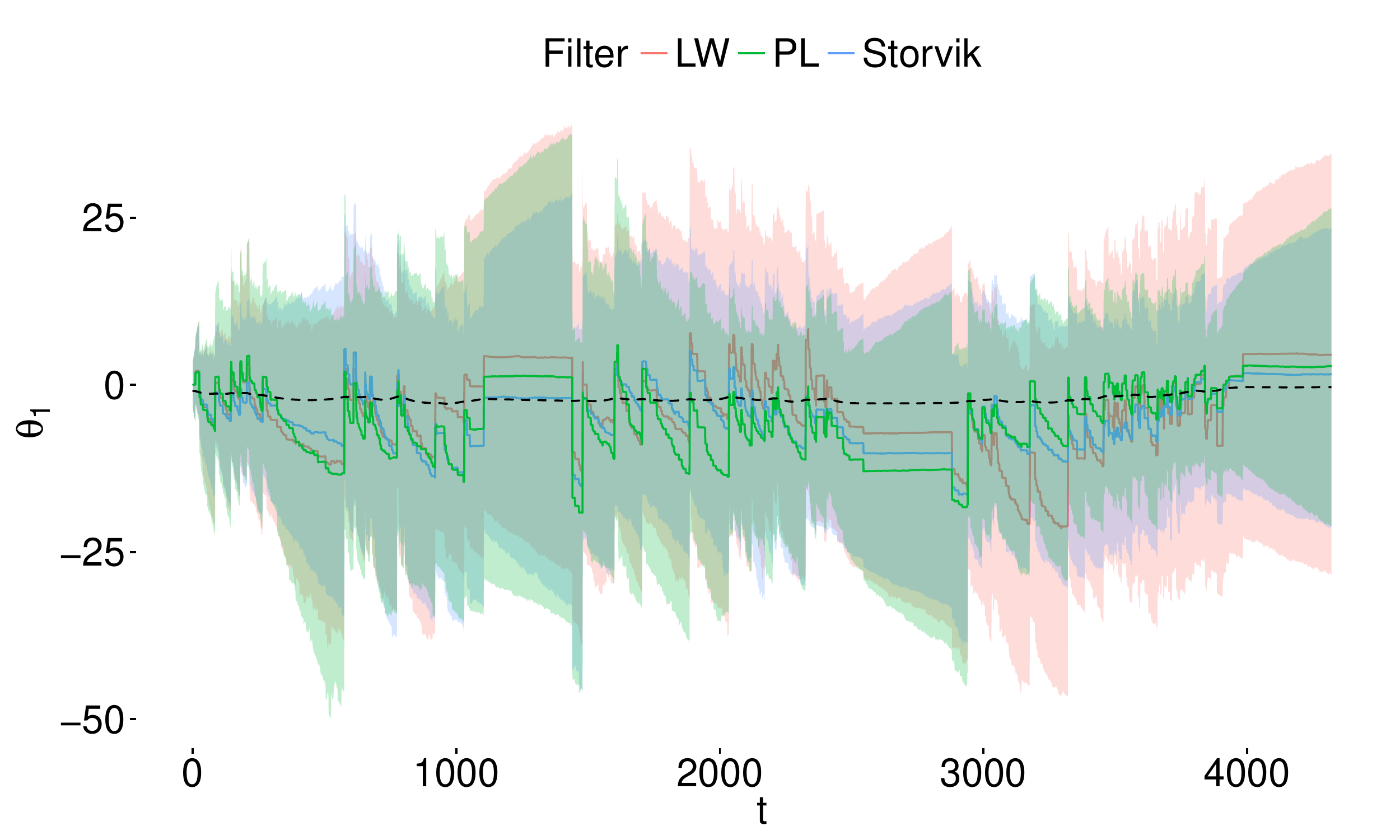}
\caption{$\theta_{1:T}^{1}$ state component estimation for the airport delay
data for Storvik, PL and LW filters. Shaded area represents 95\% coverage. Noted that the (dashed) PMMH
estimate is the smoothing estimate, and therefore not directly comparable with the filtered estimates.\label{fig:airport_state_estimation_0}}
\end{figure}

\begin{figure}[H]
\centering
\includegraphics[width=1\columnwidth]{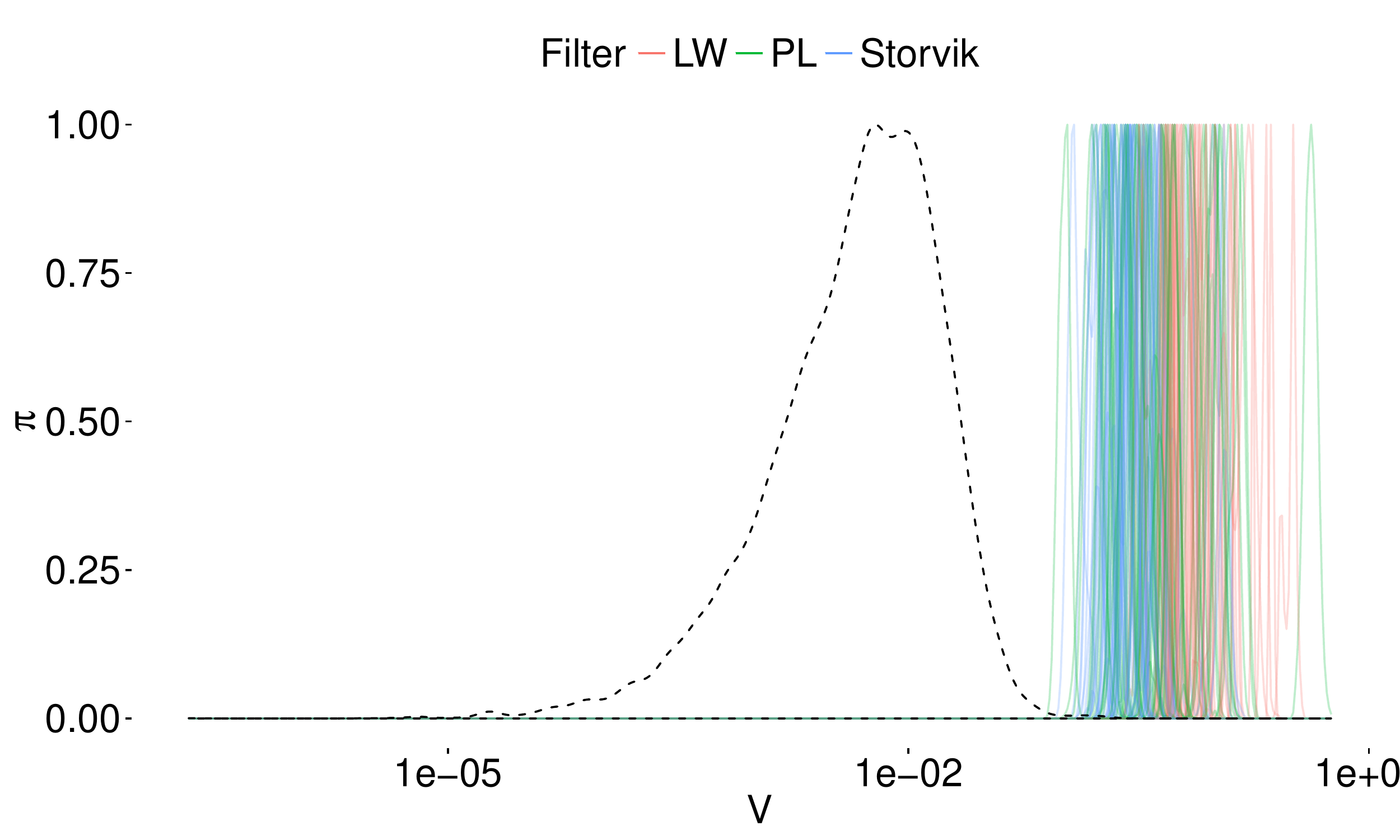}
\caption{$\mathsf{W}^{1}$ posterior at $t=N_{obs}$ (log-scale) for 50 runs 
	with the airport data with $N_{p}=5000$ (LW, Storvik, PL) compared to PMMH (dashed line)\label{fig:airport-parameter-marginals}}
\end{figure}

\begin{figure}[H]
\includegraphics[width=1\columnwidth]{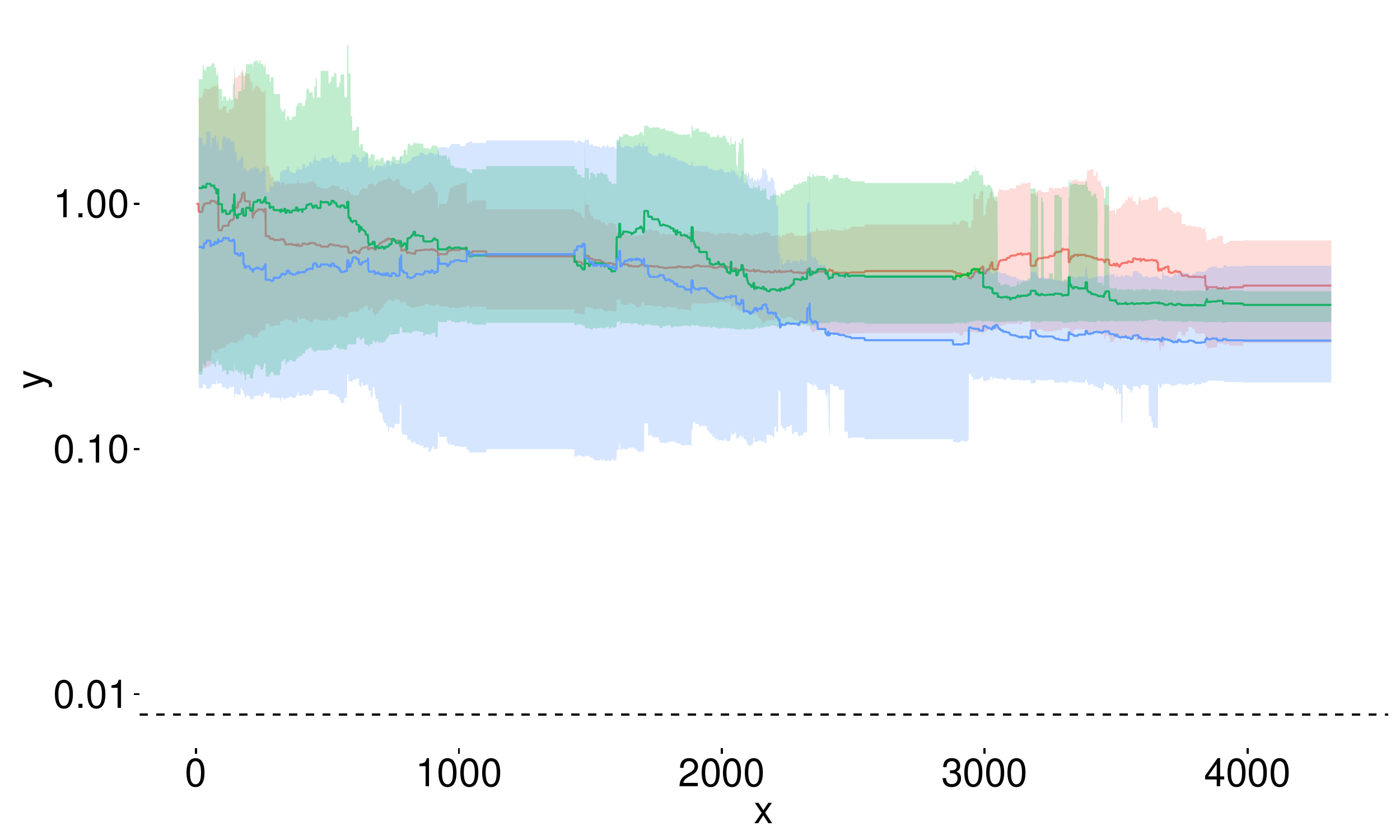} 
\caption{$\mathsf{W}^{1}$ means and 95\% coverage (shaded area) for the airport dataset using a Binomial DLM ($Np=5000$). Dashed line represents PMCMC estimation.\label{fig:airport_par_estimation_0}}
\end{figure}

Due to the high number of missing observations ($\approx88\%$), the
state estimation (Figure~\ref{fig:airport_state_estimation_0}) displays a high MSE when compared to a PMMH run
as well as poor parameter estimation (Figures~\ref{fig:airport-parameter-marginals} and~\ref{fig:airport_par_estimation_0}).
This is to be expected since for every missing observation we are
simply propagating the states forward, using~\eqref{eq:DGLM_state_model_normal}
and bypassing resampling. In these conditions, however, we can still
see (Table~\ref{tab:-MSE-compared}) that sufficient statistics
based methods perform generally better than LW, whereas LW, in terms
of computational time is less costly. 

Regarding the ESS for LW, Storvik
and PL the average value was respectively 4300.34, 3892.46 and 4708.0
for $N_{p}=5000$ and 414.7, 389.27 and 471.70 for $N_{p}=500$.

\begin{table}
\begin{centering}
\begin{tabular}{c|cccc}
\multicolumn{2}{c}{\emph{\small{}MSE}} & \multicolumn{3}{c}{\emph{\small{}Filter}}\tabularnewline
{\small{}$N_{p}$} & \begin{turn}{90}
\end{turn} & {\small{}LW} & {\small{}Storvik} & {\small{}PL}\tabularnewline
\hline 
\multirow{4}{*}{{\small{}5000}} & {\small{}$\theta^{1}$} & {\small{}39.05} & {\small{}31.15} & {\small{}14.74}\tabularnewline
 & {\small{}$\theta^{2}$} & {\small{}46.68} & {\small{}35.44} & {\small{}10.32}\tabularnewline
 & {\small{}$\theta^{3}$} & {\small{}33.16} & {\small{}27.01} & {\small{}13.14}\tabularnewline
\cline{2-5} 
 & {\small{}iteration (}\emph{\small{}ms}{\small{})} & {\small{}4.626} & {\small{}5.396} & {\small{}5.701}\tabularnewline
\hline 
\multirow{4}{*}{{\small{}500}} & {\small{}$\theta^{1}$} & {\small{}100.6} & {\small{}34.79} & {\small{}64.93}\tabularnewline
 & {\small{}$\theta^{2}$} & {\small{}71.24} & {\small{}44.58} & {\small{}33.34}\tabularnewline
 & {\small{}$\theta^{3}$} & {\small{}103.0} & {\small{}49.23} & {\small{}50.6}\tabularnewline
\cline{2-5} 
 & {\small{}iteration (}\emph{\small{}ms}{\small{})} & {\small{}0.2068} & {\small{}0.1463} & {\small{}0.1482}\tabularnewline
\cline{2-5} 
\end{tabular}
\par\end{centering}
\caption{$\theta_{1:t}^{1}$ MSE compared to PMMH for the airport data using
a Binomial DLM\label{tab:-MSE-compared}}
\end{table}

\begin{figure}[H]
\centering
\includegraphics[width=1\columnwidth]{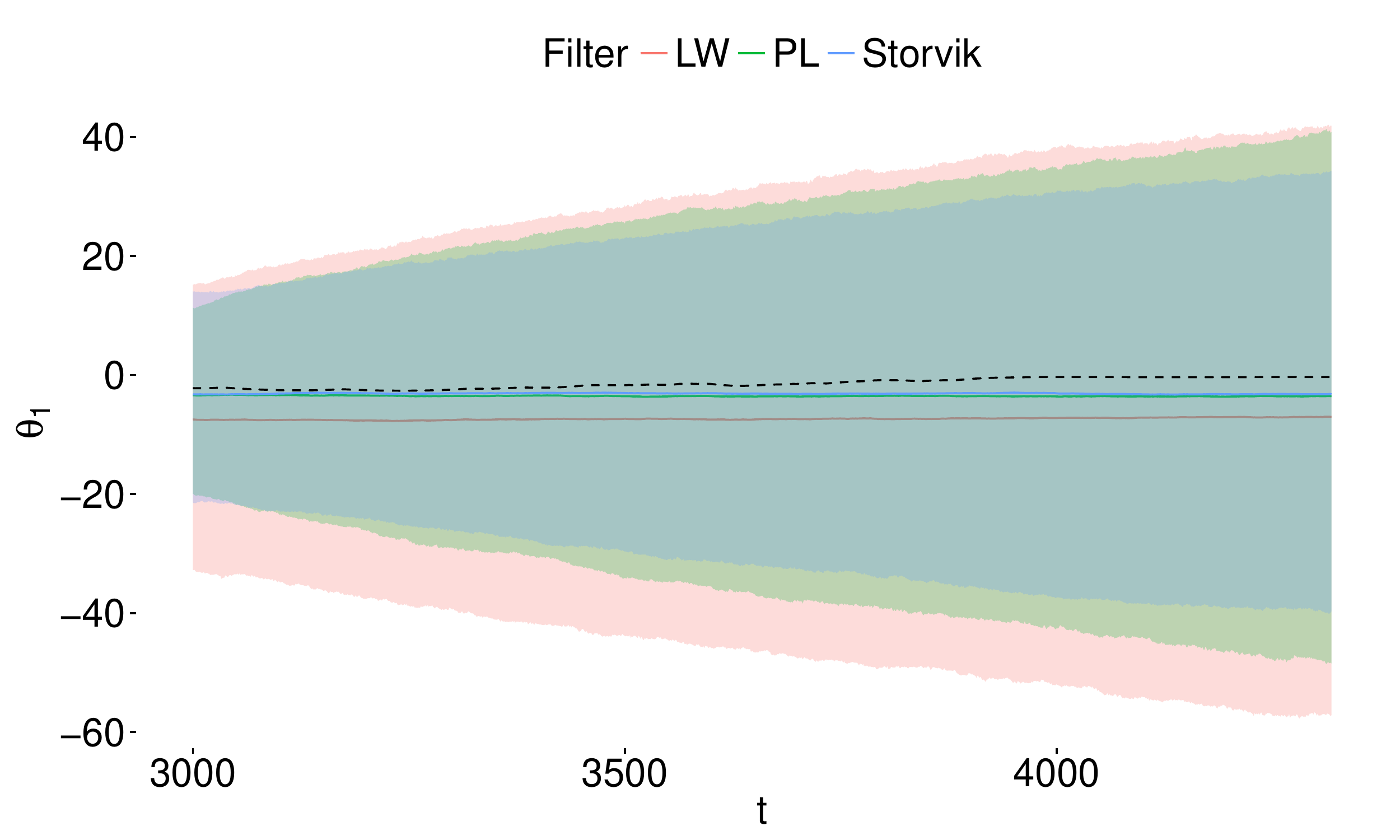}
\caption{$\theta^1$ state forecast for the airport delay compared to state estimation.
Shaded area represents 95\% coverage.~\label{fig:airport_state_forecast}}
\end{figure}

A long term state forecast was produced according to algorithm~\ref{alg:State-forecasting}
for 3360 data points, corresponding to approximately 56 hours. The
resulting forecast was then compared against the actual state estimation
for that period (Figure~\ref{fig:airport_state_forecast}). The PL
state forecast was closer to the PMMH estimation and had a smaller
variance than the remaining methods.

\subsection{World Cup 98 Web server data}

\begin{figure}[H]
	\centering
	\includegraphics[width=1\columnwidth]{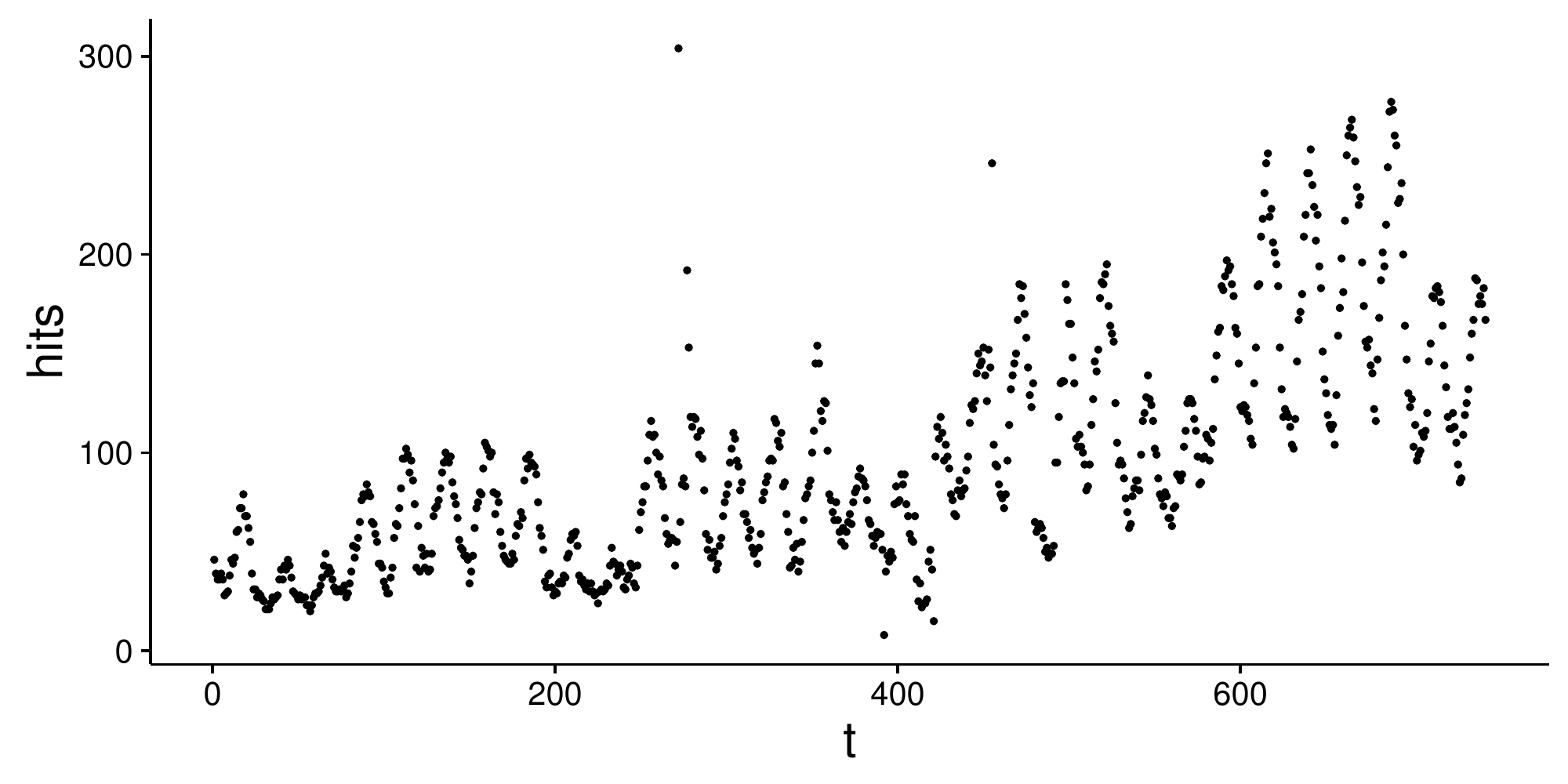} 
	\caption{WC98 server hits (May 1998)\label{fig:WC98-server-hits}}
\end{figure}

The WC98 dataset (figure \ref{fig:WC98-server-hits}) consists of
$N_{obs}=743\approx31$ days of hourly measurements and exhibits both
a daily and weekly pattern.

The estimation was performed using three of the filters (LW, Storvik
and PL) with $N_{p}=5000$. The model used was Poisson DLM, with
a structure consisting of a locally constant (LC) component with a
daily Fourier seasonal ($p_{d}=24$) and weekly
$\left(p_{w}=168\right)$ components, both with a single harmonic. Considering,
$\mathsf{J}_{2}\left(1,\omega\right)=\begin{bmatrix}\cos\left(\omega\right) & \sin\left(\omega\right)\\
-\sin\left(\omega\right) & \cos\left(\omega\right)
\end{bmatrix}$ and $\omega=2\pi/p$, the corresponding model structure is
\begin{align*}
\mathsf{F} & =\begin{bmatrix}1 & 1 & 0 & 1 & 0\end{bmatrix},\\
\mathsf{G} & =\begin{bmatrix}1 & 0 & 0\\
0 & \mathsf{J}_{2}\left(1,\omega_{d}\right) & 0\\
0 & 0 & \mathsf{J}_{2}\left(1,\omega_{w}\right)
\end{bmatrix}
\end{align*}
with a parameter set $\Phi=\left\lbrace \mathsf{W}\right\rbrace $,
where 

\begin{equation}
\text{diag}\mathsf{\left(W\right)}=\begin{bmatrix}\tau_{LC}^{2} & \tau_{d1,1}^{2} & \tau_{d2,2}^{2} & \tau_{w1,1}^{2} & \tau_{w2,2}^{2}\end{bmatrix}.
\end{equation}

The observation and state models are

\begin{align*}
p(y_{t}|\eta_{t},\Phi) & =\text{Po}\left(\eta_{t}\right)\\
\eta_{t} & =\exp\left\{ \mathsf{F}^{T}\boldsymbol{\theta}_{t}\right\} \\
p(\boldsymbol{\theta}_{t}|\boldsymbol{\theta}_{t-1},\Phi) & =\mathcal{N}\left(\mathsf{G}\boldsymbol{\theta}_{t-1},\mathsf{W}\right)
\end{align*}

The state priors were $\boldsymbol{\theta}_{0}\sim\mathcal{N}\left(\mathbf{0},5\text{\textbf{I}}_{5}\right)$
and the parameter priors where and $\mathsf{W_{0}}\sim\mathcal{IW}\left(5,\mathbf{I}_{5}\right)$.
As previously, for the LW we have used $\delta=0.98$ as a smoothing
parameter.

\begin{figure}
\centering
\includegraphics[width=1\columnwidth]{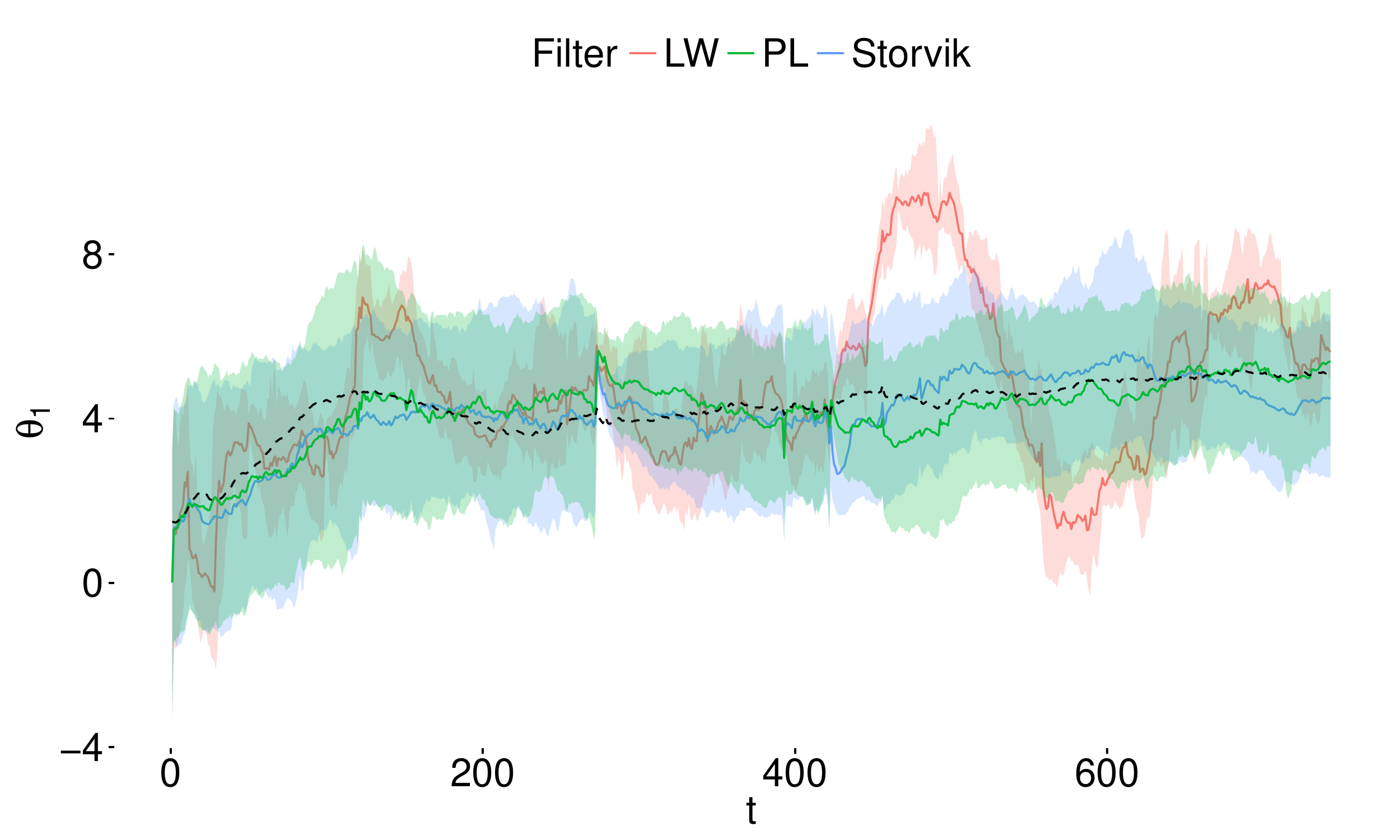}
\caption{$\theta_{1:T}^{1}$ state estimation ($N_p=5000$) for the WC98 data using a Poisson DLM. Shaded area represents 95\% coverage.
Dashed line represents the PMMH estimation.\label{fig:WC98_poisson_weekly_state_estimation_0}}
\end{figure}

\begin{table}
\begin{centering}
\begin{tabular}{c|cccc}
\multicolumn{2}{c}{\emph{\small{}MSE}} & \multicolumn{3}{c}{\emph{\small{}Filter}}\tabularnewline
{\small{}$N_{p}$} & \begin{turn}{90}
\end{turn} & {\small{}LW} & {\small{}Storvik} & {\small{}PL}\tabularnewline
\hline 
\multirow{6}{*}{{\small{}5000}} & {\small{}$\theta^{1}$} & {\small{}3.283} & {\small{}0.2721} & {\small{}0.2609}\tabularnewline
 & {\small{}$\theta^{2}$} & {\small{}0.4034} & {\small{}0.03121} & {\small{}0.02645}\tabularnewline
 & {\small{}$\theta^{3}$} & {\small{}0.3286} & {\small{}0.04591} & {\small{}0.03601}\tabularnewline
 & {\small{}$\theta^{4}$} & {\small{}3.168} & {\small{}0.2462} & {\small{}0.2342}\tabularnewline
 & {\small{}$\theta^{5}$} & {\small{}5.532} & {\small{}0.2432} & {\small{}0.2329}\tabularnewline
\cline{2-5} 
 & {\small{}iteration (}\emph{\small{}ms}{\small{})} & {\small{}4.626} & {\small{}5.396} & {\small{}5.701}\tabularnewline
\end{tabular}
\par\end{centering}
\caption{$\theta_{1:T}$ filter estimates MSE compared to PMMH for the WC98
data using a Poisson DLM.\label{tab:wc98-filter-estimates}}
\end{table}

For state estimation (Figure~\ref{fig:WC98_poisson_weekly_state_estimation_0}) we can see that the SS based
methods dominate in terms of MSE when compared to the PMMH estimation
and LW has a slight advantage in term of computational cost (Table~\ref{tab:wc98-filter-estimates}).

\begin{figure}
\centering
\includegraphics[width=1\columnwidth]{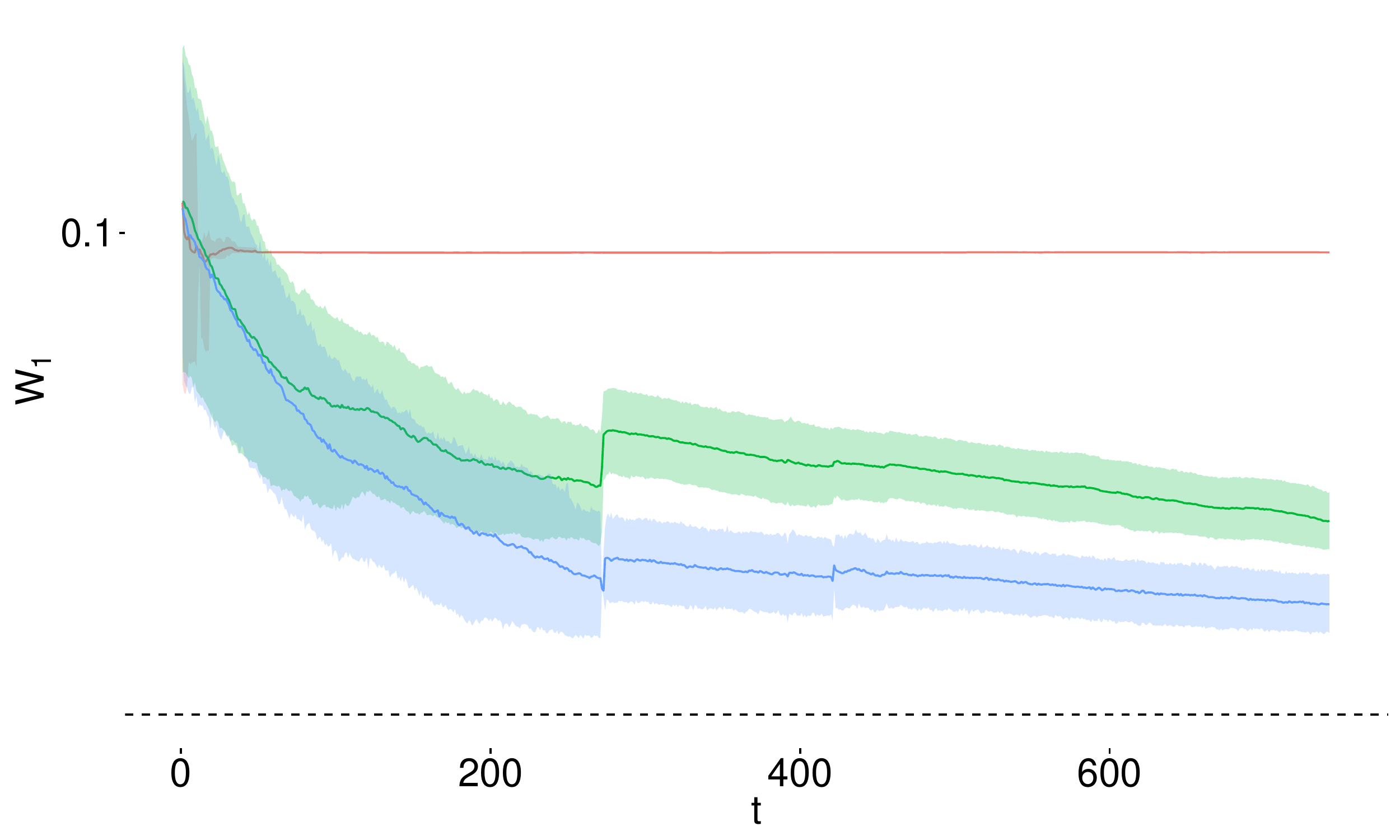}
\caption{$\mathsf{W}^1$ estimation history for the WC98 data using a Poisson DLM ($N_p=5000$).
Shaded area represents 95\% coverage.
Dashed line represents PMMH estimation.\label{fig:wc98-parameter-estimation-history}}
\end{figure}

\begin{figure}
\centering
\includegraphics[width=1\columnwidth]{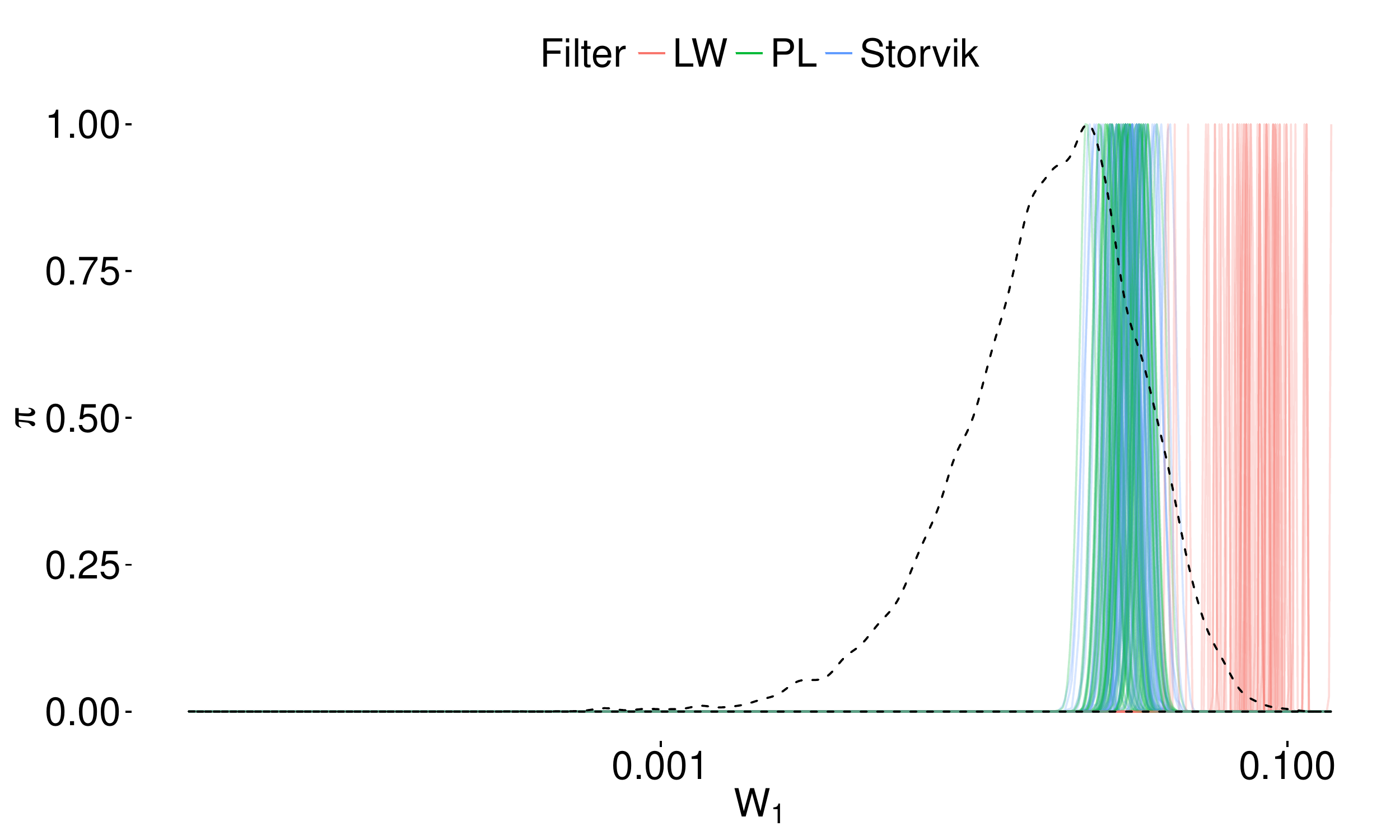}
\caption{$\mathsf{W}^{1}$ posterior at time $t=N_{\ensuremath{obs}}$ for the WC98 data using
a Poisson DLM and $N_{p}=5000$, using 50 runs. Dashed line represents PMMH estimation.\label{fig:wc98-w1-posterior}}
\end{figure}

Regarding parameter estimation, SS methods also provide a better approximation
to the PMMH result. There is a clear particle filter collapse for
Liu and West in the early stages of the estimation (Figure~\ref{fig:wc98-parameter-estimation-history})
which accounts for the poor parameter estimations (Figure~\ref{fig:wc98-w1-posterior}).

\begin{figure}
\centering
\includegraphics[width=1\columnwidth]{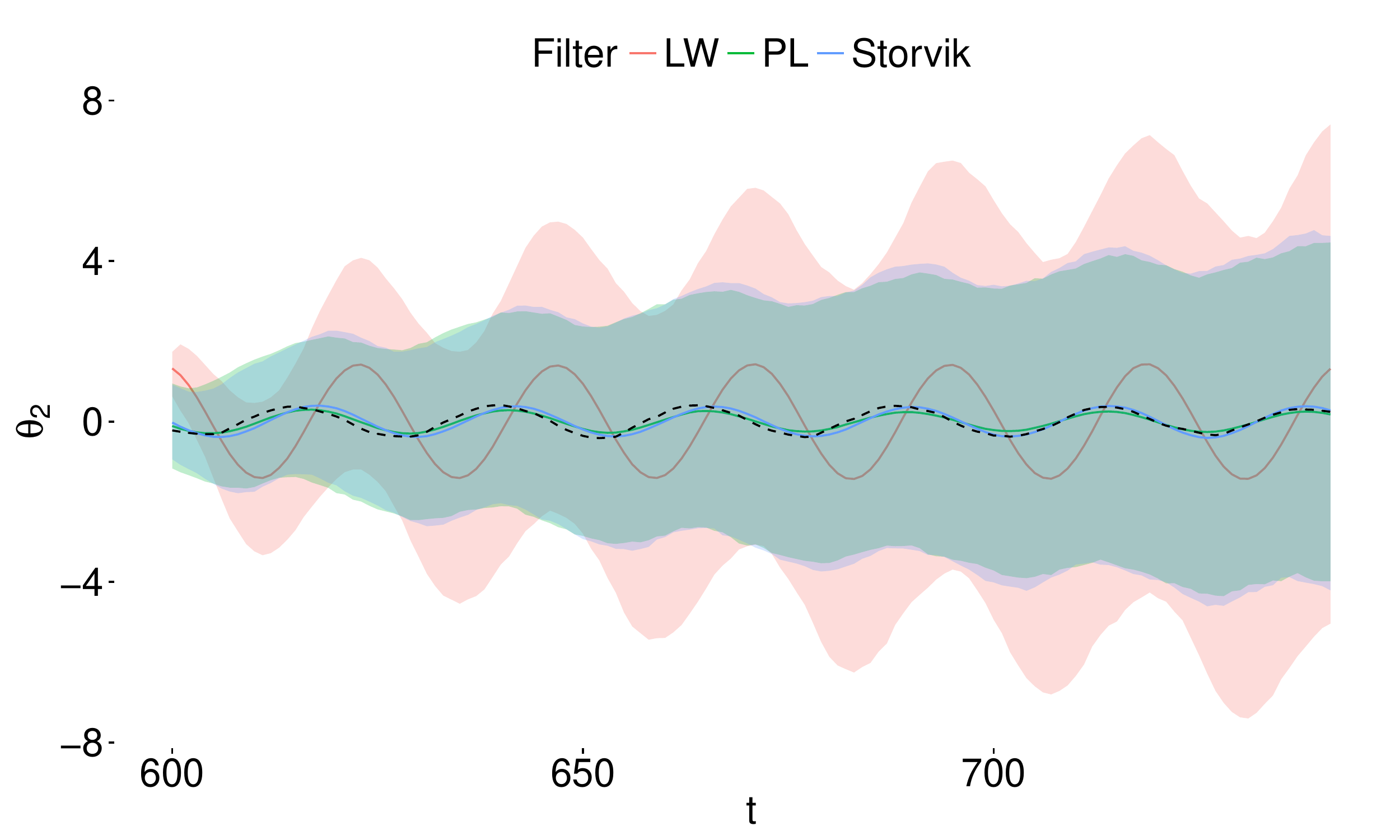}
\caption{State forecast for the WC98 data\label{fig:State-forecast-for}}
\end{figure}

When performing a state forecast for $k=600\approx6$ days, we can
see (Figure~\ref{fig:State-forecast-for}) that while sufficient
statistics based methods capture the seasonal patterns Liu and West
does not, on account of the filter's early collapse.

\begin{figure}
\centering
\includegraphics[width=1\columnwidth]{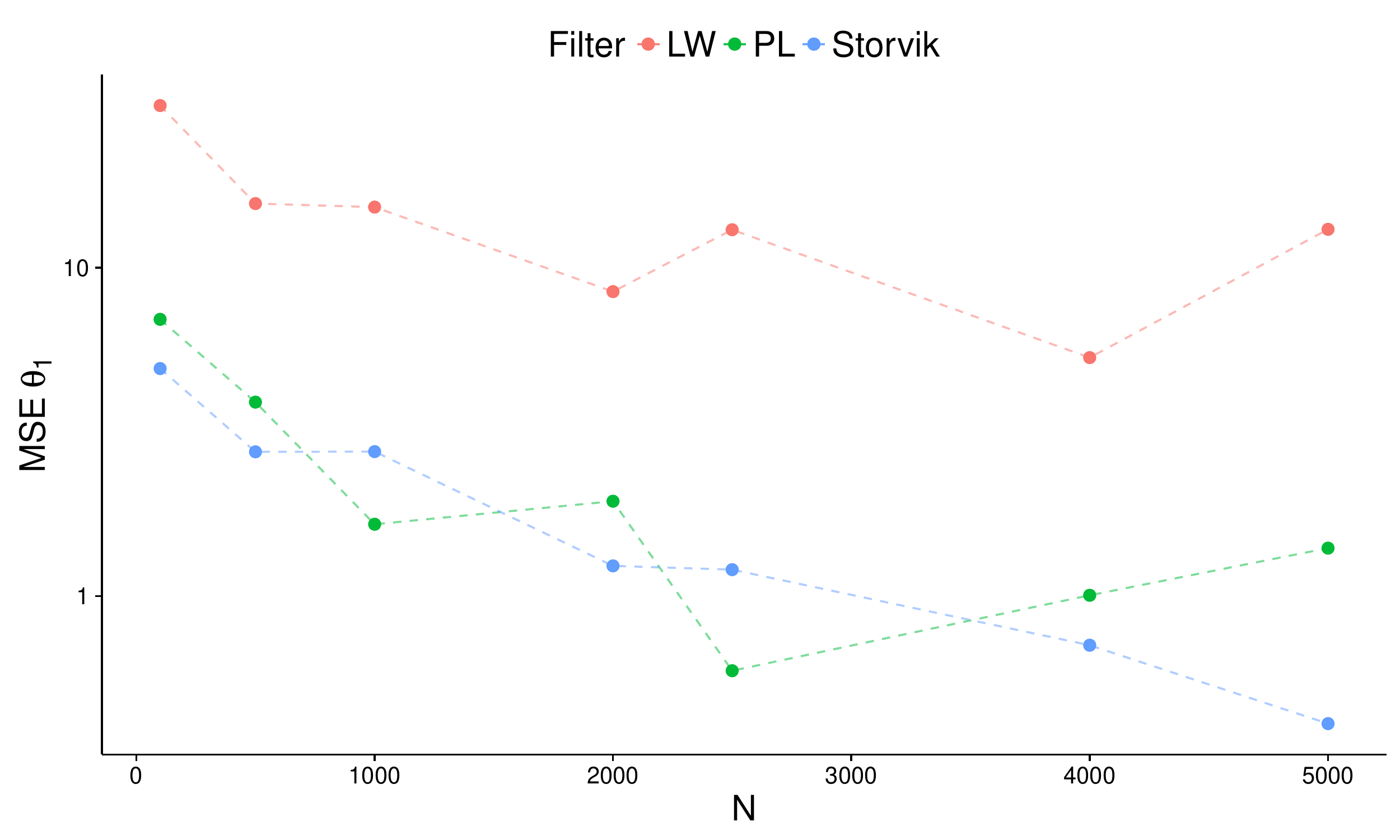}
\caption{$\theta_{1:T}^{1}$ MSE with varying $N_{p}$ for the WC98 dataset
using a Poisson DLM.\label{subsec:WC98_Number-of-particles}}
\end{figure}

\section{Conclusions}

Considering that we performed long run state forecasts (namely, in
the temperature data we forecasted $\approx2.65$ days from data sampled
every five minutes) the forecasted values were in line with the observed
data.

The Liu and West method, suffered from a problem where the parameter estimation
tended to collapse to a single value after a few hundred iterations.
This is mainly due to the depletion of particles, \emph{i.e.} particle
impoverishment. Since sufficient statistics methods did not incorporate
the parameters directly in the state space, particle impoverishment
could be delayed for much longer.

Liu and West's main appeal, the ability to perform state and parameter
estimation in all state-space models, even when no sufficient statistics
structure can be specified, does not apply to our DGLM scenario. However,
Liu and West did outperform the remaining methods when the criteria
was computation time.

Even though the parameter estimations fall within a certain range
of the PMMH estimations, they did not offer results which could justify
the substitution of standard off-line methods, such as PMMH for parameter
estimation. It is however justified in the proposed scenario of inference
for streaming data to use these methods, specifically those based on 
sufficient statistics, as it is common to try to achieve a compromise between
computational times and accuracy. 

These methods provided acceptable
results for one-step ahead, short and even medium term forecasts.
A desirable property for the application of SMC in analytics is the
ability to perform trade-offs between computational costs and estimation
accuracy. SS based methods show the capability of performing reasonable
state estimations and forecasts, parameter estimation and consequently
observation forecasts with a particle number as low as $N_{p}=100$
with a computational cost in the order of milliseconds per iteration.
This can prove extremely valuable when implementing such methods in
low powered devices.

\section*{Acknowledgement}

Rui Vieira is supported by a studentship funded by Red Hat Ltd. / JBoss.

\bibliography{paper}

\begin{thebibliography}{10}

\bibitem{Andrieu2010}
C.~Andrieu, A.~Doucet, and R.~Holenstein.
\newblock {Particle Markov chain Monte Carlo methods}.
\newblock {\em Journal of the Royal Statistical Society: Series B (Statistical
  Methodology)}, 72(3):269--342, 2010.

\bibitem{Arlitt1996}
M.~Arlitt and T.~Jin.
\newblock {A workload characterization study of the 1998 world cup web site}.
\newblock In {\em IEEE network}, volume~14, pages 30--37. IEEE, 1996.

\bibitem{Brockwell2010}
A.~Brockwell, P.~Del~Moral, and A.~Doucet.
\newblock {Sequentially interacting markov chain monte carlo methods}.
\newblock {\em The Annals of Statistics}, 38(6):3387--3411, 2010.

\bibitem{Cliffordy}
J.~Carpenter, P.~Clifford, and P.~Fearnhead.
\newblock {Improved particle filter for nonlinear problems}.
\newblock {\em IEE Proceedings - Radar, Sonar and Navigation}, 146(1):2--7,
  1999.

\bibitem{Carvalhoa}
C.M. Carvalho, M.S. Johannes, H.F. Lopes, and N.G. Polson.
\newblock {Particle Learning and Smoothing}.
\newblock {\em Statistical Science}, 25(1):88--106, 2010.

\bibitem{Chopin2010}
N.~Chopin, A.~Iacobucci, J.-M. Marin, K.~Mengersen, C.~P. Robert, R.~Ryder, and
  C.~Sch{\"{a}}fer.
\newblock {On Particle Learning}.
\newblock {\em arXiv preprint arXiv:1006.0554}, (2008):14, jun 2010.

\bibitem{Diamond2013}
H.J. Diamond, T.R. Karl, M.A. Palecki, C.B. Baker, J.E. Bell, R.D. Leeper, D.R.
  Easterling, J.H. Lawrimore, T.P. Meyers, M.R. Helfert, G.~Goodge, and P.W.
  Thorne.
\newblock {U.S. climate reference network after one decade of operations status
  and assessment}.
\newblock {\em Bulletin of the American Meteorological Society},
  94(4):485--498, apr 2013.

\bibitem{Douc2005}
R.~Douc and O.~Cappe.
\newblock {Comparison of resampling schemes for particle filtering}.
\newblock {\em ISPA 2005. Proceedings of the 4th International Symposium on
  Image and Signal Processing and Analysis, 2005}, pages 64--69, 2005.

\bibitem{Doucet2001a}
A.~Doucet, N.~de~Freitas, and N.J. Gordon.
\newblock {Sequential Monte Carlo Methods in Practice}.
\newblock {\em Technometrics}, page 583, 2001.

\bibitem{Doucet2000}
A.~Doucet, S.~Godsill, and C.~Andrieu.
\newblock {On sequential Monte Carlo sampling methods for Bayesian filtering}.
\newblock {\em Statistics and Computing}, 10(3):197--208, 2000.

\bibitem{Fearnhead2002}
P.~Fearnhead.
\newblock {Markov chain Monte Carlo, sufficient statistics, and particle
  filters}.
\newblock {\em Journal of Computational and Graphical Statistics},
  11(4):848--862, 2002.

\bibitem{Gong2012}
P.~Gong, Y.~O. Basciftci, and F.~Ozguner.
\newblock {A parallel resampling algorithm for particle filtering on
  shared-memory architectures}.
\newblock In {\em Proceedings of the 2012 IEEE 26th International Parallel and
  Distributed Processing Symposium Workshops, IPDPSW 2012}, pages 1477--1483.
  IEEE, IEEE, may 2012.

\bibitem{Gordon1993a}
N.J. Gordon, D.J. Salmond, and A.F.M. Smith.
\newblock {Novel approach to nonlinear/non-Gaussian Bayesian state estimation}.
\newblock {\em IEE Proceedings F-Radar and Signal Processing}, 140(2):107--113,
  1993.

\bibitem{Kalman1960}
R.E. Kalman.
\newblock {A new approach to linear filtering and prediction problems}.
\newblock {\em Journal of basic Engineering}, 82(1):35--45, 1960.

\bibitem{Kitagawa1996}
G.~Kitagawa.
\newblock {Monte Carlo filter and smoother for non-Gaussian nonlinear state
  space models}.
\newblock {\em Journal of computational and graphical statistics}, 5(1):1--25,
  1996.

\bibitem{Kong1994}
A.~Kong, J.S. Liu, and W.H. Wong.
\newblock {Sequencial imputations and Bayesian missing data problems}.
\newblock {\em Journal of the American statistical association},
  89(425):278--288, 1994.

\bibitem{Liu2001}
J.~Liu and M.~West.
\newblock {Combined parameter and state estimation in simulation-based
  filtering}.
\newblock In {\em Sequential Monte Carlo Methods in Practice}, pages 197--223.
  Springer, 2001.

\bibitem{Liu1996}
J.S. Liu.
\newblock {Metropolized independent sampling with comparisons to rejection
  sampling and importance sampling}.
\newblock {\em Statistics and Computing}, 6(2):113--119, 1996.

\bibitem{Liu2008}
J.S. Liu.
\newblock {\em {Monte Carlo strategies in scientific computing}}.
\newblock Springer Science \& Business Media, 2002.

\bibitem{Liu1998}
J.S. Liu and R.~Chen.
\newblock {Sequential Monte Carlo methods for dynamic systems}.
\newblock {\em Journal of the American statistical association},
  93(443):1032--1044, 1998.

\bibitem{MacEachern1999}
S.N. Maceachern, M.~Clyde, and J.S. Liu.
\newblock {Sequential importance sampling for nonparametric Bayes models: The
  next generation}.
\newblock {\em Canadian Journal of Statistics}, 27(2):251--267, 1999.

\bibitem{Murray2013}
L.M. Murray, A.~Lee, and P.E. Jacob.
\newblock {Parallel resampling in the particle filter}.
\newblock {\em Journal of Computational and Graphical Statistics}, 2015:21, jan
  2015.

\bibitem{Pitt1999}
M.K. Pitt and N.~Shephard.
\newblock {Filtering via simulation: Auxiliary particle filters}.
\newblock {\em Journal of the American statistical association},
  94(446):590--599, 1999.

\bibitem{Storvik2002a}
G.~Storvik.
\newblock {Particle filters for state-space models with the presence of unknown
  static parameters}.
\newblock {\em IEEE Transactions on signal Processing}, 50(2):281--289, 2002.

\bibitem{West1993}
M.~West.
\newblock {Mixture models, Monte Carlo, Bayesian updating, and dynamic models}.
\newblock {\em Computing Science and Statistics}, pages 325--325, 1993.

\bibitem{west1989Bayfordynmod}
M.~West and J.~Harrison.
\newblock {\em {Bayesian Forecasting and Dynamic Linear Models}}.
\newblock Springer series in statistics. Springer, New York, NY [u.a.], 1997.

\end{thebibliography}

\newpage{}

\pagebreak{}

\clearpage

\begin{appendices}

\section{Algorithms}

\raggedbottom

\begin{algorithm}[H]
	\caption{Auxiliary Particle Filter\label{alg:Auxiliary-Particle-Filter}}
	\begin{algorithmic}
	
	\State initialisation; \For{$t\leftarrow1$ to $k$}
	
	\For {$i\leftarrow1$ to $N_{p}$} 
	
	\State Calculate $\mu_{t}^{(i)}$
	
	\State Calculate $\tilde{w}_{t}^{(i)} \propto p\left(y_{t}|\mu_{t}^{(i)}\right)w_{t-1}^{(i)}$
	
	\EndFor 
	
	\State Normalise weights: $w_{t}^{(i)}=\frac{\tilde{w}_{t}^{(i)}}{\sum_{i=1}^{N_{p}}\tilde{w}_{t}^{(i)}}$ 
	
	\State Resample according to $p\left(j(i)=l\right)=w_{t}$
	(as discussed in Section~\ref{sec:Resampling}).
	
	\For{$i\leftarrow1$ to $N_{p}$}
	
	\State Draw $\boldsymbol{\theta}_{t}^{(i)}\sim p\left(\boldsymbol{\theta}_{t}|\boldsymbol{\theta}_{t-1}^{i^{(j)}}\right)$
	
	\State Calculate $\tilde{w}_{t}^{(i)}=\frac{p\left(y_{t}|\boldsymbol{\theta}_{t}^{(j)}\right)}{p\left(y_{t}|\mu_{t}^{i^{(j)}}\right)}$
	
	\EndFor
	
	\State Normalise weights: $w_{t}^{(i)}=\frac{\tilde{w}_{t}^{(i)}}{\sum_{i=1}^{N_{p}}\tilde{w}_{t}^{(i)}}$ 
	
	\EndFor \end{algorithmic} 
\end{algorithm}

\begin{algorithm}[H]
	\caption{Liu and West\label{alg:Liu-and-West}}
	\begin{algorithmic}
	
	\State initialisation; \For{$t\leftarrow1$ to $k$}
	
	\For {$i\leftarrow1$ to $N_{p}$} 
	
	\State Calculate $\mu_{t}^{(i)}$
	\State Calculate $\boldsymbol{m}_{t-1}$ according to~\ref{eq:LW_moments}
	\State Calculate $V_{t-1}$ according to~\ref{eq:LW-variance}
	\State Calculate $$\tilde{w}_{t}^{(i)}\propto p\left(y_{t}|\mu_{t}^{(i)},\boldsymbol{m}_{t-1}^{(i)}\right)w_{t-1}^{(i)}$$
	
	\EndFor 
	
	\State Normalise weights: $w_{t}^{(i)}=\frac{\tilde{w}_{t}^{(i)}}{\sum_{i=1}^{N_{p}}\tilde{w}_{t}^{(i)}}$ 
	
	\State Resample according to $p\left(j(i)=l\right)=w_{t}$
	(as discussed in Section~\ref{sec:Resampling}).
	
	\For{$i\leftarrow1$ to $N_{p}$}
	
	\State Update parameters: $$\Phi^{(i)}\sim\mathcal{N}\left(\Phi|\boldsymbol{m}_{t-1}^{i^{(j)}},h^2 V_{t-1}\right)$$
	
	\State Draw $\boldsymbol{\theta}_{t}^{(i)}\sim p\left(\boldsymbol{\theta}_{t}|\boldsymbol{\theta}_{t-1}^{i^{(j)}},\Phi^{(i)}\right)$
	
	\State Calculate $\tilde{w}_{t}^{(i)}=\frac{p\left(y_{t}|\boldsymbol{\theta}_{t}^{(j)},\Phi^{(i)}\right)}{p\left(y_{t}|\mu_{t}^{i^{(j)}},\boldsymbol{m}_{t-1}^{i^{(j)}}\right)}$
	
	\EndFor
	
	\State Normalise weights: $w_{t}^{(i)}=\frac{\tilde{w}_{t}^{(i)}}{\sum_{i=1}^{N_{p}}\tilde{w}_{t}^{(i)}}$ 
	
	\EndFor \end{algorithmic} 
\end{algorithm}

\begin{algorithm}[H]
	\caption{Storvik\label{alg:Storvik's-algorithm}}
	\begin{algorithmic}[1]
	
	\State initialisation; 
	
	\For{$t\leftarrow1$ to $k$}
	
	\For {$i\leftarrow1$ to $N_{p}$} 
	
	\State Sample $\Phi^{(i)}_t\sim p\left(\Phi|s_t^{(i)}\right)$
	
	\State Sample $\tilde{\boldsymbol{\theta}}_{t}^{(i)}\sim p\left(\boldsymbol{\theta}_{t}|\boldsymbol{\theta}_{0:t-1}^{(i)},y_{t},\Phi^{(i)}\right)$
	
	\State Calculate weights: 
		$$\tilde{w}_t \propto p\left(y_t|\theta_t^{(i)},\Phi_t^{(i)}\right)$$

	\EndFor 
	
	\State Normalise weights: $w_{t}^{(i)}=\frac{\tilde{w}_{t}^{(i)}}{\sum_{i=1}^{N_{p}}\tilde{w}_{t}^{(i)}}$ 
	
	\State Resample $\left\lbrace\theta_t^{(i)}, \Phi_t^{(i)}, s_t^{(i)}\right\rbrace_{i=1}^{N_p}$ according to $p\left(j(i)=l\right)=w_{t}$
	(as discussed in Section~\ref{sec:Resampling})
	
	\For{$i\leftarrow1$ to $N_{p}$}
	
	\State Update sufficient statistics $$s_{t}^{(i)}=\mathcal{S}\left(s_{t-1}^{\left(i^{(j)}\right)},\theta_{t},\theta_{t-1},y_{t}\right)$$
	
	\EndFor
	
	\EndFor \end{algorithmic} 
\end{algorithm}

\begin{algorithm}[H]
	\caption{Particle Learning\label{alg:Particle-Learning}}
	\begin{algorithmic}[1]
	
	\State initialisation; 
	
	\For{$t\leftarrow1$ to $k$}
	
	\For {$i\leftarrow1$ to $N_{p}$} 
	
	\State Calculate $\mu_{t}^{(i)}|\theta_{t}^{(i)},s_{t}^{(i)},\Phi_{t}^{(i)}$
	
	\State Calculate $\tilde{w}_{t}^{(i)}\propto p\left(y_{t+1}|\mu_{t}^{(i)}\right)$
	
	\EndFor 
	
	\State Normalise weights: $w_{t}^{(i)}=\frac{\tilde{w}_{t}^{(i)}}{\sum_{i=1}^{N_{p}}\tilde{w}_{t}^{(i)}}$ 
	
	\State Resample according to $p\left(j(i)=l\right)=w_{t}$
	
	\For{$i\leftarrow1$ to $N_{p}$}
	
	\State Draw $\boldsymbol{\theta}_{t+1}^{(i)}\sim p\left(\boldsymbol{\theta}_{t+1}|\mu_{t}^{(i)},y_{t+1}\right)$
	
	\State Update the sufficient statistics $$s_{t+1}^{(i)}=\mathcal{S}\left(s_{t}^{i^{(j)}},\boldsymbol{\theta}_{t+1},\boldsymbol{\theta}_{t},y_{t+1}\right)$$
	
	\State Draw $\Phi_{t+1}^{(i)}\sim p\left(\Phi_{t+1}|s_{t+1}^{(i)}\right)$
	
	\State Calculate $\tilde{w}_{t}^{(i)}=\frac{p\left(y_{t+1}|\boldsymbol{\theta}_{t+1}^{j}\right)}{p\left(y_{t+1}|\mu_{t+1}^{i^{(j)}}\right)}$
	
	\EndFor
	
	\State Normalise weights: $w_{t}^{(i)}=\frac{\tilde{w}_{t}^{(i)}}{\sum_{i=1}^{N_{p}}w_{t}^{(i)}}$ 
	
	\EndFor \end{algorithmic} 
\end{algorithm}

\clearpage

\section{One step-ahead forecast}

\begin{figure}[H]
	\centering
	\includegraphics[width=1\columnwidth]{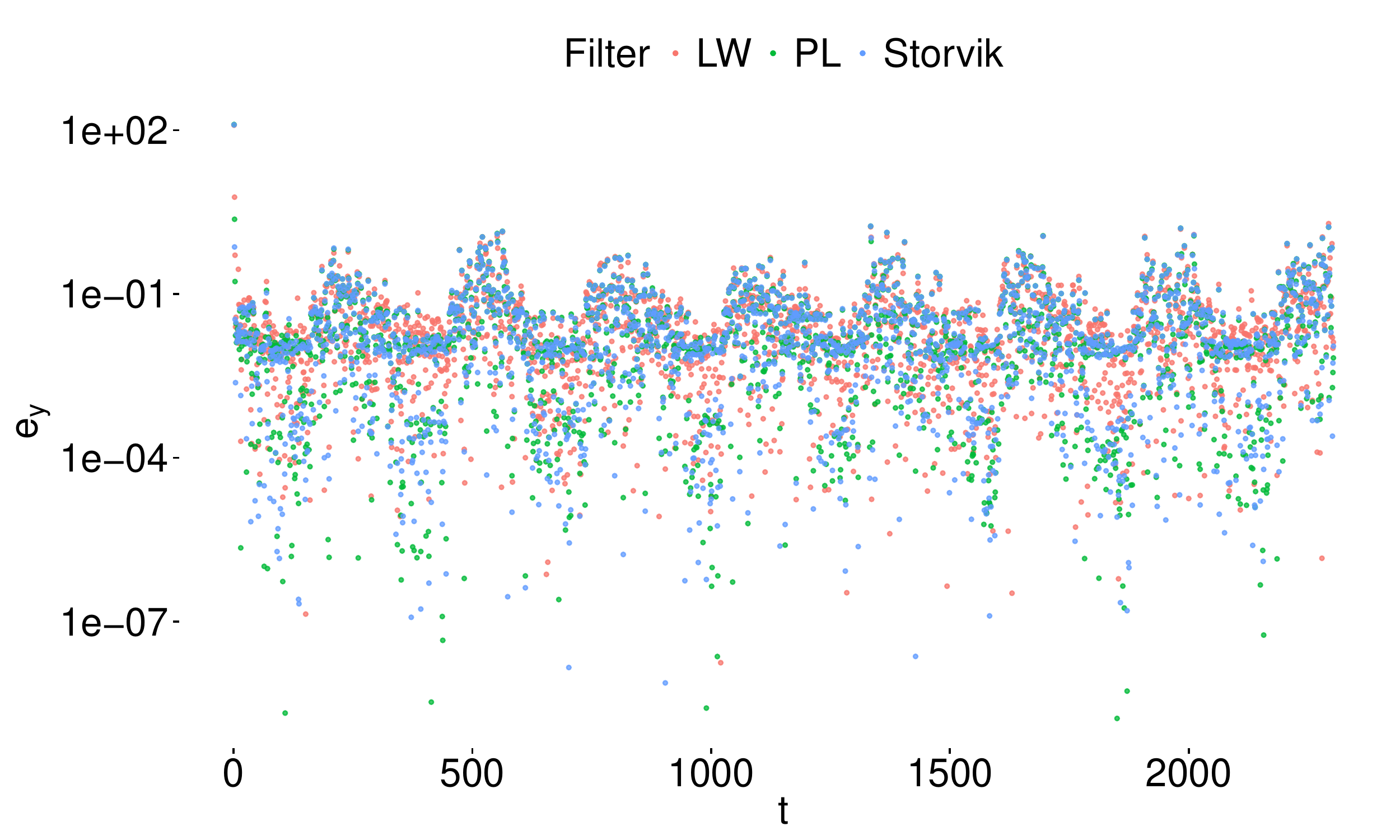}
	\caption{Temperature data one-step ahead observation forecast errors.\label{fig:Temperature-data-one-step}}
\end{figure}

\begin{figure}[H]
	\centering
	\includegraphics[width=1\columnwidth]{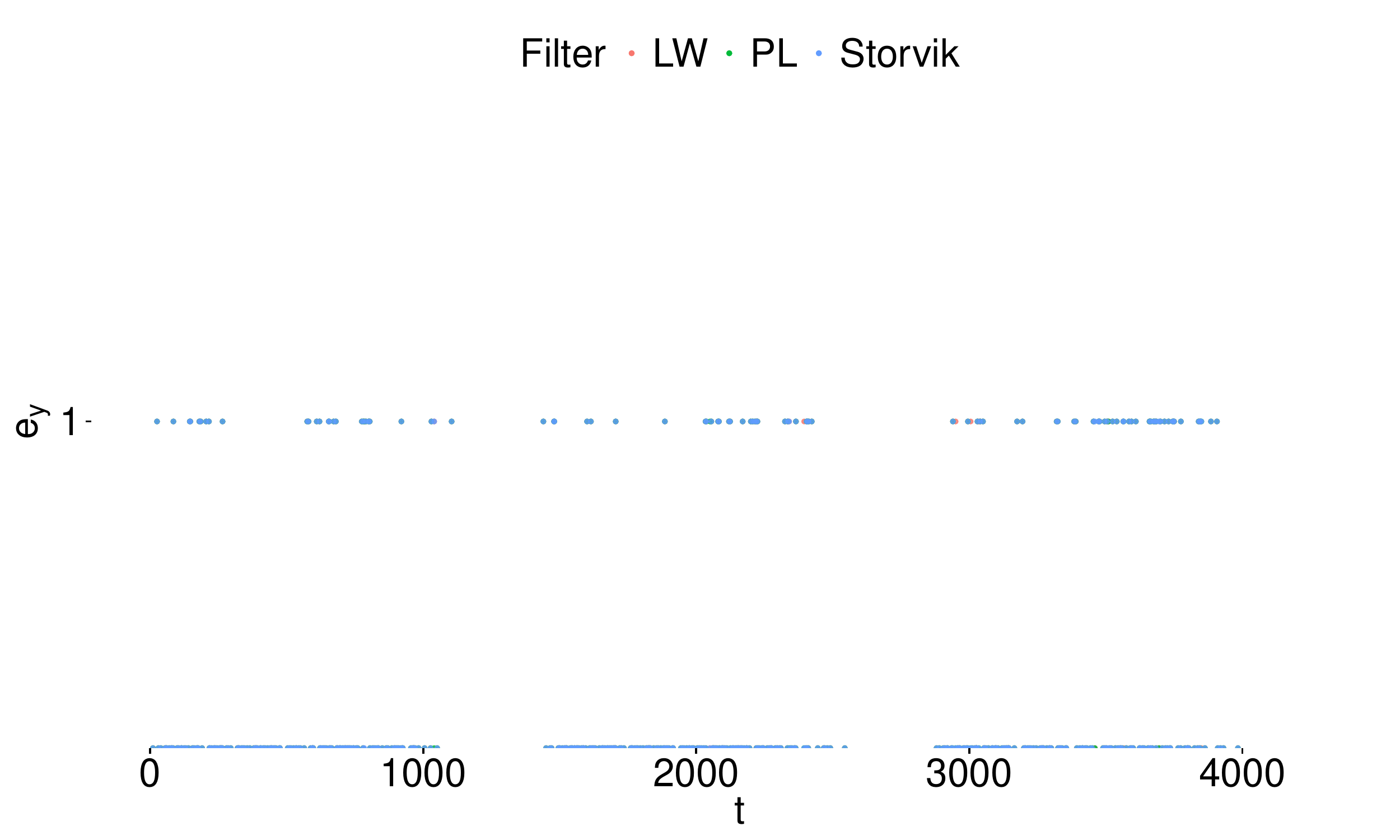}
	\caption{Airport data one-step ahead observation forecast errors.\label{fig:airport-data-one-step}}
\end{figure}

\begin{figure}[H]
	\centering
	\includegraphics[width=1\columnwidth]{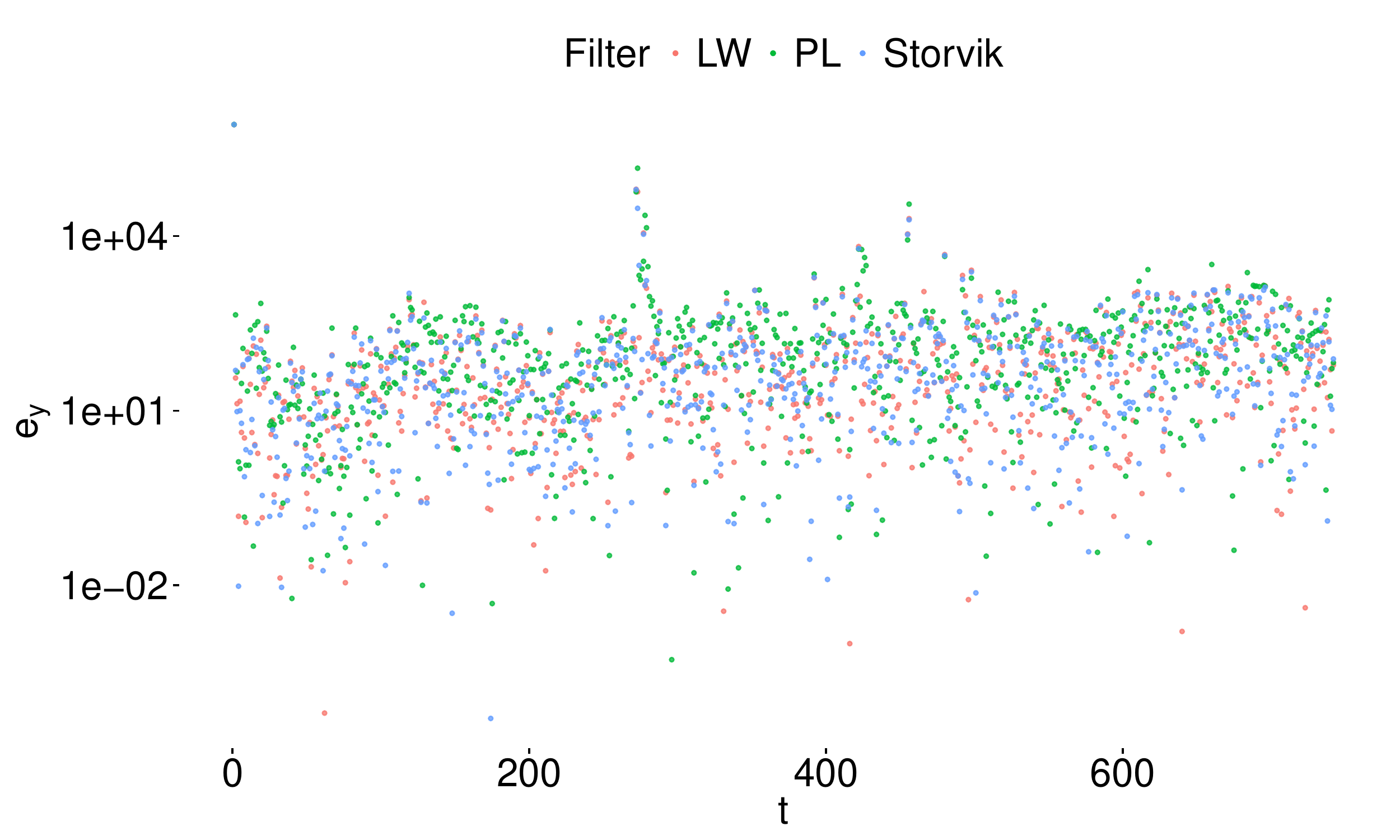}
	\caption{WC98 data one-step ahead observation forecast errors.\label{fig:wc98-data-one-step}}
\end{figure}

\clearpage

\section{State forecast}

\begin{figure}[H]
	\centering
	\includegraphics[width=0.5\columnwidth]{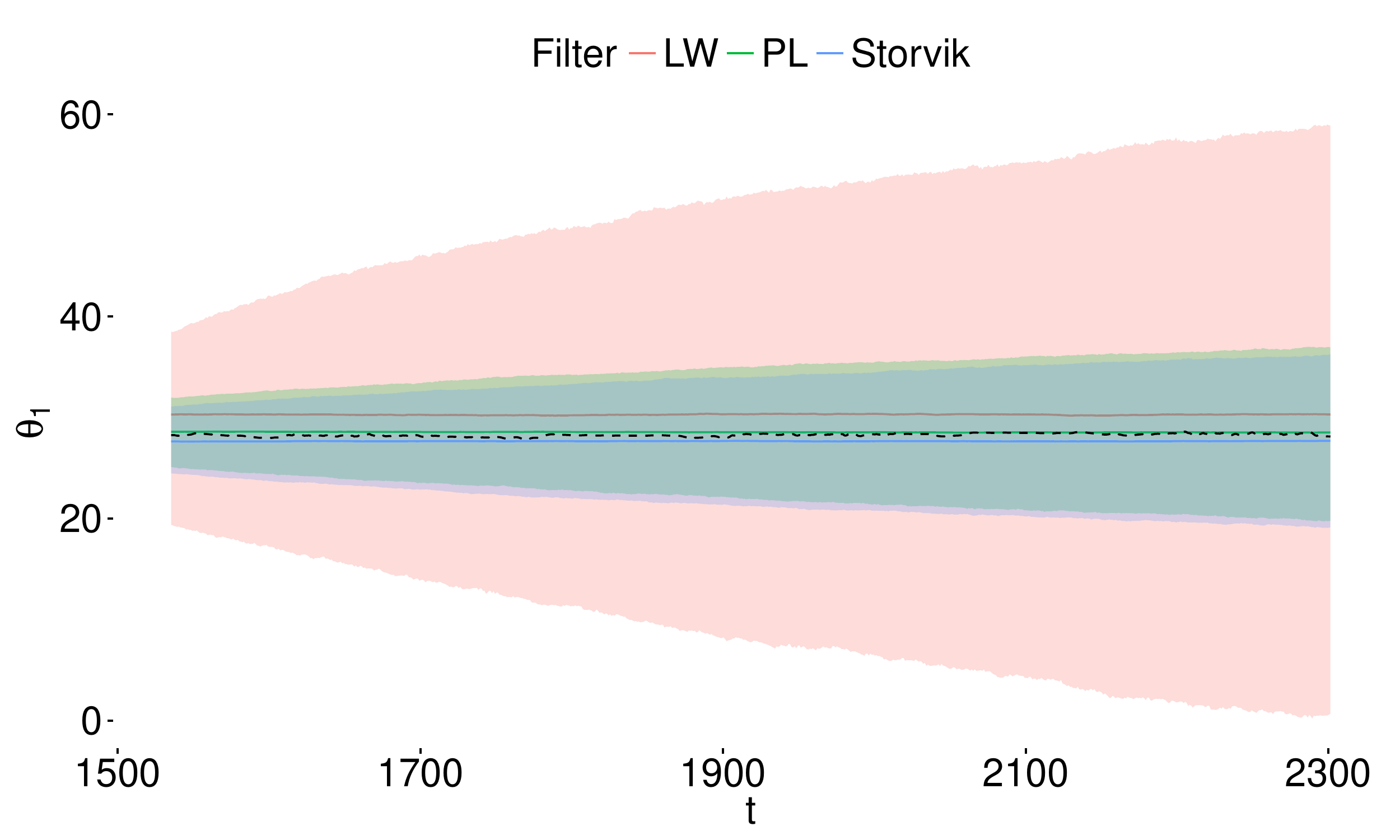}\includegraphics[width=0.5\columnwidth]{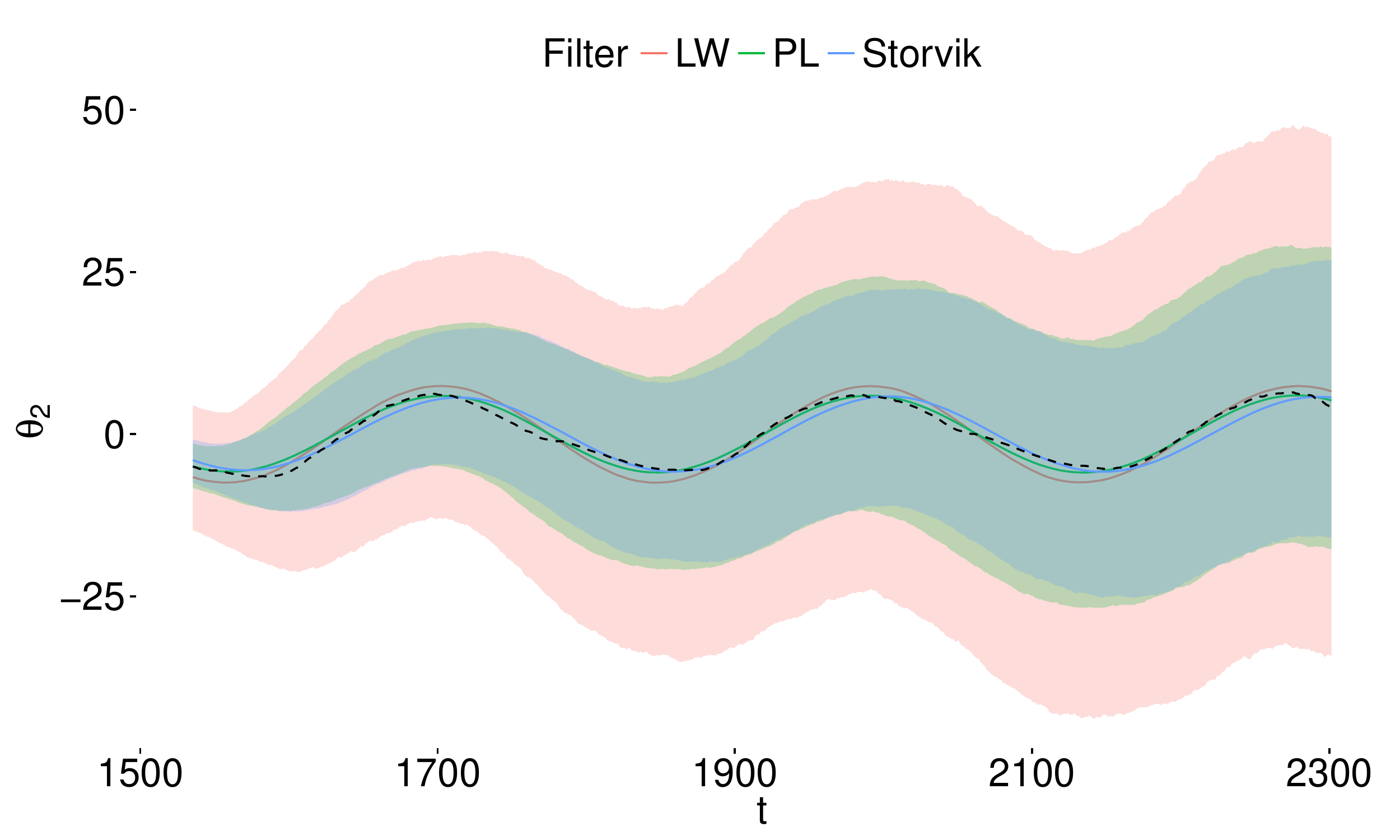}
	\raggedright\includegraphics[width=0.5\columnwidth]{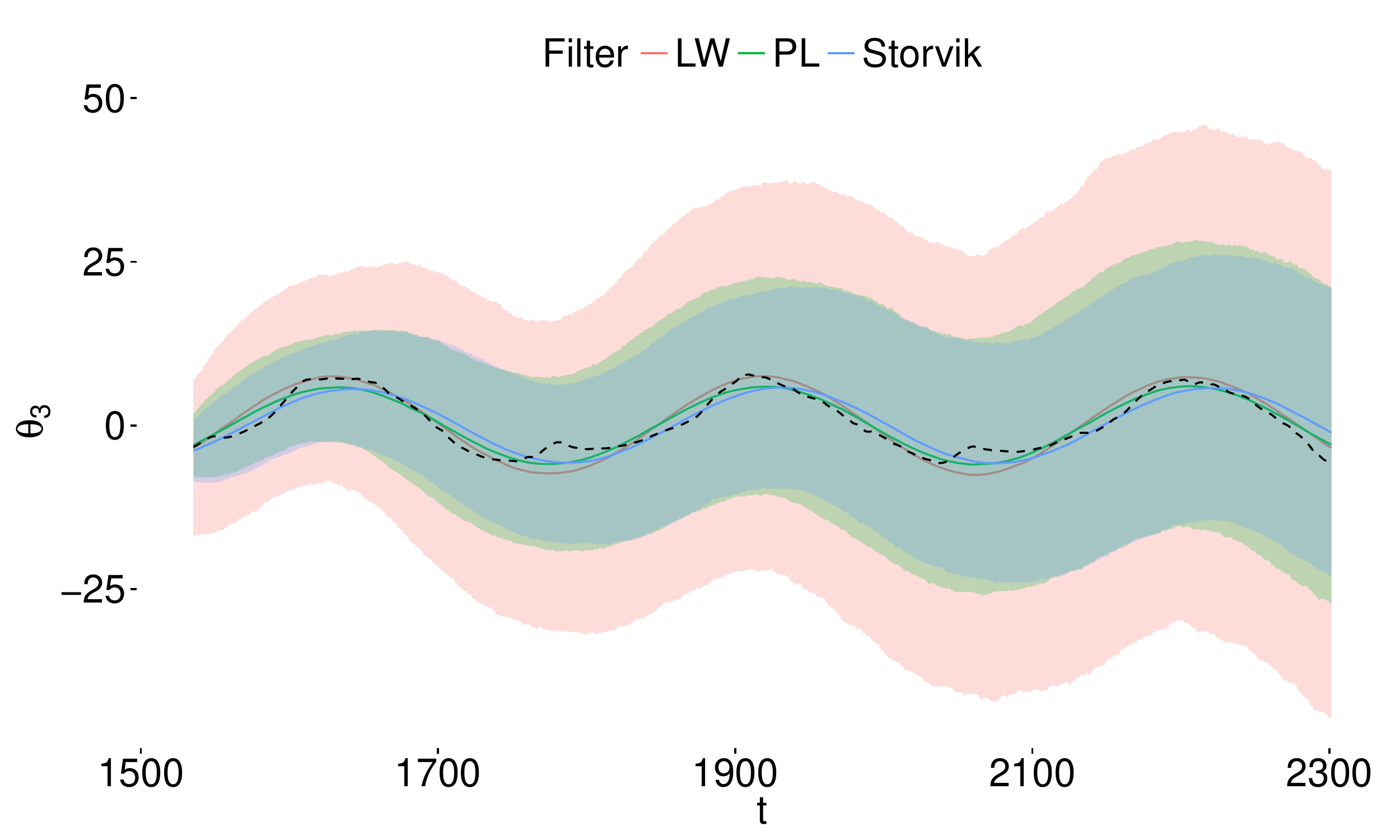}
	\caption{State forecast for the temperature data\label{fig:state-forecast-for-temperature}}
\end{figure}

\begin{figure}[H]
	\centering
	\includegraphics[width=0.5\columnwidth]{airport_state_forecast_0}\includegraphics[width=0.5\columnwidth]{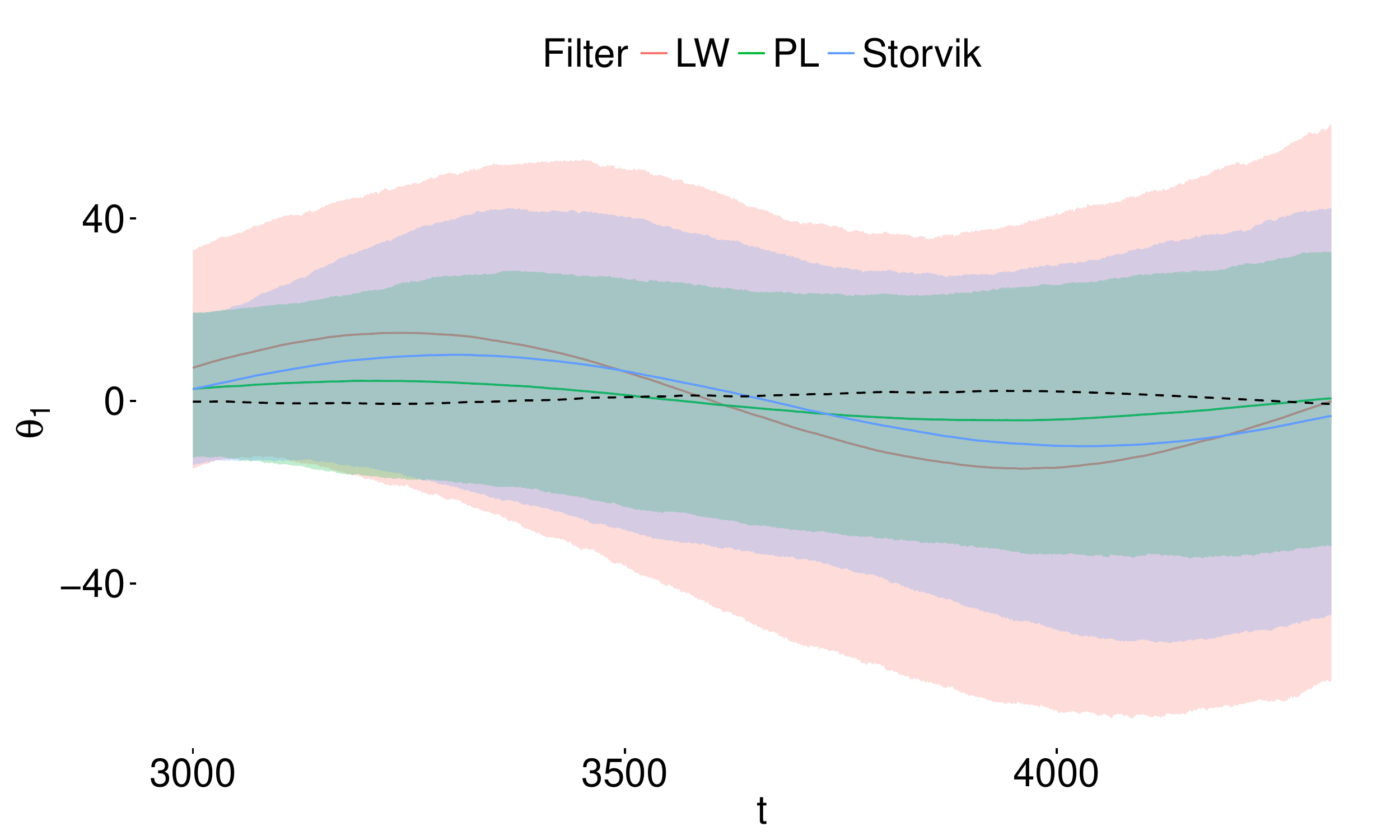}
	\raggedright\includegraphics[width=0.5\columnwidth]{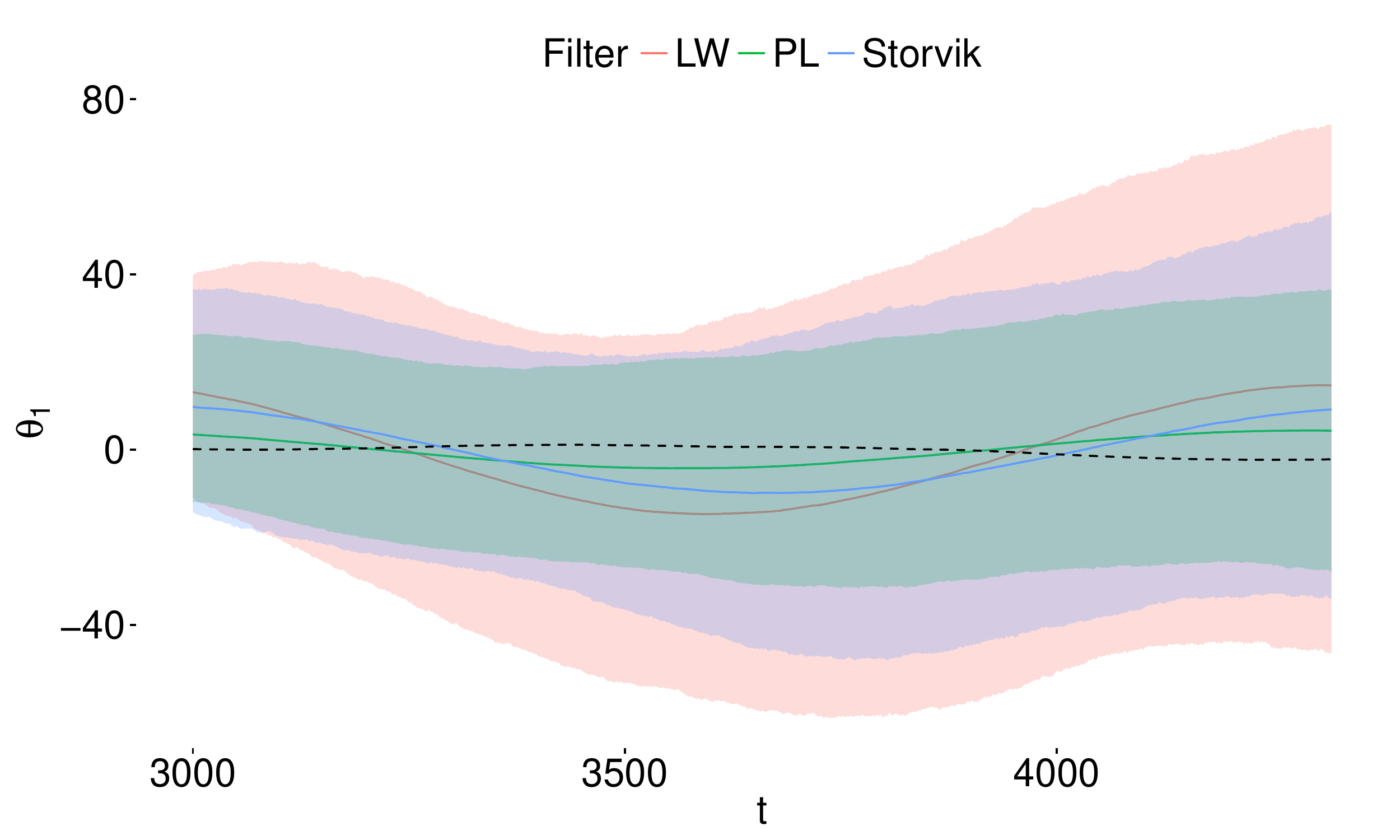}
	\caption{State forecast for the airport data\label{fig:state-forecast-for-airport}}
\end{figure}

\begin{figure}[H]
	\centering
	\includegraphics[width=0.5\columnwidth]{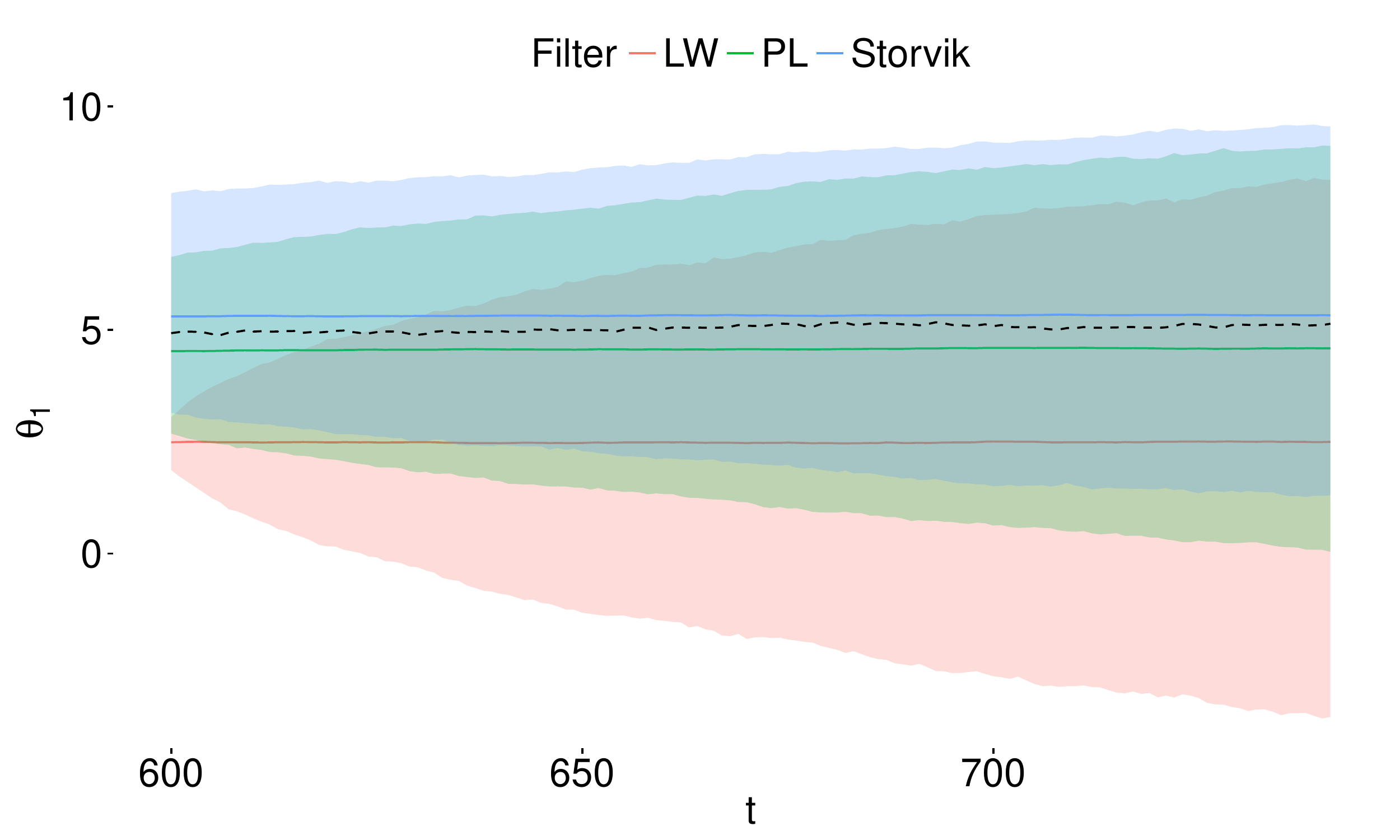}\includegraphics[width=0.5\columnwidth]{WC98_poisson_weekly_state_forecast_1}
	\includegraphics[width=0.5\columnwidth]{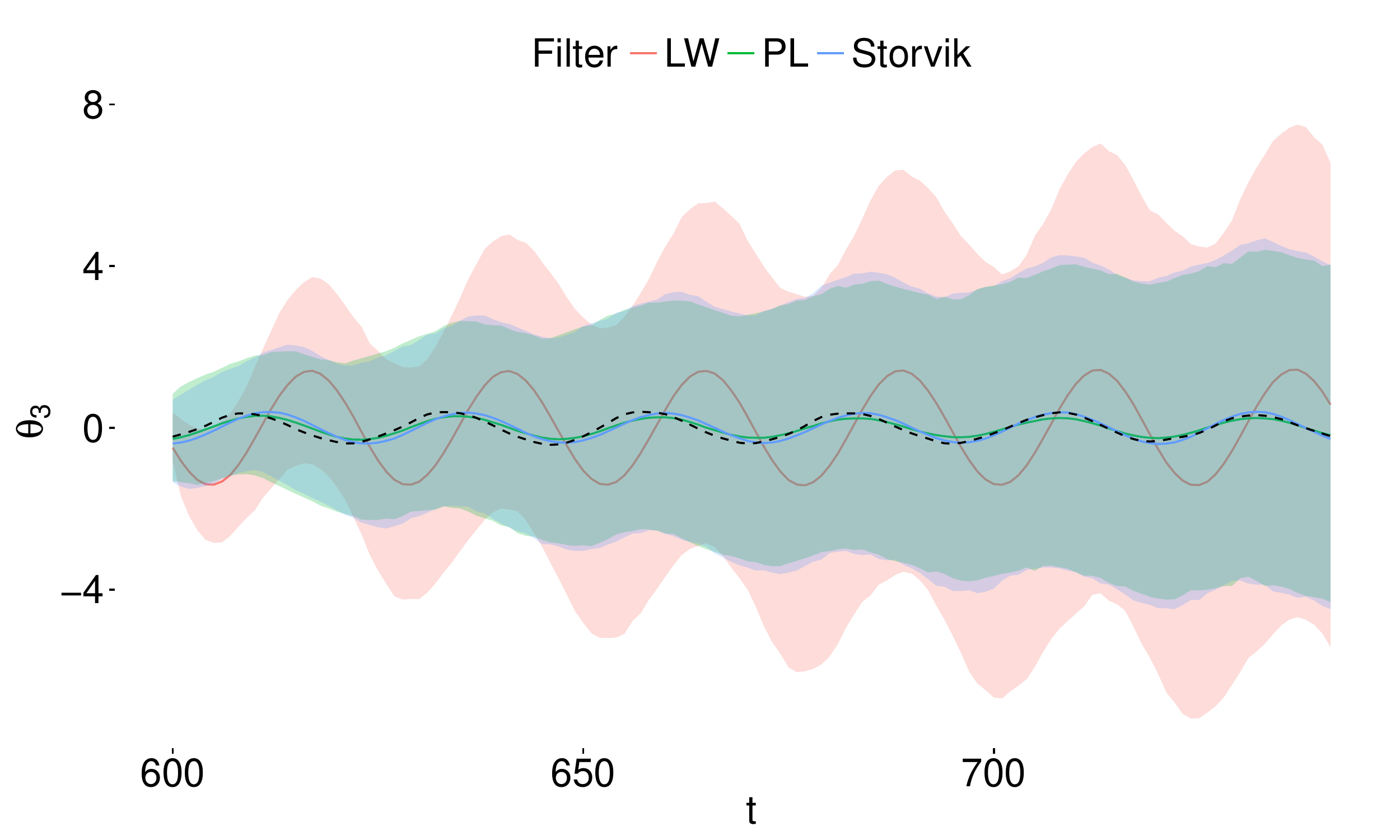}\includegraphics[width=0.5\columnwidth]{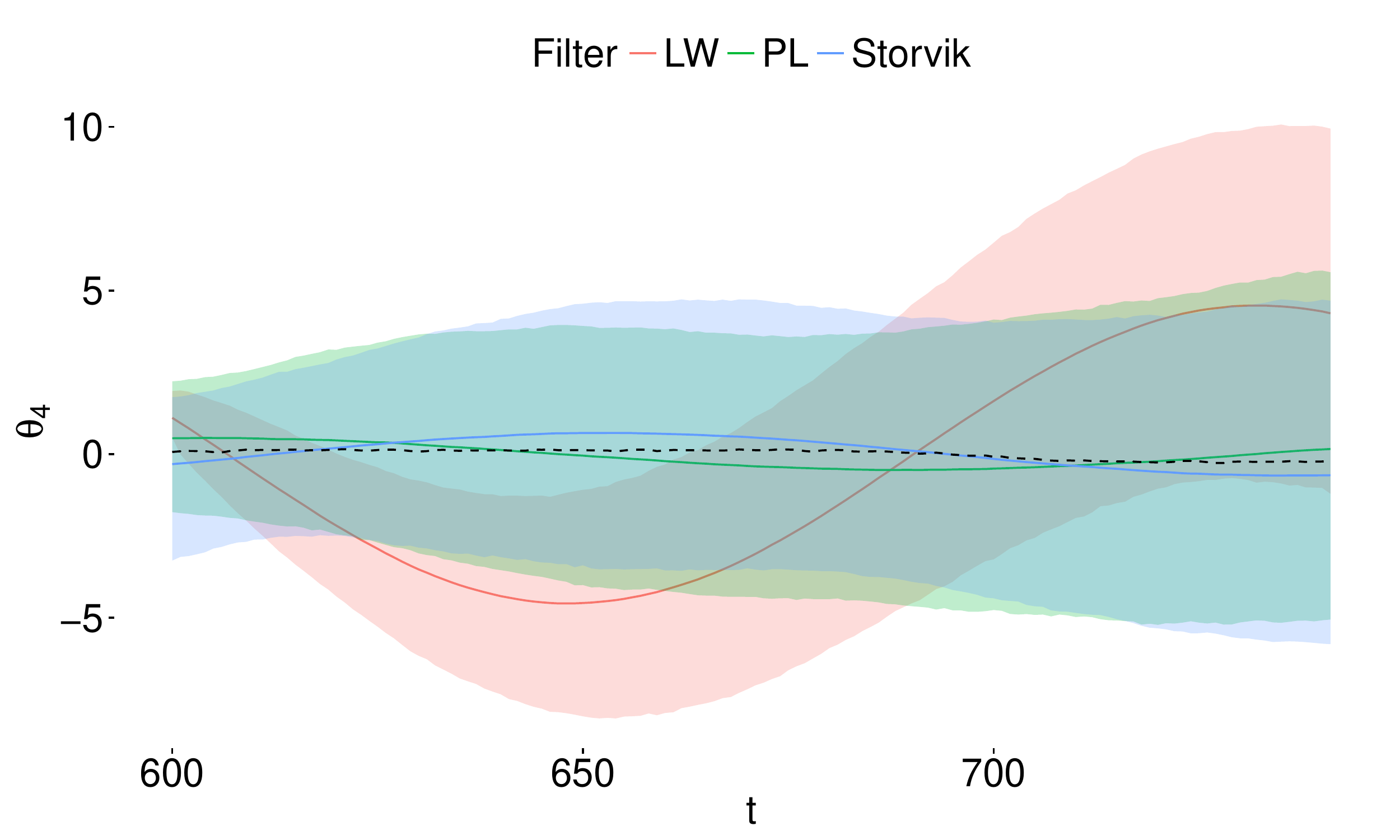}
	\raggedright\includegraphics[width=0.5\columnwidth]{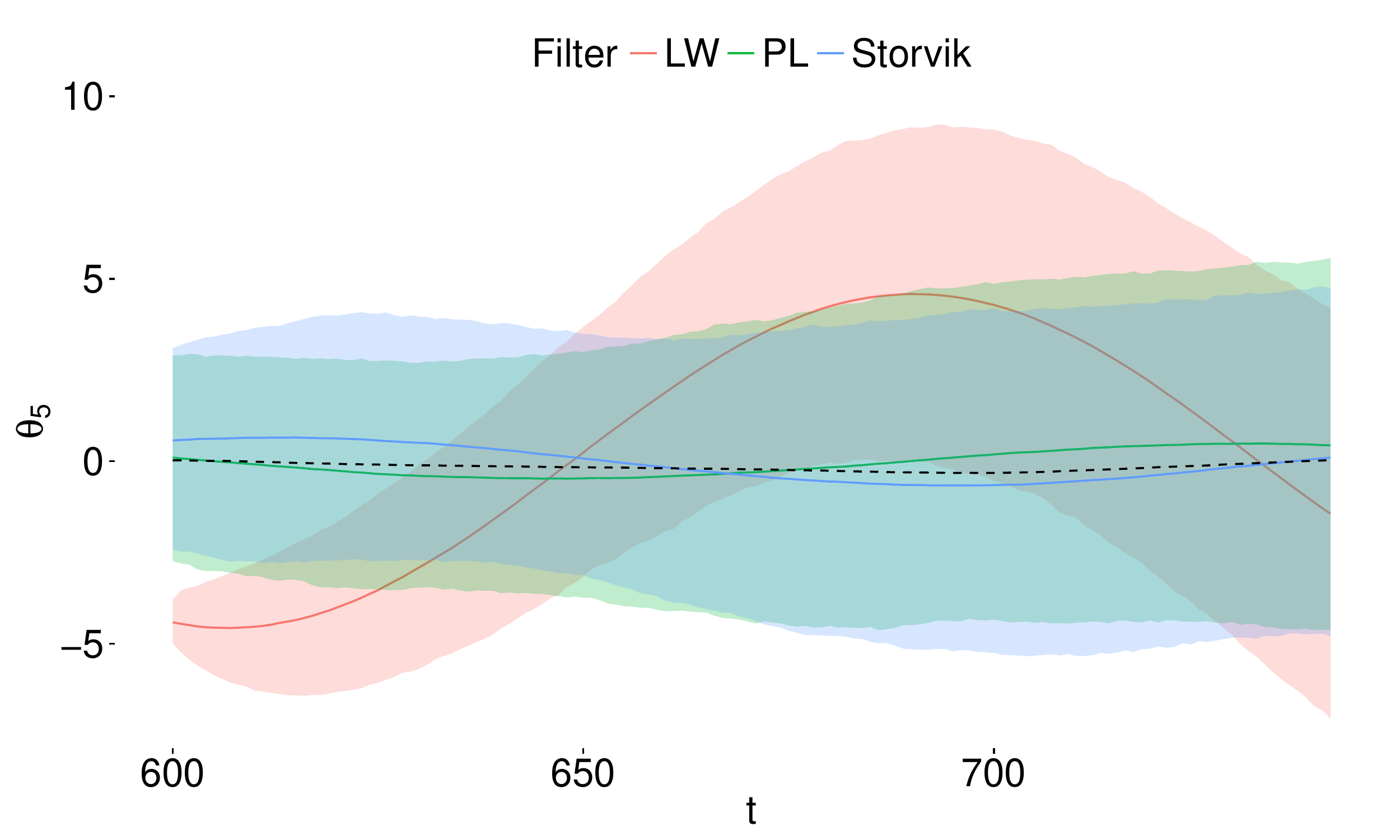}
	\caption{State forecast for the WC98 data\label{fig:state-forecast-for-wc98}}
\end{figure}

\clearpage

\section{Execution time}

\begin{figure}[H]
	\centering
	\includegraphics[width=1\columnwidth]{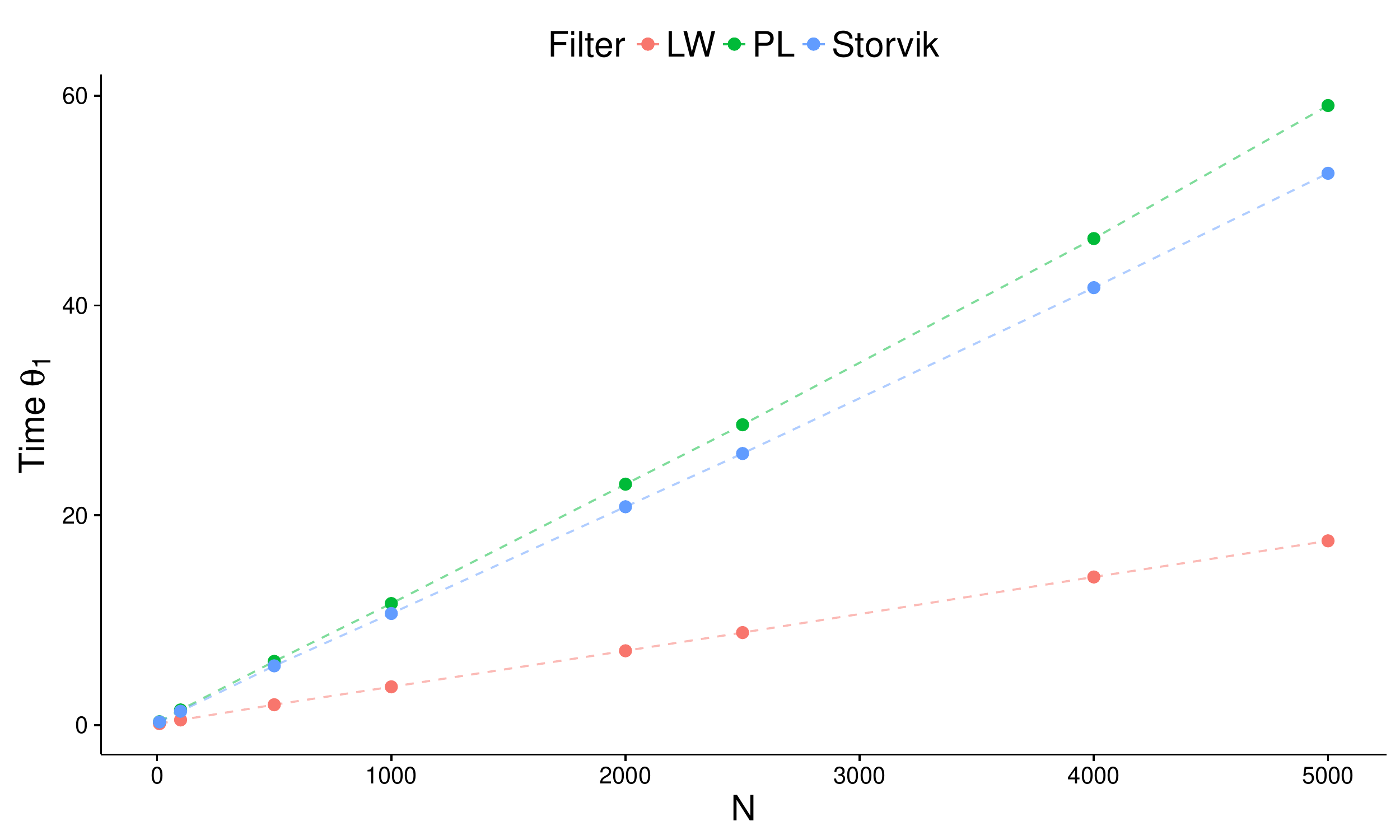}
	\caption{LW, Storvik and PL execution time (seconds) for the temperature dataset using a Normal DLM with a varying number of particles.\label{fig:temperature-n-time}}
\end{figure}

\begin{figure}[H]
	\centering
	\includegraphics[width=1\columnwidth]{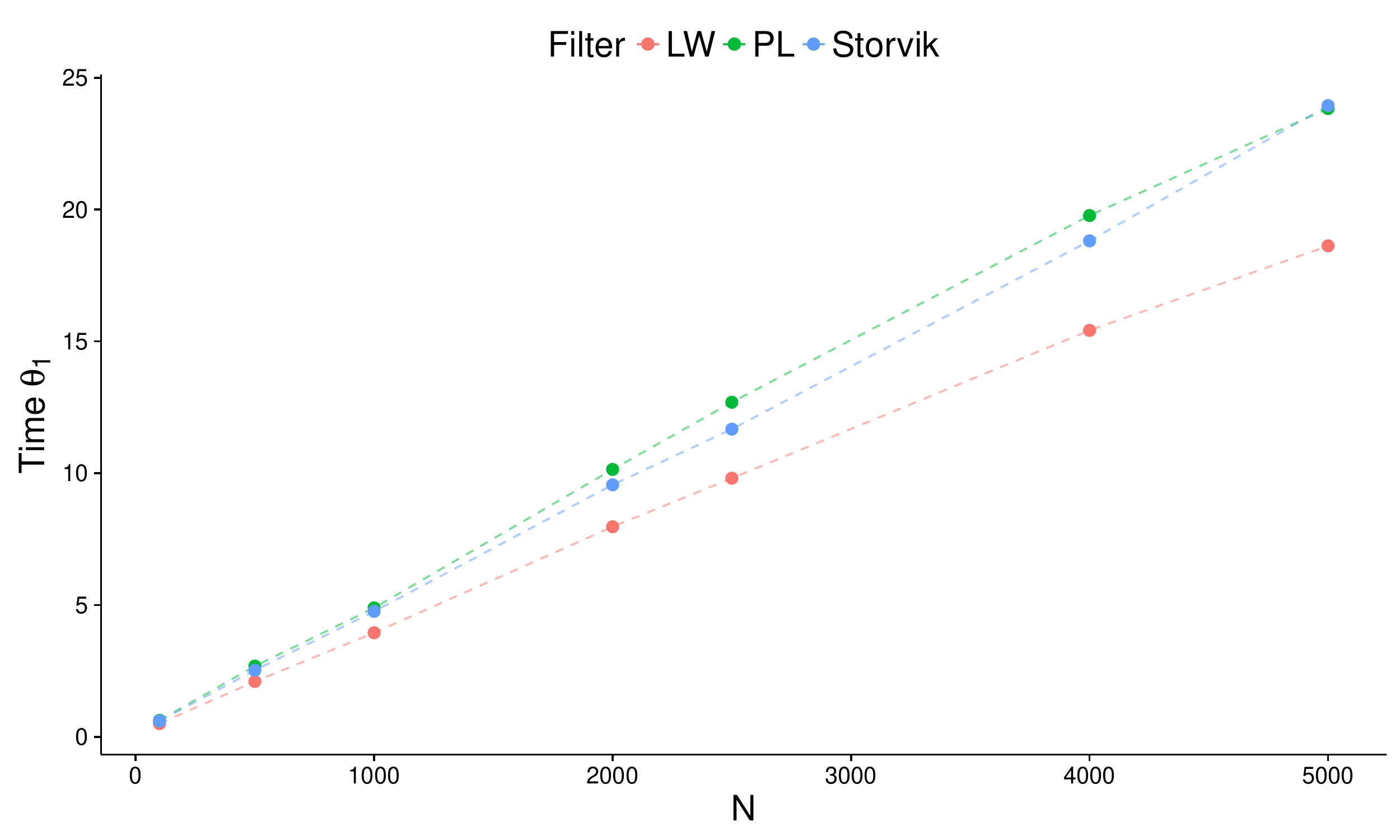}
	\caption{LW, Storvik and PL execution time (seconds) for the airport dataset using a Binomial DLM with a varying number of particles.\label{fig:airport-n-time}}
\end{figure}

\begin{figure}[H]
	\centering
	\includegraphics[width=1\columnwidth]{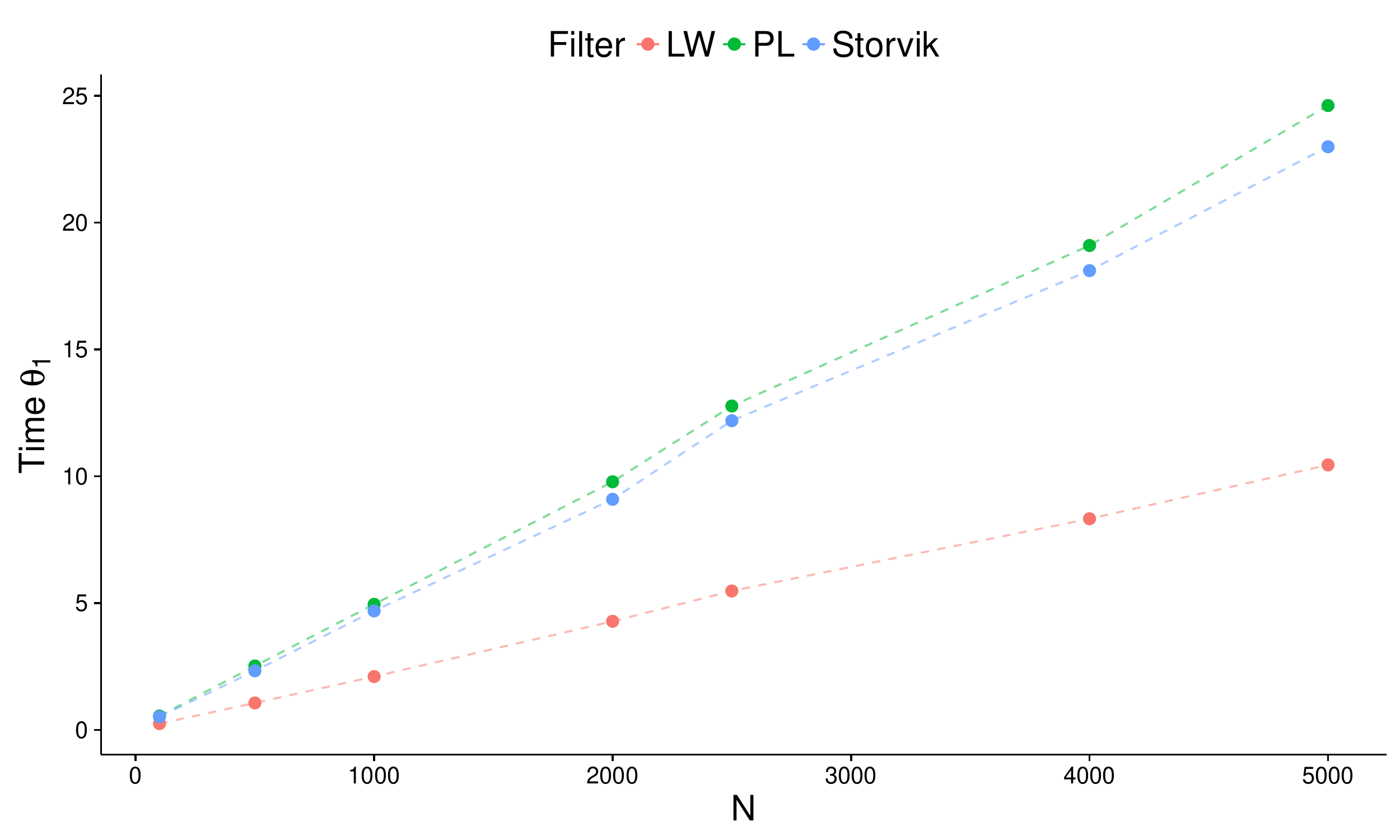}
	\caption{LW, Storvik and PL execution time (seconds) for the WC98 dataset using a Poisson DLM with a varying number of particles.\label{fig:wc98-n-time}}
\end{figure}

\end{appendices}

\end{document}